\documentclass[journal]{IEEEtran}

\ifCLASSINFOpdf
   \usepackage[pdftex]{graphicx}
   \DeclareGraphicsExtensions{.pdf,.jpeg,.png}
\else
   \usepackage[dvips]{graphicx}
   \DeclareGraphicsExtensions{.eps}
\fi

\usepackage{color}
\usepackage{cite}
\usepackage{amsmath}
\usepackage{bm}
\usepackage{amssymb}
\usepackage{array}
\usepackage{multirow}
\usepackage{makecell}
\usepackage{stfloats}
\usepackage{url}
\usepackage{booktabs}
\usepackage{color}
\usepackage{footmisc}
\usepackage{subfigure}

\newcommand{\tabincell}[2]{\begin{tabular}{@{}#1@{}}#2\end{tabular}}

\usepackage[colorlinks,
            urlcolor=red,
            hyperfootnotes = True,
            citecolor=green]{hyperref}
\def\1{\mathbf{1}}
\def\0{\mathbf{0}}

\def\mB{\mathcal{B}}
\def\mM{\mathcal{M}}

\def\mK{\mathcal{K}}
\def\mD{\mathcal{D}}

\def\mO{\mathcal{O}}
\def\mG{\mathcal{G}}

\def\mR{\mathcal{R}}

\newcommand{\PreserveBackslash}[1]{\let\temp=\\#1\let\\=\temp}
\newcolumntype{C}[1]{>{\PreserveBackslash\centering}p{#1}}
\newcolumntype{R}[1]{>{\PreserveBackslash\raggedleft}p{#1}}
\newcolumntype{L}[1]{>{\PreserveBackslash\raggedright}p{#1}}

\begin{document}
%

\title{RCDNet: An Interpretable Rain Convolutional Dictionary Network for Single Image Deraining}

\author{Hong~Wang,
        Qi~Xie,
        Qian~Zhao,
        Yuexiang~Li,
        Yong~Liang,\\
        Yefeng Zheng,~\IEEEmembership{Fellow,~IEEE},
        Deyu~Meng,~\IEEEmembership{Member,~IEEE}

\thanks{H. Wang, Y. Li, and Y. Zheng are with Tencent Jarvis Lab, Shenzhen, P.R. China. E-mail: \{hazelhwang, vicyxli, yefengzheng\}@tencent.com.}

\thanks{Q. Xie, Q. Zhao, and D. Meng are with School of Mathematics and Statistics and Ministry of Education Key Lab of Intelligent Networks and Network Security, Xi'an Jiaotong University, Shaanxi, P.R. China. D. Meng is also with Peng Cheng Laboratory, Shenzhen, China, and Macao Institute of Systems Engineering, Macau University of Science and Technology, Taipa, Macao. E-mail: \{xie.qi, timmy.zhaoqian, dymeng\}@mail.xjtu.edu.cn.}

\thanks{Y. Liang is with the Faculty of Information Technology, Macau University of Science and Technology, Macao. E-mail: yliang@must.edu.mo.}
\thanks{Q. Xie and D. Meng are corresponding authors.}
\thanks{This research is supported by the National Key R\&D Program of China (2022YFA1004100), The Major Key Project of PCL (PCL2021A12), the Macao Science and Technology Development Fund under Grant 061/2020/A2, and the China NSFC projects under contract 61721002, U1811461, 62206214.}
}

\markboth{Journal of \LaTeX\ Class Files,~Vol.~14, No.~8, August~2015}%
{Shell \MakeLowercase{\textit{et al.}}: Bare Demo of IEEEtran.cls for IEEE Journals}

\maketitle
\begin{abstract}
 As a common weather, rain streaks adversely degrade the image quality and tend to negatively affect the performance of outdoor computer vision systems. Hence, removing rains from an image has become an important issue in the field. To handle such an ill-posed single image deraining task, in this paper, we specifically build a novel deep architecture, called rain convolutional dictionary network (RCDNet), which embeds the intrinsic priors of rain streaks and has clear interpretability. In specific, we first establish a RCD model for representing rain streaks and utilize the proximal gradient descent technique to design an iterative algorithm only containing simple operators for solving the model. By unfolding it, we then build the RCDNet in which every network module has clear physical meanings and corresponds to each operation involved in the algorithm. This good interpretability greatly facilitates an easy visualization and analysis on what happens inside the network and why it works well in inference process. Moreover, taking into account the domain gap issue in real scenarios, we further design a novel dynamic RCDNet, where the rain kernels can be dynamically inferred  corresponding to input rainy images and then help shrink the space for rain layer estimation with few rain maps so as to ensure a fine generalization performance in the inconsistent scenarios of rain types between training and testing data. By end-to-end training such an interpretable network, all involved rain kernels and proximal operators can be automatically extracted, faithfully characterizing the features of both rain and clean background layers, and thus naturally leading to better deraining performance. Comprehensive experiments implemented on a series of representative synthetic and real datasets substantiate the superiority of our method, especially on its well generality to diverse testing scenarios and good interpretability for all its modules, as compared with state-of-the-art single image derainers both visually and quantitatively. Code is available at \emph{\url{https://github.com/hongwang01/DRCDNet}}.
\end{abstract}
\begin{IEEEkeywords}
Single image rain removal, dictionary learning, interpretable deep learning, generalization performance.
\end{IEEEkeywords}
%
\vspace{-0mm}\section{Introduction}\label{sec:introduction}
\IEEEPARstart{I}{mages} and videos captured in rainy scenes always suffer from noticeable visual degradations, tending to adversely affect outdoor computer vision tasks, such as automatic driving and video surveillance~\cite{Shehata2008Video}. As a hot research topic, rain removal from images and videos has brought considerable attention to the research community~\cite{wang2019a, wang2020single, benchmark}. In this work, we focus on the single image deraining task.

\begin{figure}[t]
  \begin{center}
     \includegraphics[width=1\linewidth]{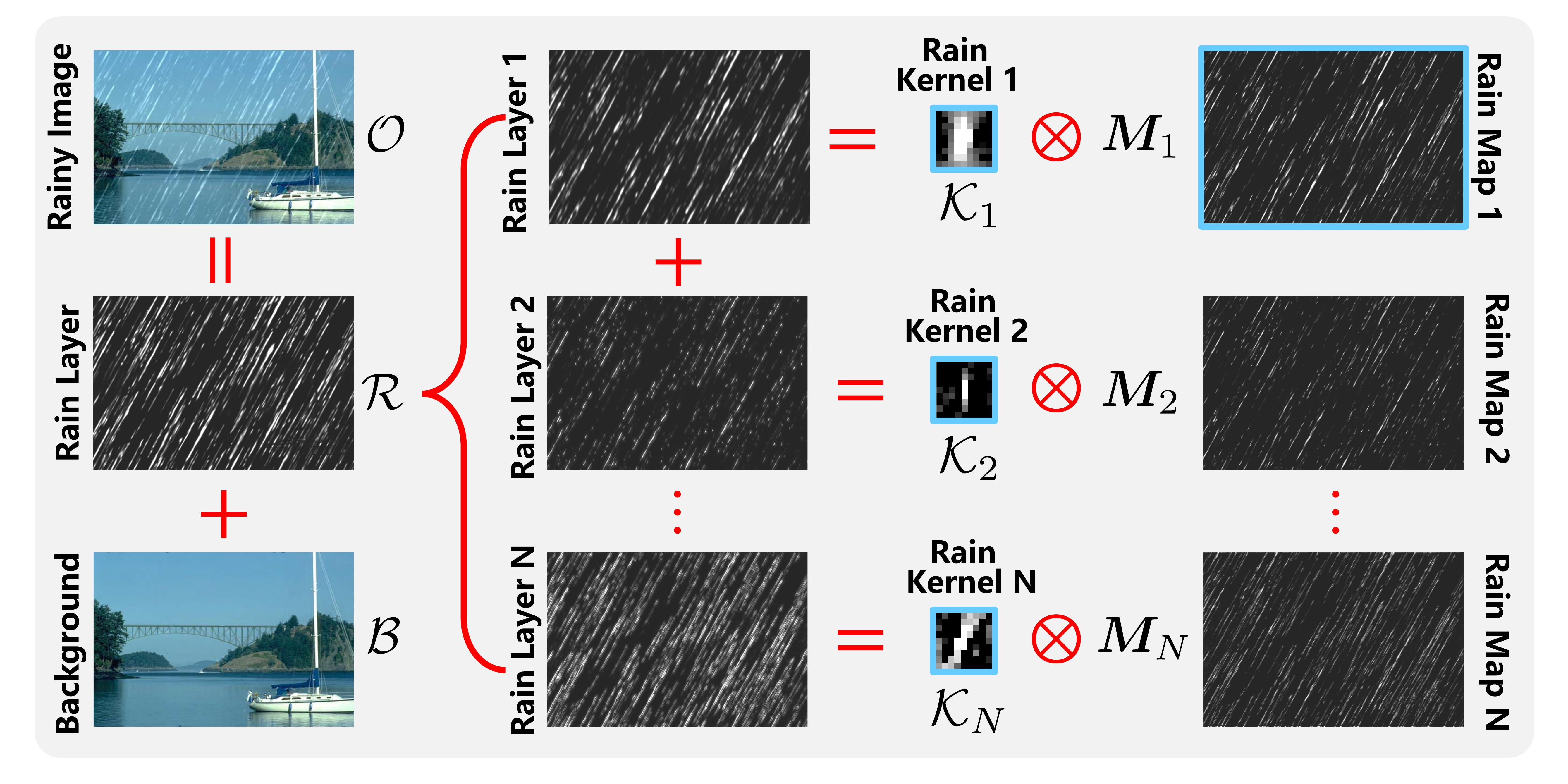}
  \end{center}
  \vspace{-6mm}
     \caption{Convolutional dictionary learning mechanism for rain layer.}
  \label{figcsc}
   \vspace{-6mm}
  \end{figure}

Recent years have witnessed significant progress in single image deraining, which can be mainly categorized into two research lines. One is the traditional unsupervised (i.e., prior-based) method, which focuses on exploring the prior structures of background and rain layers to constrain the solution space of a carefully designed optimization model. Typically presented priors include frequency information~\cite{Kang2012Automatic}, sparse representation~\cite{Yu2015Removing,Gu2017Joint,Zhu2017Joint,He2017Convolutional}, and local patch low-rankness~\cite{He2017Convolutional}. Very recently, researchers explore that rain streaks repeatedly appear at different locations over a rainy image with similar local patterns, like shape, thickness, and direction~\cite{He2017Convolutional,Huang2015Convolutional}. They formulate such an understanding (i.e., non-local self-similarity) as a convolutional dictionary learning model, where rain kernels are imposed on sparse rain maps, as intuitively depicted in Fig.~\ref{figcsc}. This idea has achieved state-of-the-art (SOTA) performance in video deraining when background frames are well extracted based on the temporal information and low-rankness prior in surveillance video sequences~\cite{li2018video}. Albeit effective in certain specific scenarios, the rationality of these conventional deraining approaches largely depend on the reliability of such manually designed prior assumptions on the unknown background and rain streaks. However, with subjective and relatively simple forms, these hand-crafted priors cannot always comprehensively and adaptively reflect the complex and variant structures underlying real rainy images collected from different resources.
  \begin{figure}[t]
  \begin{center}
  \vspace{-1.5mm}
     \includegraphics[width=1\linewidth]{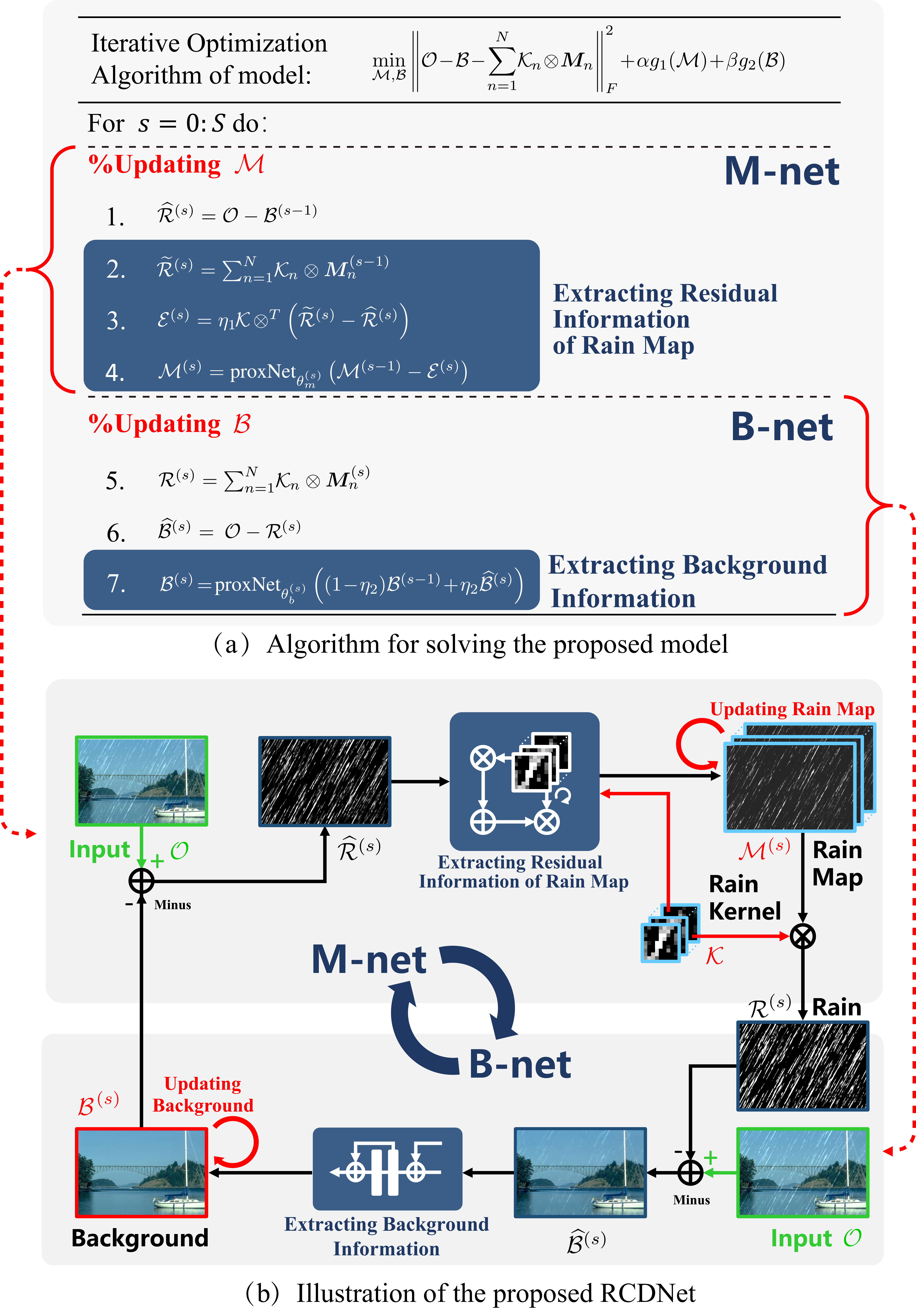}
  \end{center}
  \vspace{-5mm}
     \caption{
     (a) Rain convolutional dictionary (RCD) model and the proposed iterative solution algorithm.
     (b) Visual illustration of the proposed RCDNet corresponding to the algorithm in (a).}
  \label{figintro}
   \vspace{-5mm}
  \end{figure}

The other popular approach on this task is based on deep learning (DL). The main idea of current deep derainers is to utilize the pre-collected training samples to learn the mapping function from a rainy image to its corresponding rain-removed background layer with diverse network architectures, including CNN~\cite{Fu2017Clearing,Fu2017Removing, wang2020rain}, adversarial learning~\cite{zhang2018density,zhang2019image,wei2019deraincyclegan}, recurrent and multi-stage networks~\cite{li2018recurrent,ren2019progressive,Yang2019Joint}, multi-scale fusion architectures~\cite{fu2019lightweight,zheng2019residual,yasarla2019uncertainty, jiang2020multi, zheng2020single}, spatial attentive unit (SPANet)~\cite{wang2019spatial}, encoder-decoder network~\cite{Li2018Non,wang2019erl,hu2019depth}, and complementary sub-networks~\cite{Pan2018Learning,deng2020detail}. Due to the powerful non-linear fitting capability of deep networks, these DL based techniques can generally achieve better deraining performance than conventional prior-based methods.

Albeit attaining a huge boost in deraining performance, these DL strategies still possess evident drawbacks. As seen, although the network architectures are becoming more diverse and complicated, there is still  room  for embedding the intrinsic prior knowledge of rain streaks so that one can design more interpretable network architecture and make the network output finely comply with expected prior properties.
For example, the rain layers extracted by current DL based methods often contain some unexpected background details, which causes the over-smoothness of the derained results to a certain extent~(as shown in Fig.~\ref{figrainl}). In fact, rational explicit constraints (e.g., sparsity and non-local similarity) on the rain layer should be helpful for alleviating this problem, which is however neglected by most of current deep single image derainers.


Another important issue lies in the generalization capability. Since the existing paired training sets are pre-collected and synthesized manually, it is inevitable that there is a bias about rain types between synthetic training data and real testing data. In this case, most of current deep deraining methods are prone to suffer from the over-fitting issue, since they generally adopt complicated and diverse network modules and put less emphasis on embedding the intrinsic prior constraint about rain layer. Thus, it is meaningful and also necessary to design a DL regime capable of finely fitting testing samples even when their rain types are different from training samples.

\begin{figure}[t]
  \begin{center}
  \vspace{-2mm}
     \includegraphics[width=1\linewidth]{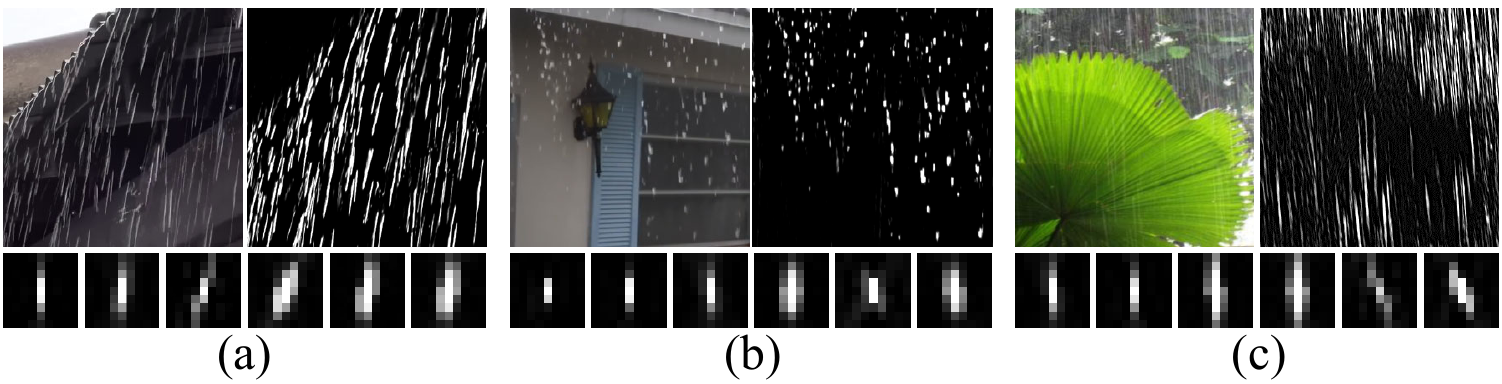}
  \end{center}
  \vspace{-6mm}
     \caption{Illustration of the rain kernels and rain layers estimated by DRCDNet for (a)-(c) three different testing samples from the \textcolor{black}{real SPA-Data} dataset. Here DRCDNet is trained on the \textcolor{black}{synthetic Rain100L} dataset.}
  \label{dynamickernel}
    \vspace{-5mm}
\end{figure}

To address the aformentioned issues, a rational way is to embed the prior knowledge of rain layers into deep networks. This is because that prior structures can rationally regularize and constrain the solution space, which not only {\textcolor{black}{helps}} avoid unexpected image details to be estimated as rain streaks, but also {\textcolor{black}{helps}} alleviate the over-fitting problem of the network. 
In this paper, we explore the way to embed a well-studied prior model of rain layer (as shown in Fig.~\ref{figcsc}) into deep networks, and propose a novel network architecture with fine interpretability and generalization ability.\footnote{As compared with our conference paper~\cite{wang2020model}, the work has made substantial extensions. Specifically, a novel network with fine interpretability and generalization ability is designed. More model analysis, methodology expansions, visualization verifications, and experimental evaluations are provided. Especially, a core extension is that rain kernels (see Fig.~\ref{dynamickernel}) are adaptively inferred according to input rainy image. This dynamic prediction mechanism makes it possible to achieve better generalization performance even when the rain patterns are different between training and testing samples.} Specifically, our main contributions are summarized as follows:



Firstly, we utilize the intrinsic convolutional dictionary learning mechanism to encode rain shapes, and propose a concise rain convolutional dictionary (RCD) model for single rainy image. To solve it, we adopt the proximal gradient technique~\cite{beck2009fast} to develop an optimization algorithm. Different from conventional solvers made up of complicated operators (e.g., Fourier transform), the proposed algorithm only consists of simple computations (see Fig.~\ref{figintro}~(a)) easy to be implemented by general network modules. This novel manner not only explicitly incorporates the intrinsic prior structures of rain streaks, but also facilitates us to easily unfold this algorithm into a deep network architecture.

Secondly, by unfolding every step of the algorithm, we construct an interpretable network for single image deraining, named as RCDNet. Every module in this network corresponds to the implementation operator of the proposed algorithm, and thus all network modules have clear physical interpretability as demonstrated in Fig.~\ref{figintro}. 
Specifically, the RCDNet successively consists of M-net and B-net, updating the rain map $\mathcal{M}$ and the background layer $\mathcal{B}$, respectively. All the operators in these two sub-networks are easy to understand and suitable for extracting rain layers, since they are consistent with the corresponding algorithm.\footnote{This interpretable network design greatly facilitates us to analyze what happens during the network training, and understand the implementation mechanisms (see model visualization in Sec.~\ref{visualanalysis}).} Moreover, the rain layer extracted by RCDNet naturally complies with the prior constraints and can better exclude the background details as shown in Fig.~\ref{figrainl}.

Thirdly, we further construct a dynamic rain convolutional dictionary network, called DRCDNet, for better generalization capability. Unlike RCDNet which estimates a large rain dictionary $\mathcal{K}$ for the entire dataset, DRCDNet dynamically infers the rain kernel $\mathcal{K}$ for each rainy sample. In this way,  the number of the to-be-estimated rain map $\mathcal{M}$ can be greatly reduced, and the hidden solution space for estimating rain layer is also  greatly shrunk, which naturally improves the generalization ability. As presented in Fig.~\ref{dynamickernel}, in the cross-domain testing scenario, rain kernels are adaptively inferred according to the rain types of testing rainy images. To the best of our knowledge, we are the first to fully incorporate the intrinsic generative mechanism of rain layer into network design and also the first to design such a dynamic and flexible rain kernel prediction scheme.


Fourthly, under two kinds of testing settings (i.e., training/test domain match/mismatch settings), comprehensive experimental results substantiate the superiority of RCDNet and DRCDNet beyond conventional methods. Especially, attributed to the fine interpretability, not only the underlying rationality of such an interpretable network can be intuitively understood by general users through visualizing the amelioration process (e.g., the gradually rectified rain maps and background layers) over the network layers at all stages, but also the network can yield diverse rain kernels for describing rain shapes and proximal operators for delivering the priors of background and rain maps for a rainy image, facilitating their general {\textcolor{black}{applicability}} to more real rainy images.


The paper is organized as follows. {\textcolor{black}{Secs.}}~\ref{notation} and~\ref{relatedwork} review the necessary notations and related works, respectively. \textcolor{black}{Sec.}~\ref{rcdmodel} presents the RCD model for rain removal as well as its optimization algorithm. \textcolor{black}{Sec.}~\ref{crcdnet} constructs the interpretable RCDNet, where the rain kernels are shared among the entire training samples, mainly usable under similar training-testing rain types. \textcolor{black}{Sec.}~\ref{drcdnet} further builds the DRCDNet to adaptively infer rain kernels for diverse input rainy images, applicable to \textcolor{black}{the case that training-testing rain patterns mismatch}. \textcolor{black}{Sec.}~\ref{train} states the training details. \textcolor{black}{Sec.}~\ref{exp} demonstrates the experimental evaluations to validate the superiority of the proposed network. The paper is finally concluded ~\textcolor{black}{with Sec.~\ref{sec:conclusion}}.

\vspace{-1mm}\section{Notations and Preliminaries}\label{notation}
For ease of understanding, we introduce some necessary notations and preliminaries as follows.

Denote $\mathcal{A}\in \mathbb{R}^{I_{1}\times{I_{2}}\times\cdots\times{I_{N}}}$  as a tensor of order $N$. The unfolding matrix $U^{f}_{n}(\mathcal{A})\in \mathbb{R}^{I_{n}\times \left(I_{1}\cdots I_{n-1} I_{n+1} \cdots I_{N}\right)}$ is composed by taking the mode-$n$ vector of $\mathcal{A}$ as its columns. This matrix can also be seen as the mode-$n$ flattening of $\mathcal{A}$. The vectorization of $\mathcal{A}$ is $\text{vec}\left(\mathcal{A}\right) \in \mathbb{R}^{I_{1}{I_{2}}\cdots{I_{N}}}$. All these can be easily achieved by the ``reshape" function in PyTorch~\cite{paszke2017automatic}.

The symbol $\otimes$ represents the 2-dimensional (2D) convolutional operation. {\textcolor{black}{It can be extended to the convolution in the form of tensor in deep networks as:}}
\begin{equation}\label{conv}
\mathcal{Y} = \mathcal{C} \otimes \mathcal{X},
\end{equation}
where $\mathcal{C}\in\mathbb{R}^{k\times k\times N_{o} \times N_{i}}$, $\mathcal{X}\in\mathbb{R}^{H\times W\times N_{i}}$, and $\mathcal{Y}\in\mathbb{R}^{H\times W\times N_{o}}$. At the mode-$3$ of  $\mathcal{Y}$,
$\mathcal{Y}[:,:,j] = \sum ^{N_{i}}_{n=1}\mathcal{C}[:,:,j,n]\otimes\mathcal{X}[:,:,n]$, $j=1,2,\cdots,N_{o}$. The notation $\otimes$ between $\mathcal{C}[:,:,i,n]$ and $\mathcal{X}[:,:,n]$ is a 2D convolutional computation. The convolutional operation in Eq.~(\ref{conv}) can be easily achieved by the off-the-shelf function ``torch.nn.Conv2d" in PyTorch.

The symbol $\otimes^{d}$ represents the depthwise convolutional operation. Specifically,
\begin{equation}\label{depconv}
\mathcal{Z} = \mathcal{C} \otimes^{d} \mathcal{M},
\end{equation}
where $\mathcal{M}\in\mathbb{R}^{H\times W\times N}$ and $\mathcal{Z}\in\mathbb{R}^{H\times W\times  N_{o} \times N_{i} \times N}$. Specifically,
$\mathcal{Z}[:,:,j,k,n]=\mathcal{C}[:,:,j,k]\otimes \mathcal{M}[:,:,n]$, $j=1,2,\cdots,N_{o}, k=1,2,\cdots,N_{i}, n = 1,2,\cdots,N$. The depthwise convolutional operation in Eq.~(\ref{depconv}) can be easily performed through the group convolution by setting the parameter ``group" in the function ``torch.nn.Conv2d''.

\vspace{-0mm}\section{Related Work}\label{relatedwork}
In this section, we review the most related work, including video deraining and single image deraining.

\vspace{-3mm}\subsection{Video Deraining Methods}
\noindent\textbf{Traditional \textcolor{black}{Prior Based Methods}.} Garg \emph{et al.}~\cite{Garg2004Detection} made early attempt to study the visual effect of rain streaks on imaging systems, and proposed to adopt motion blur model and space-time model to describe the photometry and dynamics of rain streaks, respectively. Later, many physical properties of rain were investigated, including chromatic, temporal, spatial, and frequency domain characteristics~\cite{zhang2006rain,park2008rain,bossu2011rain,barnum2010analysis}. In the past few years, researchers formulated intrinsic priors of rainy videos into model design, and adopted some iterative optimization algorithms for rain detection and removal. For example, the frequently-adopted prior knowledge include low-rankness among multi-frames~\textcolor{black}{\cite{Jin2015Video,wei2017should,li2018video,kim2015video}}, smoothness of background frame in the rain-perpendicular direction and that of rain streaks in the direction of raindrops~\cite{Jiang2017A,Jiang2018FastDeRain}, sparsity and repeatability of rain~\cite{Chen2013A,Ren2017Video,li2018video}. Recently, Wei \emph{et al.}~\cite{wei2017should} proposed to stochastically encode the rain layer patch as a mixture of Gaussian model (P-MoG) for adapting a wide range of rain streaks. Li \emph{et al.}~\cite{li2018video} further investigated the characteristics of rain streaks, i.e., non-local self-similarity and multi-scale, and formulated them as a multi-scale convolutional sparse coding (MSCSC) model, achieving good performance on this video deraining task.

\noindent\textbf{Deep Learning Based Methods.} Recently, deep learning has attained tremendous success in various low-level vision tasks, such as image super-resolution~\cite{fu2022kxnet,dong2015image}, low-light enhancement~\cite{liu2022low}, and CT artifact reduction~\cite{wang2022adaptive, wang2022osc}. For video deraining, the early work~\cite{Jie2018Robust} presented a convolutional neural network (CNN) architecture where superpixels were utilized as the basic element for content alignment. To improve the rain removal performance, Liu~\emph{et al.}~\cite{liu2018erase} explored the wealth of temporal redundancy of videos and proposed J4R-Net, which integrates rain degradation classification, spatial texture knowledge based rain removal, and temporal relevance based background reconstruction. To handle dynamic video contexts, the authors further designed a dynamic routing residue recurrent network~\cite{Liu2018D3R}. Very recently, they \textcolor{black}{embedded} a dual-level flow regularization into a two-stage recurrent network~\cite{Wenhan2020Frame}. Although these approaches perform well, they generally cannot be finely applied to the single image deraining task which has no temporal information.

\vspace{-3mm}\subsection{Single Image Deraining Methods}

\noindent\textbf{Traditional Unsupervised Methods.} To reconstruct background from a rainy image, the early attempts mainly focused on extracting rain-removed high frequency part (HFP) with various filtering strategies, such as guided filters~\cite{Jing2012Removing}, bilateral filtering~\cite{Kang2012Automatic}, multiple guided filtering~\cite{zheng2013single}, guided $L_{0}$ smoothing filtering~\cite{Ding2016Single}, and nonlocal means filtering~\cite{Kim2014Single}. During the past decades, researchers devoted themselves to designing prior terms for regularizing the to-the-estimated background/rain layer. For example, Luo \emph{et al.}~\cite{Yu2015Removing} proposed an image patch-based discriminative sparse coding scheme. Li \emph{et al.}~\cite{Li2016Rain} adopted \textcolor{black}{Gaussian mixture models (GMM)} to separate the background from rain streaks. Wang \emph{et al.}~\cite{Wang2017A} developed a 3-layer hierarchical scheme to categorize the HFP into rain/snow and rain/snow-free parts. Gu \emph{et al.}~\cite{Gu2017Joint} encoded the rain-free and rain parts as analysis and synthesis sparse representation models, respectively. The main drawback of these traditional model-based methods is that the hand-crafted prior assumptions are always subjective and limited, which might be possibly not able to faithfully reflect the complicated and diverse rain types collected from practice.

\noindent\textbf{Deep Learning Based Methods.} Recently, DL has achieved promising performance in this single image deraining task, showing evident superiority to conventional methods, \textcolor{black}{such as~\cite{yu2021single,yadav2021deraingan,yuan2021single,wei2021semi,wei2019coarse,gou2020clearer,li2022all, wang2021rain, shao2021selective}}. In the early period, Fu~\emph{et al.}~\cite{Fu2017Clearing} proposed a CNN to extract discriminative features of rain in the HFP of a single rainy image and further developed a deep detail network which introduced the residual learning to speed up the training process~\cite{Fu2017Removing}. Later, Zhang~\emph{et al.}~\cite{zhang2018density} designed a rain density classifier aided multi-stream dense network. Further, the authors proposed a conditional generative adversarial network for better visual quality~\cite{zhang2019image}. Recently, recurrent architectures have been intensively studied for rain removal in a stage-wise manner~\cite{li2018recurrent, ren2019progressive}. There are also some works incorporating the multi-scale learning by analyzing the self-similarity both in the same scale or across different scales~\cite{fu2019lightweight,zheng2019residual,yasarla2019uncertainty,wang2020structural,jiang2020multi}. In~\cite{hu2019depth} and~\cite{li2019heavy}, physical formulations were merged into the entire network design. The work~\cite{wang2019erl} formulated an entangled representation learning model made up of a {\textcolor{black}{two-branch}} encoder. In~\cite{deng2020detail}, a detail-recovery image deraining network {\textcolor{black}{was}} proposed where rain removal and detail reconstruction {\textcolor{black}{were}} viewed as two separate tasks. A few researchers explored the rain imaging process and created some more realistic rainy images~\cite{benchmark,halder2019physics,hu2019depth,wang2019spatial}. 
To reduce the cost of pre-collecting abundant paired training samples and bridge the domain gap between synthetic and real data, semi-/un- supervised learning is also attracting much attention recently~\cite{wei2019semi,yasarla2020syn2real,jin2019unsupervised,wei2019deraincyclegan}.  {\textcolor{black}{Currently, there is a new research line where researchers propose to utilize the high-level semantic information for better rain removal, such as~\cite{wei2022sginet}. Besides, few researchers begin to focus on more complicated task settings, including synchronous rain streaks and raindrops removal~\cite{wei2022robust}.}}


Albeit attaining significant success, most of these deep networks are assembled with some off-the-shelf modules in current DL toolkits and have less specific interpretability to this practical deraining task. Especially, they have not explicitly embedded sufficient prior knowledge underlying rain streaks into the network design. Hence, there is still large room for further performance improvement for this task. Besides, most of these deep derainers tend to suffer from the overfitting issue due to the training-testing bias about rain distributions.

\section{RCD Model for Single Image Deraining}\label{rcdmodel}
\subsection{Model Formulation}\label{property}
Given $\mO\in\mathbb{R}^{H\times W\times 3}$ as an observed color rainy image, we can rationally separate it as:
\begin{equation}\label{e1}
\mO=\mB+\mR,
\end{equation}
where $H$ and $W$ are the height and width of the image, respectively;  $\mB$ and $\mR$ are the clear background and rain layers, respectively.\footnote{Note that Eq.~(\ref{e1}) is an approximate model, which provides a rough direction for network learning. During the network implementation in Sec.~\ref{crcdnet}, we add an adjustment module to flexibly deal with complicated rainy images. Sec.~\ref{exp} validates the effectiveness of our  method in diverse rain scenarios.} To recover the background, most of current deep derainers focus on establishing complex network architectures to learn the mapping function between $\mO$ and $\mB$ (or $\mR$).

Instead of designing complex networks heuristically, we first consider the traditional rain generation model which reflects the intrinsic prior structures of rain streaks~\cite{Gu2017Joint, He2017Convolutional,li2018video}. In specific, with the rain convolutional dictionary (RCD) physical mechanism as visually illustrated in Fig.~\ref{figcsc},  the rain layer can be rationally expressed as:
\begin{equation}\label{e2}
  \mR^c =  \sum_{n=1}^{N} \bm{K}^c_{n}  \otimes \bm{M}_{n},~~~ c=1,2,3,
\end{equation}
where $\mR^c$ denotes the $c^{\text{th}}$ color channel of $\mR$; $\left\{\bm{K}^c_{n}\right\}_{n,c} \subset \mathbb R^{k \times k}$ is a set of rain kernels with the size $k \times k$ representing the repetitive local patterns of rain streaks; $\left\{\bm{M}_{n}\right\}_{n} \subset \mathbb R^{H \times W}$ is the rain maps representing the locations where local patterns repeatedly appear; $N$ is the number of rain kernels; and $\otimes$ is the 2D convolutional operation. For simplicity, throughout the paper, we rewrite Eq.~(\ref{e2}) as:
\begin{equation}\label{e3}
\mR = \sum^{N}_{n=1} \mK_{n}\otimes \bm{M}_{n}=\mathcal{K} \otimes \mathcal{M},
\end{equation}
where $\mathcal{K}_{n}\in\mathbb{R}^{k\times k\times 3}$, $\mK\in\mathbb{R}^{k\times k\times 3\times N}$, and $\mM\in\mathbb{R}^{H\times W\times N}$ are stacked by $\bm{K}^c_n$s, $\mK_{n}$s, and $\bm{M}_n$s, respectively.
The 2D convolutional operation $\otimes$ between $\mK_{n}$ and $\bm{M}_{n}$  is executed in the channel-wise manner, and the computation $\mK \otimes \mM$ is the extension of $\otimes$ from 2D to tensor form.

We can rewrite the single rainy image model in Eq.~(\ref{e1}) as:
\begin{equation}\label{e5}
{\mO}=\mB+ \mK \otimes \mM.
\end{equation}
Clearly, our goal is to estimate the $\mK$, $\mM$, and $\mB$ from $\mO$. With sparse constraints on $\mathcal{M}$, it is easy to see that Eq.~(\ref{e3}) can well model the sparsity and non-local similarity of rains.

The rain kernel $\mK$ can be viewed as a set of convolutional dictionary \cite{Huang2015Convolutional} for representing the repetitive and similar local patterns underlying rain streaks. In the training-testing domain match scenario where the rain patterns between training data and testing data are similar, a small number of rain kernels can finely represent a wide range of rain shapes~\cite{li2018video}. Thus, they are the common knowledge for representing different rain types across all rainy images, and can be learned from abundant training samples by virtue of the strong learning ability of CNN with an end-to-end training manner (see more details in Sec.~\ref{crcdnet}). Thus, for predicting the clean background from an input rainy image, the key issue is to output $\mM$ and $\mB$ from $\mO$ with $\mK$ fixed. Correspondingly, the optimization problem is:
\begin{equation}\label{o1}
\min_{{\mM,\mB}}\left\|\mO-\mB-\mK \otimes \mM\right\|_{F}^{2}+\lambda_{1} p_{1}(\mM)+\lambda_{2} p_{2}(\mB),
\end{equation}
where $\lambda_{1}$ and $\lambda_{2}$ are trade-off parameters; $p_{1}(\cdot)$ and $p_{2}(\cdot)$ denote the penalty functions (i.e., regularizers) to deliver the prior structures of $\mM$ and $\mB$, respectively.



The first term of the problem (\ref{o1}) is a rational approximate model for rain streak generation, which well encodes the sparsity and non-local similarity of rain layer. Motivated by this, we believe that based on the solver of the problem (\ref{o1}), the constructed deep network modules are able to embed the prior of rain streak and constrain the space for rain layer estimation.

\vspace{-4mm}\subsection{Optimization Algorithm}\label{solvemb}
Deep unfolding technique is an intuitive way to combine the solver of optimization models with deep learning methods. This technique releases us from manually designing penalty terms, but also brings new challenges. The first one is {\textcolor{black}{how}} to 
develop an optimization algorithm which only contains simple computations easy to be transformed to network modules. 

Confronted with the convolutional dictionary representation model~(\ref{o1}), the traditional solvers usually contain complex computations, e.g., the Fourier transform and inverse Fourier transform~\cite{Huang2015Convolutional, WohlbergEfficient,li2018video}, tending to make this unfolding task difficult. We thus prefer to build a new algorithm where the to-the-estimated variables $\mM$ and $\mB$ are alternately updated by the proximal gradient technique~\cite{beck2009fast}. In this way, the solution process only consists of simple computations, making it possible to easily achieve the transformation from the algorithm to network architectures. The details are as follows:

\noindent\textbf{Updating $\mM$}: At the $s^{\text{th}}$ iteration, the rain map $\mM$ can be updated by solving the quadratic approximation~\cite{beck2009fast} of the problem (\ref{o1}) with regard to $\mM$ as:
\begin{equation}\label{minm}
  \!\min_{\mM} \frac{1}{2} \left\| \mM \!-\! \left(\! \mM^{(s-1)}\!\!-\! \eta_{1}\nabla f\left(\mM^{(s-1)}\right)  \!\right) \right\|_F^2 + \lambda_{1}\eta_{1} p_{1}\left( \mM \right),
\end{equation}
where $\mM^{(s-1)}$ is the updating result of the last iteration, $\eta_{1}$ is the stepsize parameter, and  $f\left(\mM^{(s-1)}\right)\!=\! \left\|\mO\!-\!\mB^{(s-1)}\!- \!\mK\otimes\mM^{(s-1)}\right\|_{F}^{2}$. Under general regularization terms \cite{donoho1995noising}, the solution of Eq. (\ref{minm}) is expressed as:
\begin{equation}\label{solvem}
 \mM^{(s)} = \mbox{prox}_{\lambda_{1}\eta_{1}}\left(\! \mM^{(s-1)} \!-\! \eta_{1}\nabla f\left(\mM^{(s-1)}\right)  \!\right),
\end{equation}
{\textcolor{black}{where $\mbox{prox}_{\lambda_{1}\eta_1}(\cdot)$ is the proximal operator dependent on the regularization term $p_{1}(\cdot)$ with respect to $\mM$.}} By  substituting
\small
\begin{equation}\label{gradm}
  \nabla f\left(\mM^{(s-1)}\right)=\mK\otimes^{T}\left(\mathcal{K} \otimes \mathcal{M}^{(s-1)} +\mB^{(s-1)}-\mO\right),
\end{equation}
\normalsize
where $\otimes^T$ denotes the transposed convolution,\footnote{The operation $\otimes^T$ can be directly executed by the function as ``torch.nn.ConvTransposed2d" in PyTorch~\cite{paszke2017automatic}.} we can obtain the updating formula for $\mM$ as:\footnote{It can be proved that, with  small enough $\eta_1$ and $\eta_2$,  Eq. (\ref{updatem}) and Eq. (\ref{updateb}) can both lead to the decrease of the objective function in (\ref{o1})~\cite{beck2009fast}. \label{fn:repeat}}
\small
\begin{equation}\label{updatem}
  \mM^{(s)}=
  \mbox{prox}_{\lambda_{1}\eta_{1}}\left(\mM^{(s-1)}
  - \eta_{1}\mK\!\otimes^{T}\!\left(\mathcal{K}\! \otimes \!\mathcal{M}^{(s-1)}\!+\!\mB^{(s-1)}\!-\!\mO\!\right) \!\right)\!.
\end{equation}
\normalsize
Instead of being derived from manually-designed regularizer as in traditional methods, the form of the implicit proximal operator $\mbox{prox}_{\lambda_{1}\eta_1}(\cdot)$ can be expressed through a convolutional network module and automatically adapted from training data in an end-to-end manner, which is described in Sec.~\ref{crcdnet} below.

\noindent\textbf{Updating $\mB$}:
Similarly, the quadratic approximation of the problem (\ref{o1}) with respect to $\mB$  is:
\small
\begin{equation}\label{minb}
  \min_{\mB}\frac{1}{2}\left\| \mB - \left( \mB^{(s-1)} \!-\! \eta_{2}\nabla g\left(\mB^{(s-1)}\right)  \!\right) \right\|_F^2 + \lambda_{2}\eta_{2} p_{2} \left( \mB \right),
\end{equation}\normalsize
where $g\left(\mB^{(s-1)}\right)\!=\! \left\|\mO\!-\!\mB^{(s-1)}- \mK\otimes\mM^{(s)}\right\|_{F}^{2}$. By substituting
$\nabla g\left(\!\mB^{(s-1)}\!\right) = \mathcal{K} \otimes \mathcal{M}^{(s)}+\mB^{(s-1)}-\mO$,
it is easy to deduce that  the final updating rule for $\mB$ is:\footref{fn:repeat}
\small
\begin{equation}\label{updateb}
\begin{split}
\mB^{(s)} \!=\!\mbox{prox}_{\lambda_{2}\eta_{2}}\left(\left(1-\eta_2\right) \mB^{(s-1)}
  \!+\!\eta_{2}\left(\!\mO\! -\mathcal{K}\otimes \mathcal{M}^{(s)}\right)\right),
\end{split}
\end{equation}
\normalsize
where $\mbox{prox}_{\lambda_{2}\eta_2}(\cdot)$ is the  proximal operator correlated to the regularization term $p_{2}(\cdot)$ with respect to $\mB$.

Based on this iterative algorithm, we can then construct our deep unfolding network as follows.

\vspace{-0mm}\section{Rain convolutional dictionary network}\label{crcdnet}
Inspired by the recent deep unfolding techniques in various tasks, e.g., deconvolution~\cite{zhang132017learning}, compressed sensing~\cite{yang2017admm}, image super-resolution~\cite{zhang2020deep}, \textcolor{black}{CT metal artifact reduction~\cite{wang2021indudonet, wang2021indudonet+, wang2021dicdnet, wang2022osc}, low light enhancement~\cite{liu2022low}, and pansharpening~\cite{cao2021pancsc}}, we build a novel network structure for this single image deraining task by separating and transforming each iterative step of the aforementioned algorithm as a specific form of network connection. Its specificity is that all network modules correspond to the algorithm operators, and thus the entire network has clear interpretability.

\begin{figure*}[t]
  \begin{center}
  \vspace{-2mm}
     \includegraphics[width=0.95\linewidth]{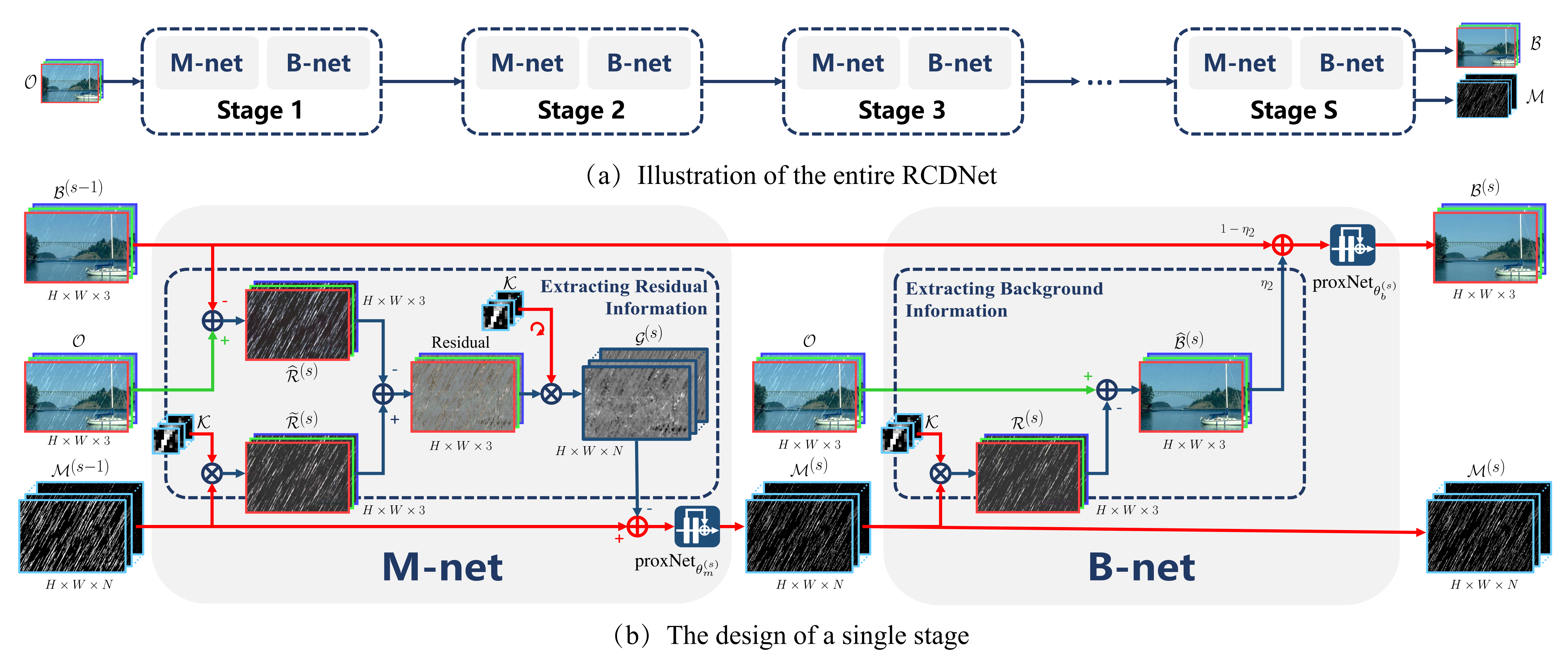}
  \end{center}
  \vspace{-5.5mm}
     \caption{(a) The proposed RCDNet with $S$ stages. The network takes a rainy image $\mathcal{O}$ as input and outputs the learned rain map $\mathcal{M}$ and background image $\mathcal{B}$. (b) Illustration of the network architecture at the $s^{\text{th}}$ stage. Each stage consists of M-net and B-net to accomplish the update of rain map $\mathcal{M}$ and background layer $\mathcal{B}$, respectively. The images are better to be observed by zooming in on screen.}
  \label{figflow}
    \vspace{-5mm}
  \end{figure*}

As shown in Fig.~\ref{figflow}~(a), the proposed network consists of $S$ stages, representing $S$ iterations of the algorithm for solving ~(\ref{o1}). Each stage achieves sequential updates of $\mM$ and $\mB$ by the M-net and the B-net, respectively. Specifically, as displayed in Fig.~\ref{figflow}~(b), in each stage of the network, the M-net takes the rainy image $\mO$ and the previous outputs $\mB^{(s-1)}$ and $\mM^{(s-1)}$ as inputs, and outputs an updated $\mM^{(s)}$, and then the B-net takes $\mO$ and $\mM^{(s)}$ as inputs, and outputs an updated $\mB^{(s)}$.

From the updating rules (\ref{updatem}) and (\ref{updateb}), it is easily understood that the involved concise iterative computations can be naturally performed with commonly-used operators in normal networks~\cite{paszke2017automatic}. The key issue of unrolling the algorithm is how to represent the two proximal operators $\mbox{prox}_{\lambda_{1}\eta_1}(\cdot)$ and $\mbox{prox}_{\lambda_{2}\eta_2}(\cdot)$. In this work, we adopt the deep residual network (ResNet) \cite{he2016deep} to construct the operator as many other works~\cite{yang2018proximal,xie2019multispectral,wang2020model} did.\footnote{Please refer to the supplementary materials for more analysis.}
Then, we can separately decompose the updating rules for $\mM$ and $\mB$ into sub-steps and achieve the following procedures for the $s^{\text{th}}$ stage of the proposed rain convolutional dictionary network (RCDNet):
\begin{eqnarray}\label{unfoldm}
\hspace{-11mm}\text{M-net}: \left\{\begin{matrix}
\hspace{-28mm}\widehat{\mR}^{(s)}=\mO-\mB^{(s-1)},\\
\hspace{-26mm}\widetilde{\mR}^{(s)}=\mathcal{K} \otimes \mathcal{M}^{(s-1)},\\
\hspace{-10mm}\mathcal{G}^{(s)}=\eta_{1}\mK\otimes^{T}\left(\widetilde{\mR}^{(s)}-\widehat{\mR}^{(s)}\right),\\
\mM^{(s)}=\text{proxNet}_{\theta_m^{(s)}}\left(\mM^{(s-1)}-\mathcal{G}^{(s)}\right),
\end{matrix}\right.
\end{eqnarray}
\vspace{-4mm}
\begin{eqnarray}\label{unfoldb}
\text{B-net}: \left\{\begin{matrix}
 \hspace{-42mm}{\mR}^{(s)}=\mathcal{K}\otimes \mathcal{M}^{(s)},\\
 \hspace{-43mm}\widehat{\mB}^{(s)}=\mO-{\mR}^{(s)},\\
\mB^{(s)}=\text{proxNet}_{\theta_b^{(s)}}\left((1-\eta_2)\mB^{(s-1)}+\eta_2\widehat{\mB}^{(s)}\right),
\end{matrix}\right.
\end{eqnarray}
where $\text{proxNet}_{\theta_m^{(s)}}(\cdot)$ and $\text{proxNet}_{\theta_b^{(s)}}(\cdot)$  are two ResNets consisting of several Resblocks with the parameters ${\theta_m^{(s)}}$ and ${\theta_b^{(s)}}$ at the $s^{\text{th}}$ stage, respectively.

We can then design the network architecture, as shown in Fig. \ref{figflow}, by transforming the operators in (\ref{unfoldm}) and (\ref{unfoldb}) step-by-step. {\textcolor{black}{All the parameters involved can be automatically fit from training data (\emph{i.e.}, the paired clean image $\mathcal{B}$ and the rainy image $\mathcal{O}$) in an end-to-end manner, including $\{\theta_m^{(s)},\theta_b^{(s)}\}_{s=1}^{S}$, rain kernels $\mK$, $\eta_1$, and $\eta_2$. }}Considering that in some scenarios, the composition of rainy images is complicated. Thus we further refine the reconstructed result $\mB^{(S)}$ by feeding it into an extra ResNet which has the same structure as $\text{proxNet}_{\theta_b^{(s)}}(\cdot)$.


It should be indicated that every module has its specific physical meanings. As shown in Fig. \ref{figflow}~(b), at every stage, the M-net accomplishes the learning of the ameliorative gradient direction $\mG^{(s)}$ of rain maps and further helps rectify the $\mM$. Specifically, $\widehat{\mR}^{(s)}$ is the rain layer estimated with the previous background $\mB^{(s-1)}$, and $\widetilde{\mR}^{(s)}$ is the rain layer achieved by the generative model (\ref{e3}) with the estimated $\mM^{(s-1)}$. Then the M-net calculates the residual information between the two rain layers obtained in this way, and extracts the gradient updating direction $\mG^{(s)}$ of rain maps with the transposed convolution of rain kernels to update the rain map. Next, the B-net recovers the background $\widehat{\mB}^{(s)}$ estimated with current rain kernel and rain maps $\mM^{(s)}$, and fuses such estimated $\widehat{\mB}^{(s)}$ with the previously estimated $\mB^{(s-1)}$ by weights $\eta_2$ and ($1-\eta_2$), \textcolor{black}{respectively,} to get the updated background ${\mB}^{(s)}$. Clearly, such an interpretable network design makes it easy to intuitively observe what happens inside the network flow and understand the intrinsic implementation mechanisms.\footnote{More details about network design are described in supplementary file.}

\noindent \textit{\textbf{Remark 1:}} \textcolor{black}{As analyzed in Sec.~\ref{sec:introduction}, it is a common phenomenon that the reconstructed background images may lose some textures. However, our proposed RCD model~(\ref{e3}) can help alleviate this issue. Specifically, the intrinsic prior structures (\emph{e.g.}, sparsity and non-local self-similarity) considered in the model~(\ref{e3}) are unique to rain streaks not to background textures. Thus, this model can regularize the extracted rain layer and help distinguish rain streaks from background textures, which guarantees the texture fidelity of the reconstructed images. Besides, intensity fidelity can also be guaranteed, which is mainly attributed to the sparsity regularization on the feature map $\mathcal{M}$ via the ReLU activation function in ResNet. Such sparsity regularization can ensure that the most region of the extracted rain layer is with zero elements and then the intensity of background images corresponding to this region is enforced to be the same as that of the input rainy image, leading to the intensity fidelity. These are finely validated by Fig.~\ref{figrainl} below.}

\vspace{-3mm}
\section{Dynamic RCDNet}\label{drcdnet}
As seen, the large rain dictionary $\mathcal{K}$ in RCDNet is {\textcolor{black}{shared}} among the entire dataset. Such settings would be more applicable for the consistent case that training and testing datasets are with similar rain patterns. To further enhance the generalization capability, we construct a dynamic rain convolutional dictionary network, called DRCDNet. Specifically, in DRCDNet, the rain kernel $\mathcal{K}$ is dynamically inferred for each rainy image. In this way, the number of the to-be-estimated rain map $\mathcal{M}$ can be greatly reduced, and the hidden solution space for estimating rain layer is also greatly shrunk, which naturally improves the generalization ability. 
{\textcolor{black}{For clarity, we also refer the RCDNet in Sec.~\ref{crcdnet} as consistent RCDNet (CRCDNet) whenever necessary.}} The details of DRCDNet are as follows.

\begin{figure}[t]
  \begin{center}
  \vspace{-2mm}
     \includegraphics[width=0.9\linewidth]{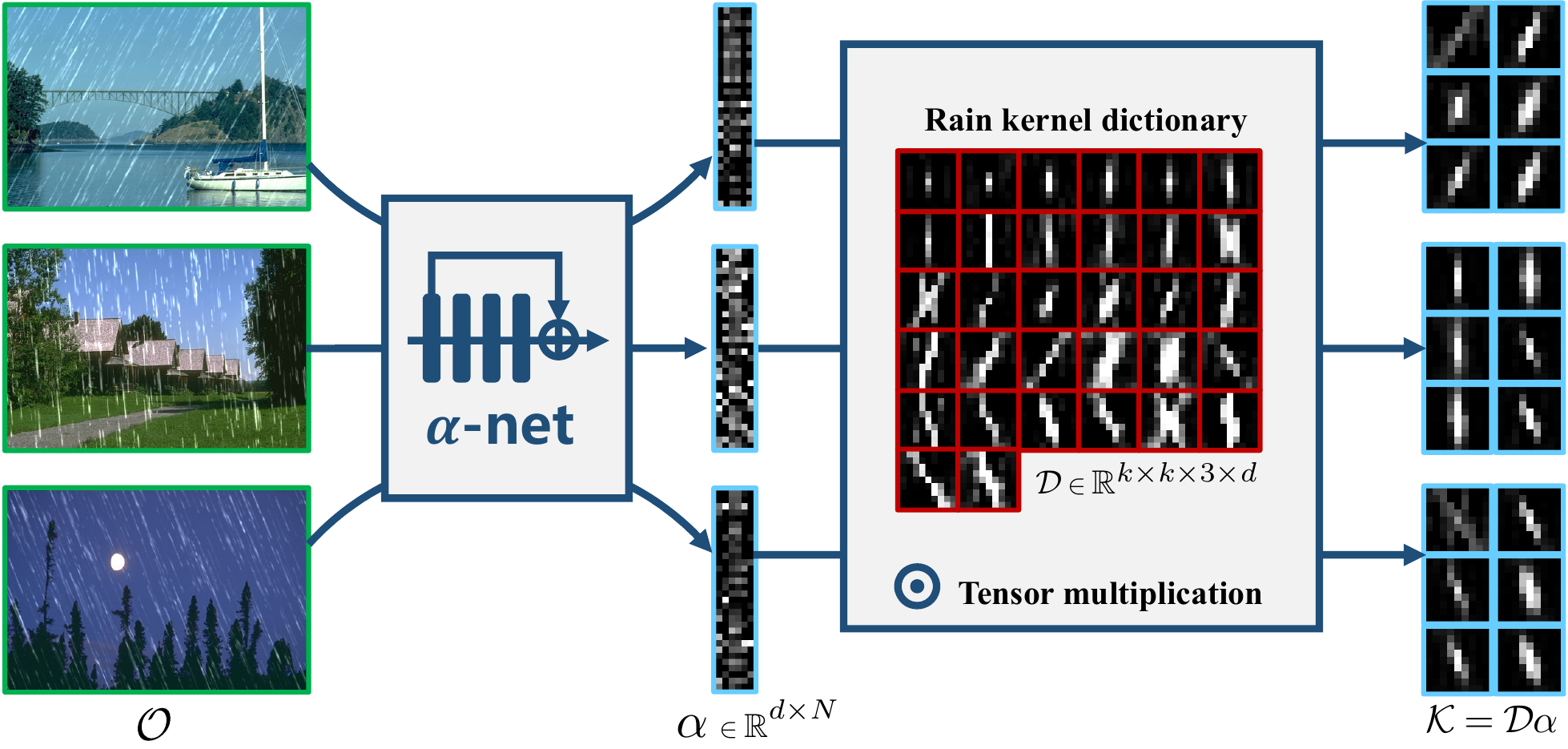}
  \end{center}
  \vspace{-5mm}
     \caption{Dynamic rain kernel inference in DRCDNet. The dictionary $\mD$ for DRCDNet is pre-learned like the kernel $\mK$ for RCDNet. But the involvement of $\alpha$ helps the DRCDNet possess adaptive learning capability,
    with dynamically predicted rain kernels $\mK$ for different testing rainy images $\mO$.}
  \label{kernelcompare}
    \vspace{-6mm}
\end{figure}

\noindent\textbf{Model Formulation.} For DRCDNet, we reformulate the rain kernel $\mK_{n}$ in Eq.~(\ref{e3}) as:
\begin{equation}\label{e20}
\mK_{n}=\mD\alpha_{n},
\end{equation}
where $\mD\in\mathbb{R}^{k\times k\times 3 \times d}$ is rain kernel dictionary representing common knowledge for conveying variant rain types across the entire training set; $d$ is the number of rain kernels in this dictionary; and $\alpha_{n}\in\mathbb{R}^{d}$ denotes the weighting coefficient.\footnote{$\mD\alpha_{n}$ is computed between the tensor $\mD \in\mathbb{R}^{k\times k\times 3 \times d}$  and the vector $\alpha_{n}\in\mathbb{R}^{d}$, expressed as $\mK_{n}= \sum^{d}_{i=1} \mD[:,:,:,i]\odot \alpha_{n}[i]$, which can be easily achieved by combining the ``reshape'' operation and the function as ``torch.matmul'' in {\textcolor{black}{PyTorch}}. $\odot$ is point-wise multiplication.} Instead of pre-training and then fixing rain kernels $\mK_{n}$s for any testing rainy image as CRCDNet does, DRCDNet can flexibly infer the rain kernels $\mK_{n}$s for every rainy sample by dynamically updating $\alpha_{n}$s. One can refer to Fig.~\ref{kernelcompare} for easy understanding. This motivation is finely verified in Sec.~\ref{visualanalysis}. 

Then the rainy image in Eq.~(\ref{e5}) is rewritten as:
\begin{equation}\label{dcsc}
\mO=\mB+\mD\alpha \otimes\mM,
\end{equation}
where  $\alpha \in\mathbb{R}^{d \times N}$ is stacked by $\alpha_{n}$s and $\mD\alpha\in\mathbb{R}^{k \times k \times 3 \times N}$. 

Similar to the dictionary $\mK$ in CRCDNet, the common dictionary $\mD$ in this dynamic case can be automatically learned from training samples in an end-to-end manner by virtue of the strong non-linear fitting ability of deep network. Our goal is to estimate the unknown $\mM$, $\mB$, and $\alpha$ from $\mO$. Thus the corresponding optimization problem is formulated as:
\begin{equation}\label{o3}\small
\begin{split}
&\min_{{\mM,\mB,\alpha}}\left\|\mO\!-\!\mB\!-\!\mD\alpha\otimes\mM\right\|_{F}^{2}\!\!+\!\!\lambda_{1} p_{1}(\mM)\!\!+\!\!\lambda_{2} p_{2}(\mB)\!\!+\!\!\lambda_{3} p_{3}(\alpha),\\
& s.t. \left\|\alpha_{n}\right\|_{2}=1, n =1,2,\ldots, N,
\end{split}
\end{equation}
\normalsize
where the explicit constraint, i.e., $\left\|\alpha_{n}\right\|_{2}=1$, is used to control the energy of weighting coefficient $\alpha_{n}$ so as to avoid affecting the learning of rain kernels. Similar to $p_{1}(\cdot)$ and $p_{2}(\cdot)$, we also prefer to automatically fit the regularizer $p_{3}(\cdot)$ for $\alpha$ via deep unrolling network modules.

\noindent\textbf{Optimization Algorithm.} With the similar algorithm for the problem ~(\ref{o1}) given in Sec.~\ref{solvemb}, we can easily derive the updating rules of $\mM$ and $\mB$ for the problem ~(\ref{o3}) as:
\begin{equation}\label{updatem2}\footnotesize
\begin{split}
  &\mM^{(s)}= \!\\&\!
  \mbox{prox}_{\lambda_{1}\eta_{1}}\!\!\left(\! \mM^{(s-1)}
  \!-\! \eta_{1}\!\left(\!\mathcal{D}\alpha^{(s-1)}\!\right)\!\otimes^{T}\!\!\!\left(\! \mathcal{D}\alpha^{(s-1)}\!\!\otimes\! \mM^{(s-1)}\!\!+\!\mB^{(s-1)}\!\!\!-\!\mO\!\right) \!\right)\!,
\end{split}
\end{equation}
\begin{equation}\label{updateb2}\footnotesize
\begin{split}
  \mB^{(s)} =\mbox{prox}_{\lambda_{2}\eta_{2}}\left(\!\left(1-\eta_2\right) \mB^{(s-1)}
  +\eta_{2}\left(\mO - \mathcal{D}\alpha^{(s-1)} \otimes \mM^{(s)}\right)\right).
\end{split}
\end{equation}
\normalsize
As for $\alpha$, the quadratic approximation of the problem (\ref{o3}) with respect to $\alpha$ is derived as:
\begin{equation}\label{minalpha}
  \!\min_{\alpha \in \Omega} \frac{1}{2} \left\| \alpha \!-\!\! \left(\! \alpha^{(s-1)}\!\!-\! \eta_{3}\nabla h\left(\alpha^{(s-1)}\right)  \!\right) \right\|_F^2 \!+ \lambda_{3}\eta_{3} p_{3}\left( \alpha \right),
\end{equation}
where $\Omega=\{\alpha|\left\|\alpha_{n}\right\|_{2}=1, n=1,2,\cdots,N\}$; $h\left(\alpha^{(s-1)}\right)= \left\|\mO\!-\!\!\mB^{(s)}\!-\mathcal{D}\alpha^{(s-1)} \otimes \mM^{(s)}\right\|_F^2$. Then, we can derive that
\begin{equation}
\small
\begin{split}
\frac{\partial h\left(\alpha^{(s-1)}\right)}{\partial \alpha_{n}}\!=\!\left(\!U^{f}_{4}\!\left(\!\mathcal{D}\otimes^{d} \bm{M}_{n}^{(s)}\right)\!\right)\!\text{vec}\!\left(\!\mathcal{D}\alpha^{(s-1)}\!\! \otimes \!\mM^{(s)}\!+\!\mB^{(s)}\!-\!\mO\!\right),
\end{split}
\end{equation}
\normalsize
where the computed result of $\mathcal{D}\otimes^{d} \bm{M}_{n}^{(s)}$ has the size of $H\times W \times 3 \times d$; $U^{f}_{4}(\cdot)$ represents unfolding the result at the $4^{th}$ mode and the resulted shape is $d \times 3HW$.

Clearly, the updating rule for $\alpha$ is finally derived as:
\small
\begin{equation}\label{updatealpha}
\alpha^{(s)}=
  \mbox{prox}_{\lambda_{3}\eta_{3}}\!\!\left(\alpha^{(s-1)}
  - \eta_{3}\nabla h\left(\alpha^{(s-1)} \right)\right),
\end{equation}
\normalsize
where
$\nabla h\left(\!\alpha^{(s-1)}\!\right) \!\!= \!\!\left[\!
\frac{\partial h\left(\alpha^{(s-1)}\right)}{\partial \alpha_{1}},
\frac{\partial h\left(\alpha^{(s-1)}\right)}{\partial \alpha_{2}},\ldots, \frac{\partial h\left(\alpha^{(s-1)}\right)}{\partial \alpha_{N}}\!\right]$.

Such concise iterative rules (\ref{updatem2}), (\ref{updateb2}), and (\ref{updatealpha}) facilitate us to unfold this iterative algorithm into a deep interpretable network as follows. Note that the constraint space $\Omega$ can be easily achieved by embedding a normalization operation into the implicit proximal operator $\mbox{prox}_{\lambda_{3}\eta_{3}}(\cdot)$.

\noindent\textbf{Network Design.} Similar to Sec.~\ref{crcdnet}, we subsequently decompose these updating rules (\ref{updatem2}), (\ref{updateb2}), and (\ref{updatealpha}) into sub-steps and achieve the following procedures for the $s^{\text{th}}$ stage of the proposed DRCDNet:
\begin{eqnarray}\label{unfoldm2}
\hspace{-10mm}\text{M-net}: \left\{\begin{matrix}
\hspace{-27mm}\widehat{\mR}^{(s)}=\mO-\mB^{(s-1)},\\
\hspace{-16mm}\widetilde{\mR}^{(s)}=\mathcal{D}\alpha^{(s-1)} \otimes \mathcal{M}^{(s-1)},\\
\hspace{-0.5mm}\mathcal{G}^{(s)}=\eta_{1}\mathcal{D}\alpha^{(s-1)}\otimes^{T}\left(\widetilde{\mR}^{(s)}-\widehat{\mR}^{(s)}\right),\\
\mM^{(s)}=\text{proxNet}_{\theta_m^{(s)}}\left(\mM^{(s-1)}-\mathcal{G}^{(s)}\right),
\end{matrix}\right.
\end{eqnarray}
\vspace{-3mm}
\begin{eqnarray}\label{unfoldb2}
\text{B-net}: \left\{\begin{matrix}
 \hspace{-32mm}{\mR}^{(s)}=\mathcal{D}\alpha^{(s-1)}\otimes \mathcal{M}^{(s)},\\
 \hspace{-43mm}\widehat{\mB}^{(s)}=\mO-{\mR}^{(s)},\\
\mB^{(s)}=\text{proxNet}_{\theta_b^{(s)}}\left((1-\eta_2)\mB^{(s-1)}+\eta_2\widehat{\mB}^{(s)}\right),
\end{matrix}\right.
\end{eqnarray}
\vspace{-3mm}
\begin{eqnarray}\label{unfoldalpha}
\alpha\text{-net}: \left\{\begin{matrix}
 \hspace{-40mm}\widehat{\mR}^{(s)}=\mO-\mB^{(s)},\\
\hspace{-28mm}\widetilde{\mR}^{(s)}=\mathcal{D}\alpha^{(s-1)} \otimes \mM^{(s)},\\
\hspace{0.5mm}\mathcal{G}_{\alpha_{n}}^{(s)}=\eta_{3}\!\left(\!U^{f}_{4}\!\left(\!\mathcal{D}\otimes^{d}\! \bm{M}_{n}^{(s)}\right)\right)\!\text{vec}\!\left(\!\widetilde{\mR}^{(s)}\!-\!\widehat{\mR}^{(s)}\!\right),\\
\hspace{-13.5mm}\alpha^{(s)}\!=\!\text{proxNet}_{\theta_a^{(s)}}\left(\alpha^{(s-1)}-\mathcal{G}_\alpha^{(s)}\right),
\end{matrix}\right.
\end{eqnarray}
where $\mathcal{G}_\alpha^{(s)}=\left[\mathcal{G}_{\alpha_{1}}^{(s)}, \mathcal{G}_{\alpha_{2}}^{(s)}, \ldots, \mathcal{G}_{\alpha_{N}}^{(s)} \right]$. The parameters in Eqs.~(\ref{unfoldm2}) and (\ref{unfoldb2}) have been explained in Eqs.~(\ref{unfoldm}) and (\ref{unfoldb})\textcolor{black}{, respectively}. For $\text{proxNet}_{\theta_a^{(s)}}(\cdot)$, it is a ResNet only consisting of one Resblock with the parameters ${\theta_a^{(s)}}$. Specifically, the Resblock simply contains two linear layers followed by a normalization operation at the second dimension of $\alpha$.\footnote{Please refer to the supplementary file for more details about DRCDNet.}

\begin{figure}[t]
  \begin{center}
     \includegraphics[width=1.0\linewidth]{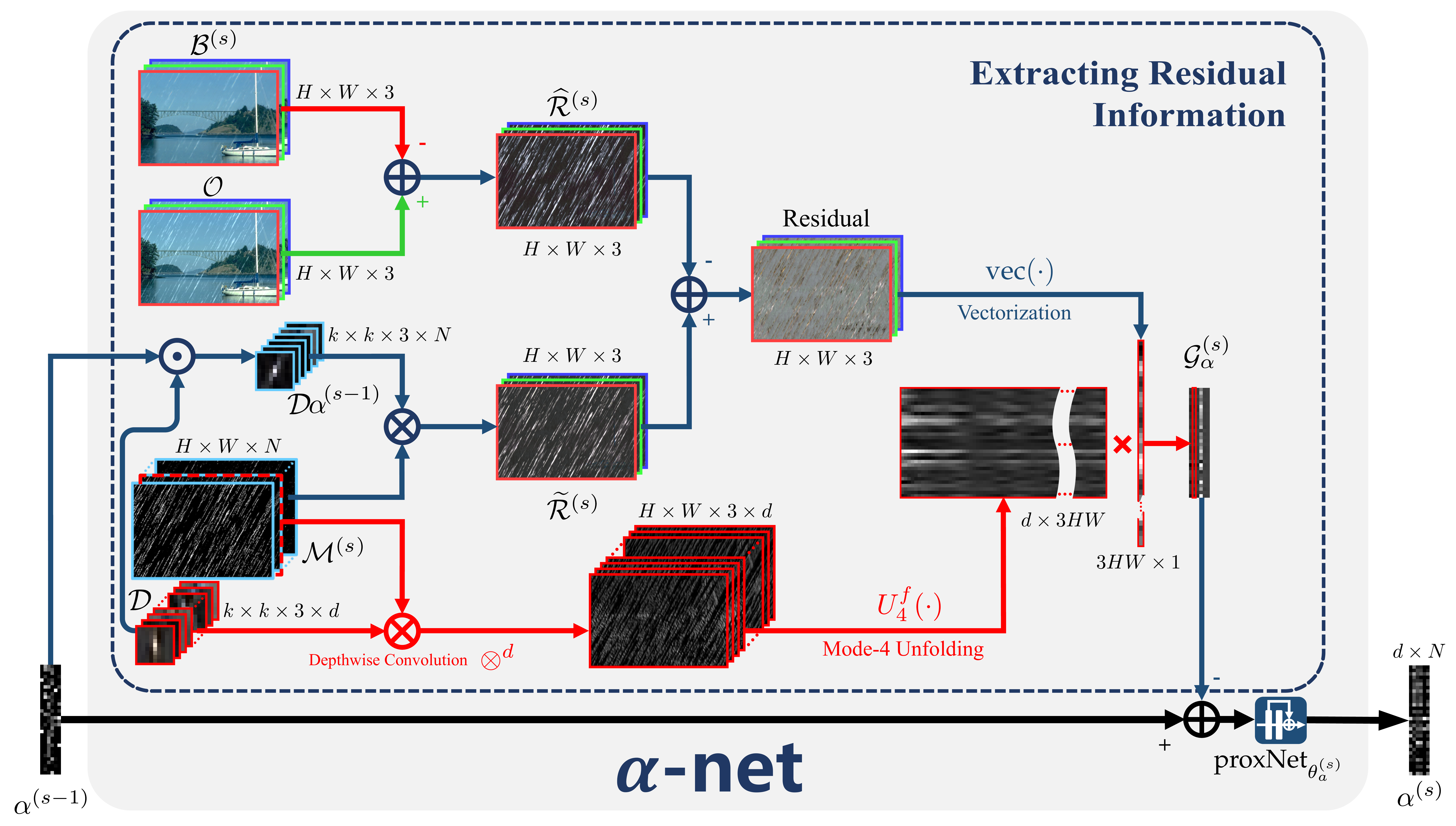}
  \end{center}
  \vspace{-5mm}
     \caption{The $\alpha$-net of the proposed DRCDNet at the $s^{\text{th}}$ stage. Every network module is correspondingly constructed based on Eq.~(\ref{unfoldalpha}).}
  \label{alphanet}
   \vspace{-7mm}
  \end{figure}

Then, by transforming the operators in (\ref{unfoldm2}), (\ref{unfoldb2}), and (\ref{unfoldalpha}) step-by-step, we can construct the DRCDNet. Clearly, at each stage, the DRCDNet is composed of three sub-networks, i.e., M-net, B-net, and $\alpha$-net. Specifically, by comparing (\ref{unfoldm}), (\ref{unfoldb}) with (\ref{unfoldm2}), (\ref{unfoldb2}), respectively, we can directly construct the M-net and B-net by replacing the rain kernel  $\mK$ in Fig.~\ref{figflow}(b) with $\mD\alpha^{(s-1)}$. For $\alpha$-net, its structure is built as shown in Fig.~\ref{alphanet}. {\textcolor{black}{In this DRCDNet, all the involved parameters, including $\{\theta_m^{(s)},\theta_b^{(s)},\theta_a^{(s)}\}_{s=1}^{S}$, rain kernel dictionary $\mD$, $\eta_1$, $\eta_2$, and $\eta_3$, can be automatically learned from training data (\emph{i.e.}, the paired clean background image $\mathcal{B}$ and the rainy image $\mathcal{O}$) in an end-to-end manner.}}

\noindent\textit{\textbf{Remark 2:}} Similar to the CRCDNet, all the network modules in DRCDNet are correspondent to the iterative computations~(\ref{updatem2}), (\ref{updateb2}), and (\ref{updatealpha}) and thus the DRCDNet also has clear interpretability. Compared with CRCDNet, DRCDNet has specific merits. First, at the testing phase, although the common rain kernel dictionary $\mD$ is pre-trained and fixed, the dynamic inference of $\alpha$ makes it possible to achieve the flexible prediction of rain kernel $\mD\alpha$ according to the rain types of variant testing rainy images. {\textcolor{black}{Besides, in CRCDNet, the rain kernels $\mathcal{K}$ are utilized to represent the entire dataset. As compared with the entire dataset which contains more rain types, depicting a specific rainy image should need fewer rain kernels $\mathcal{D}\alpha$. Hence, we can choose smaller $N$ for DRCDNet.}} Equivalently, the channel number of rain map $\mathcal{M}$ is also smaller than that in CRCDNet. Under this setting, the hidden space for estimating rain layer is greatly shrunk, which naturally helps improve the generalization ability. This is comprehensively substantiated in Sec.~\ref{mismatch}.

\noindent\textit{\textbf{Remark 3:}} Compared with the general channel attention mechanism, the $\alpha$-net has specific characteristics. First, instead of weighting feature maps on the channel dimension, we focus on weighting the rain kernel dictionary $\mD$, which would save the computational cost. Second, the $\alpha$-net is built based on an optimization algorithm and thus it has clear physical interpretability. Third, as shown in Fig.~\ref{kernelcompare}, the obtained rain kernel $\mK$ has obvious physical meanings, which validates the effectiveness of such weighting operators.

\vspace{-2mm}
\section{Network Training}\label{train}
\noindent\textbf{Training Loss.} For simplicity, we adopt the mean \textcolor{black}{squared} error (MSE)~\cite{Jing2012Removing} for the learned background and the rain layer at every stage as the training objective function:
\begin{equation}\label{Loss}
  L = \sum_{s=0}^{S}\rho_{s}\left\|\mathcal{B}^{(s)}\!-\!\mathcal{B} \right\|_F^2\!+\!\sum_{s=1}^{S}\gamma_{s}\left\| {\mathcal{O}\!-\!\mathcal{B}\!-\!\mathcal{R}}^{(s)} \right\|_F^2,
\end{equation}
where $\mathcal{B}^{(s)}$ and $\mathcal{R}^{(s)}$ separately denote the derained result and extracted rain layer at the $s^{\text{th}}$ stage ($s=0,1,\cdots,S$), as expressed in Eq.~(\ref{unfoldb}) for CRCDNet and Eq.~(\ref{unfoldb2}) for DRCDNet. $\mB^{(0)}$ is initialized by a convolutional operator on $\mO$. $\rho_{s}$ and $\gamma_{s}$ are tradeoff parameters, and simply set as $\rho_{S}=\gamma_{S}=1$ and others as $0.1$ for all experiments to make the outputs at the final stage play a dominant role. More parameter settings are discussed in the supplementary file.

\noindent\textbf{\textcolor{black}{Implementation} Details.}
We use PyTorch~\cite{paszke2017automatic} to {\textcolor{black}{implement our method and the network is trained based on an}} NVIDIA GeForce GTX 1080Ti GPU. For both CRCDNet and DRCDNet, we adopt the Adam optimizer~\cite{Kingma2014Adam} with the batch size of 10 and the patch size of 64$\times$64. The initial learning rate is 0.001 and divided by 5 every 25 epochs. The total epoch number is 100. It is worth mentioning that {\textcolor{black}{we use these same parameter settings for all experiments}}. This would show the favorable robustness and generality of our method.

\vspace{-1mm}\section{Experimental Results}\label{exp}
We first conduct model verification to verify the working mechanisms of the proposed network. Then we evaluate the superiority of CRCDNet by comparing it with other SOTA single image derainers based on synthetic datasets. Finally, the performance of DRCDNet is verified by generalization experiments where rain patterns are obviously different between training samples and testing ones.

\vspace{-3mm}\subsection{Details Explanations}
\noindent\textbf{Benchmark Datasets.} Eight datasets are adopted as listed in Table~\ref{tab:dataset}, including five synthesized ones and three real ones. {\textcolor{black}{Similar to other supervised methods, during the training process, what we explicitly need are the paired clean image $\mathcal{B}$ and the rain-affected image $\mathcal{O}$.}}\footnote{Detailed explanations are included in supplementary material.}
\begin{table}[t]
\centering
\caption{Benchamark datasets with different rain types.}\vspace{-2mm}
\label{tab:dataset}
\setlength{\tabcolsep}{1.7mm}{
\begin{tabular}{c|c|c|c}
\Xhline{0.9pt}
Dataset & {\textcolor{black}{\#Training pairs}} & {\textcolor{black}{\#Testing pairs}} & Scenario\\
\Xhline{0.9pt}
Rain100L~\cite{Yang2019Joint}  & 200 & 100 &Synthetic \\
\hline
Rain100H~\cite{Yang2019Joint} &1,800 &100 & Synthetic \\
\hline
Rain1400~\cite{Fu2017Removing} &12,600 &1,400  &Synthetic\\
\hline
Dense10~\cite{wei2019semi} & 0&10 & Synthetic\\
\hline
Sparse10~\cite{wei2019semi} & 0&10 &Synthetic\\
\hline
SPA-Data~\cite{wang2019spatial} &638,492&1,000 &Real\\
\hline
Internet-Data~\cite{wang2019spatial} &0&146 (no label) & Real\\
\hline
MPID\_Rain+Mist(R)~\cite{benchmark} & 0 & 30 (no label) & Real \\
\Xhline{0.9pt}
\end{tabular}}
\vspace{-3mm}
\end{table}

\begin{table}[t]
\centering
\caption{Effect of stage number $S$ on the performance of CRCDNet.}\vspace{-2mm}
\label{tabS}
\begin{tabular}{c|c|c|c|c|c|c}
\Xhline{0.9pt}
{\textcolor{black}{\#Stage}}  & $S$=0 & $S$=2 & $S$=8 & $S$=11 & $S$=17 & $S$=20\\
\Xhline{0.9pt}
PSNR & 35.93&38.46 &39.60  &39.81 &40.00 &39.91 \\
\hline
SSIM  &0.9689 &0.9813 &0.9850 &0.9855 &0.9860 &0.9858\\
\Xhline{0.9pt}
\end{tabular}
\vspace{-5mm}
\end{table}

\begin{figure}[t]
  \begin{center}
  \vspace{-3mm}
     \includegraphics[width=0.95\linewidth]{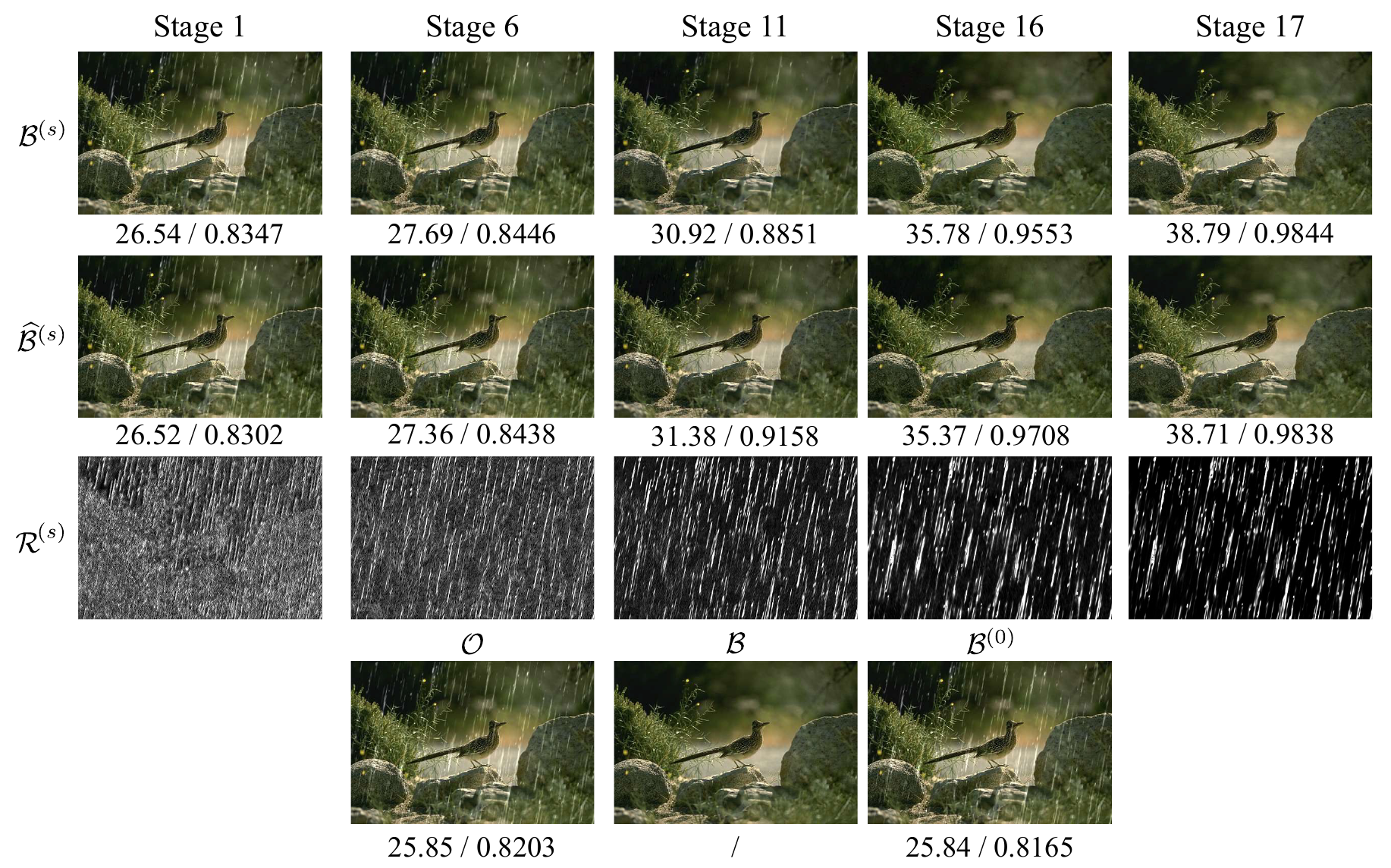}
  \end{center}
  \vspace{-4mm}
     \caption{Visualization of the recovery background $\mB^{(s)}$, $\widehat{\mB}^{(s)}$ of CRCDNet and the rain layer ${\mR}^{(s)}$ at different stages. The total stage number $S$ is 17. PSNR/SSIM are listed below the corresponding results for easy reference.}
  \label{figverstage}
    \vspace{-2mm}
\end{figure}
\begin{figure}[t]
  \begin{center}
     \hspace{-1mm}\includegraphics[width=0.95\linewidth]{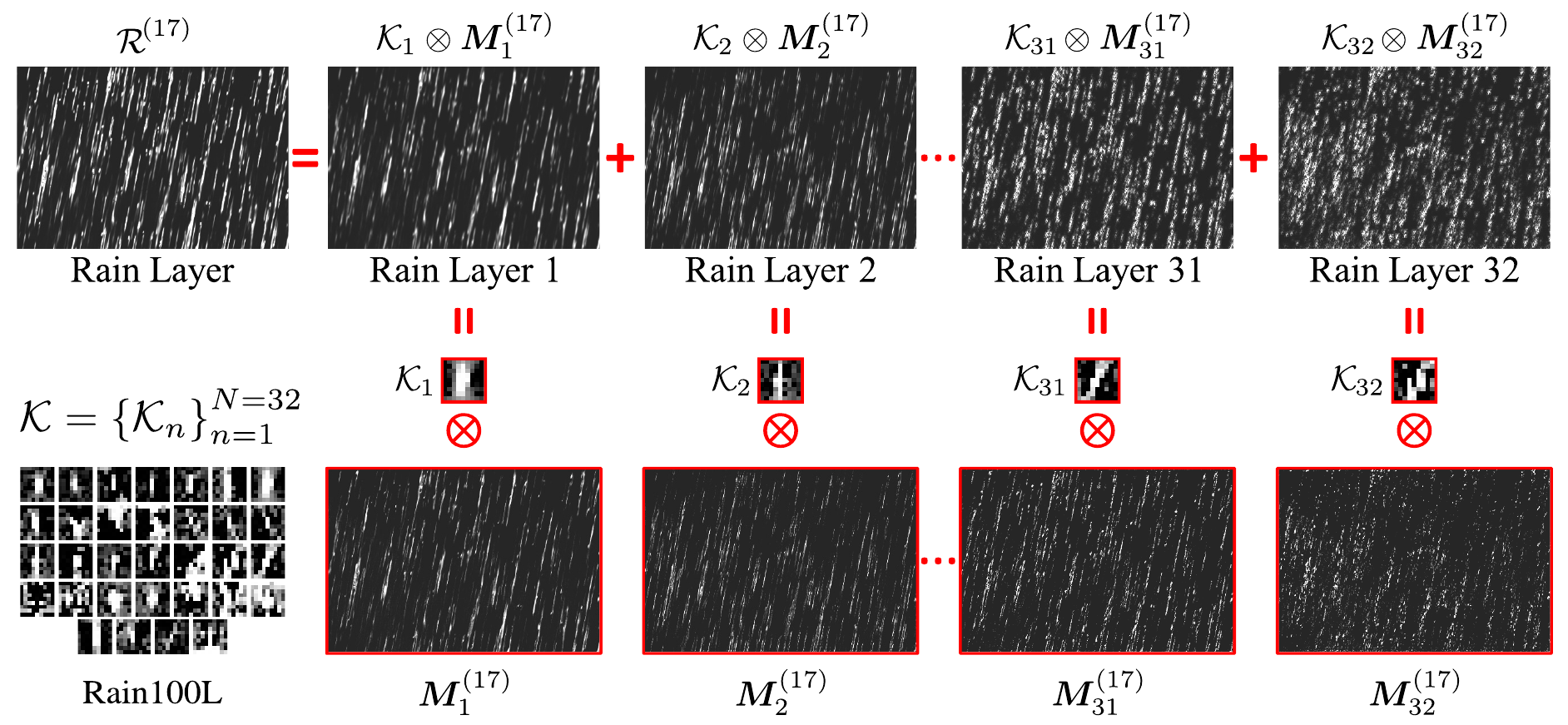}
  \end{center}
  \vspace{-4mm}
     \caption{At the final stage $s=17$, the extracted rain layer, rain kernels ${\mK_{n}}$, and rain maps $\bm{M_{n}}$ for the input $\mO$ in Fig.~\ref{figverstage}. The lower left is the rain kernels ${\mK}$ learned by CRCDNet based on Rain100L training pairs. In CRCDNet, $N=32$.}
  \label{figvercm}
    \vspace{-4mm}
\end{figure}
%
%

\noindent\textbf{Comparison Methods.} We compare our network with current SOTA single image derainers, including:\footnote{The code/project links can be found from \url{https://github.com/hongwang01/Video-and-Single-Image-Deraining}.}

1) Prior-based methods: DSC~\cite{Yu2015Removing} and JCAS~\cite{Gu2017Joint};

2) Deep learning methods: Clear~\cite{Fu2017Clearing}, DDN~\cite{Fu2017Removing}, RESCAN~\cite{li2018recurrent}, PReNet~\cite{ren2019progressive}, SPANet~\cite{wang2019spatial}, and JORDER\_E~\cite{Yang2019Joint};

3) Semi-supervised method: SIRR~\cite{wei2019semi}.

\noindent\textbf{Performance Metrics.}
For paired data, the classical metrics are PSNR~\cite{Huynh2008Scope} and SSIM~\cite{wang2004image}. Since the human visual system is sensitive to the luminance channel (Y) channel of a color image in the YCbCr space, similar to ~\cite{Li2016Rain,ren2019progressive,zhang2019image}, we also compute PSNR and SSIM based on the luminance channel. While for unlabel data, we adopt the non-reference indicators, i.e., naturalness image quality evaluator (NIQE)~\cite{mittal2012making} and blind/referenceless image spatial quality evaluator (BRISQUE)~\cite{mittal2012no}. \textcolor{black}{Specifically, PSNR and SSIM generally measure the intensity fidelity and structure fidelity relative to a reference image, respectively. NIQE and BRISQUE aim to quantify the quality of a distorted image in a way which matches the human judgments of visual quality as closely as possible. Higher PSNR and SSIM, as well as lower NIQE and BRISQUE, indicate better result.}

\vspace{-4mm}\subsection{Model Verification}\label{visualanalysis}
Here we utilize Rain100L to execute the model verification.

\noindent\textbf{Stage Number $S$.}
Table~\ref{tabS} reports the effect of stage number $S$ on deraining performance of the proposed CRCDNet. Here, $S=0$ represents the fact that without adopting the RCD mechanism, the initialization $\mathcal{B}^{(0)}$ is directly regraded as the final rain-removed result.
Taking $S=0$ as a baseline, it is easily seen that with only two stages, our method already achieves significant rain removal performance improvement, substantiating the essential role of the constructed M-net and B-net. We also find that when $S=20$, there are no further obvious performance gains, since larger $S$ would make the gradient propagation more difficult. Based on this observation, we set $S$ as 17 for the CRCDNet throughout all our experiments. More discussions are listed in the supplemental file.

\noindent\textbf{Network Visualization.}
We then visually show how the interpretability of CRCDNet facilitates an easy analysis on the working mechanism inside the network modules. Fig.~\ref{figverstage} presents the extracted background $\mB^{(s)}$ ($1^{\text{st}}$ row), $\widehat{\mB}^{(s)}$($2^{\text{nd}}$ row) that represents the role of M-net in helping restore clean background, and rain layer ${\mR}^{(s)}$ ($3^{\text{rd}}$ row) at different stages of CRCDNet. We can find that with the increase of $s$, ${\mR}^{(s)}$ covers more rain streaks and fewer image details, and $\widehat{\mB}^{(s)}$ and $\mB^{(s)}$ are also gradually ameliorated. These should be attributed to the proper guidance of the RCD prior for rain streaks and the mutual promotion of M-net and B-net that enables the CRCDNet to be evolved to a right direction.

\begin{figure*}[h]
  \begin{center}
  \vspace{-2mm}
     \includegraphics[width=0.93\linewidth]{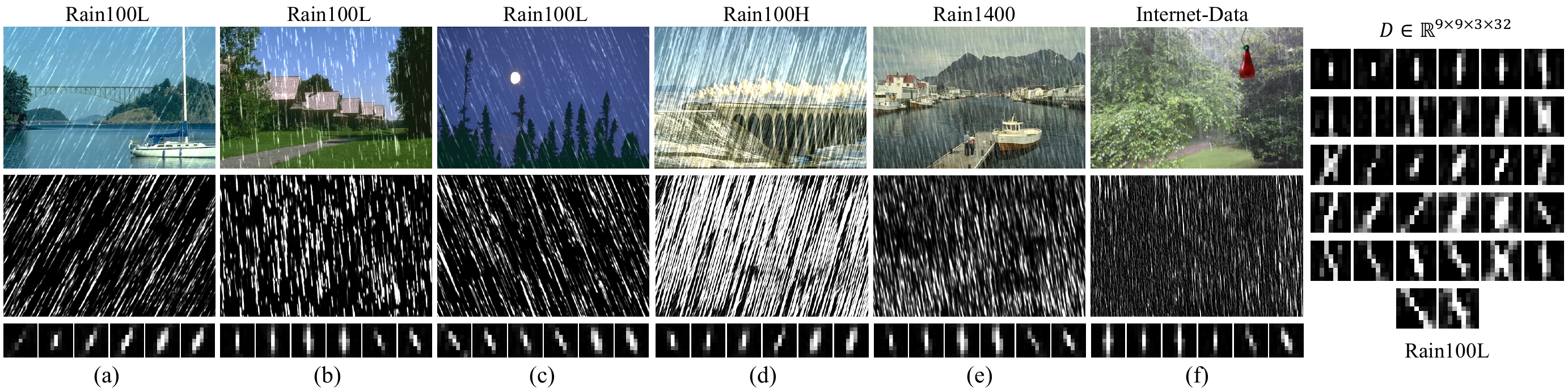}
  \end{center}
  \vspace{-5mm}
     \caption{The left: for each column in (a)-(f), ($1^{\text{st}}$ row) rainy images, ($2^{\text{nd}}$ row)  the extracted rain layers $\mathcal{R}^{(11)}= \sum_{n=1}^{N=6}\mD\alpha_{n}^{(11)}\otimes\bm{M}_{n}^{(11)}$,  and ($3^{\text{rd}}$ row)  the corresponding rain kernels $\mathcal{D}\alpha^{(11)}=\{\mD\alpha_{n}^{(11)}\}^{N=6}_{n=1}$ dynamically predicted by DRCDNet. The right:  the rain kernel dictionary ${\mD}$ learned by DRCDNet based on Rain100L training set. Especially, the rainy images in (a)-(c) are from Rain100L testing set and the ones in (d)-(f) are from another testing set. In DRCDNet, $S=11$, $d=32$, and $N=6$. More explanations are included in the supplemental file.}
  \label{figverdrcd}
    \vspace{-4mm}
\end{figure*}
\noindent\textbf{RCD Model Visualization.}
For the input $\mO$ in Fig.~\ref{figverstage}, the rain kernels and the rain maps learned by CRCDNet are presented in Fig.~\ref{figvercm}.
Clearly, the CRCDNet finely extracts proper rain layers explicitly complying with the RCD model (\ref{e3}). This not only verifies the reasonability of our method but also manifests the peculiarity of our proposal. On one hand, we utilize an M-net to learn sparse rain maps instead of directly learning rain streaks that makes learning process easier. On the other hand, we exploit training data to automatically learn rain kernels representing general repetitive local patterns of rain with diverse shapes. This facilitates their general applicability to more real-world rainy images.

\noindent\textbf{Rain Kernel Visualization.} By training DRCDNet on Rain100L, the learned rain kernel dictionary $\mathcal{D}$ is shown at the right of Fig.~\ref{figverdrcd}. As compared with the dictionary $\mK$ learned by CRCDNet in Fig.~\ref{figvercm}, we can easily find that the rain kernels in $\mathcal{D}$ are fairly diverse. With the trained model on Rain100L, we test typical rainy samples from different sources, including training/test domain match cases (a)-(c) and mismatch cases (d)-(f). As shown in each column of (a)-(f), the extracted rain layers ($2^{\text{nd}}$ row) contain {\textcolor{black}{fewer}} background details, and the inferred rain kernels ($3^{\text{rd}}$ row) are finely in accordant with the rain patterns (e.g., directions, scales, and thickness) in input rainy images ($1^{\text{st}}$ row). Besides, we can also observe that the rain kernels ($3^{\text{rd}}$ row) for every testing sample are not simply selected from $\mathcal{D}$, and they are adaptively inferred by DRCDNet, even with new rain patterns not in $\mathcal{D}$. This not only validates the effectiveness of the dynamic RCD modelling manner (\ref{dcsc}) for rain layer, but also reflects the advantages of the DRCDNet over adaptive inference. 

\begin{figure*}[t]
  \begin{center}
     \includegraphics[width=0.93\linewidth]{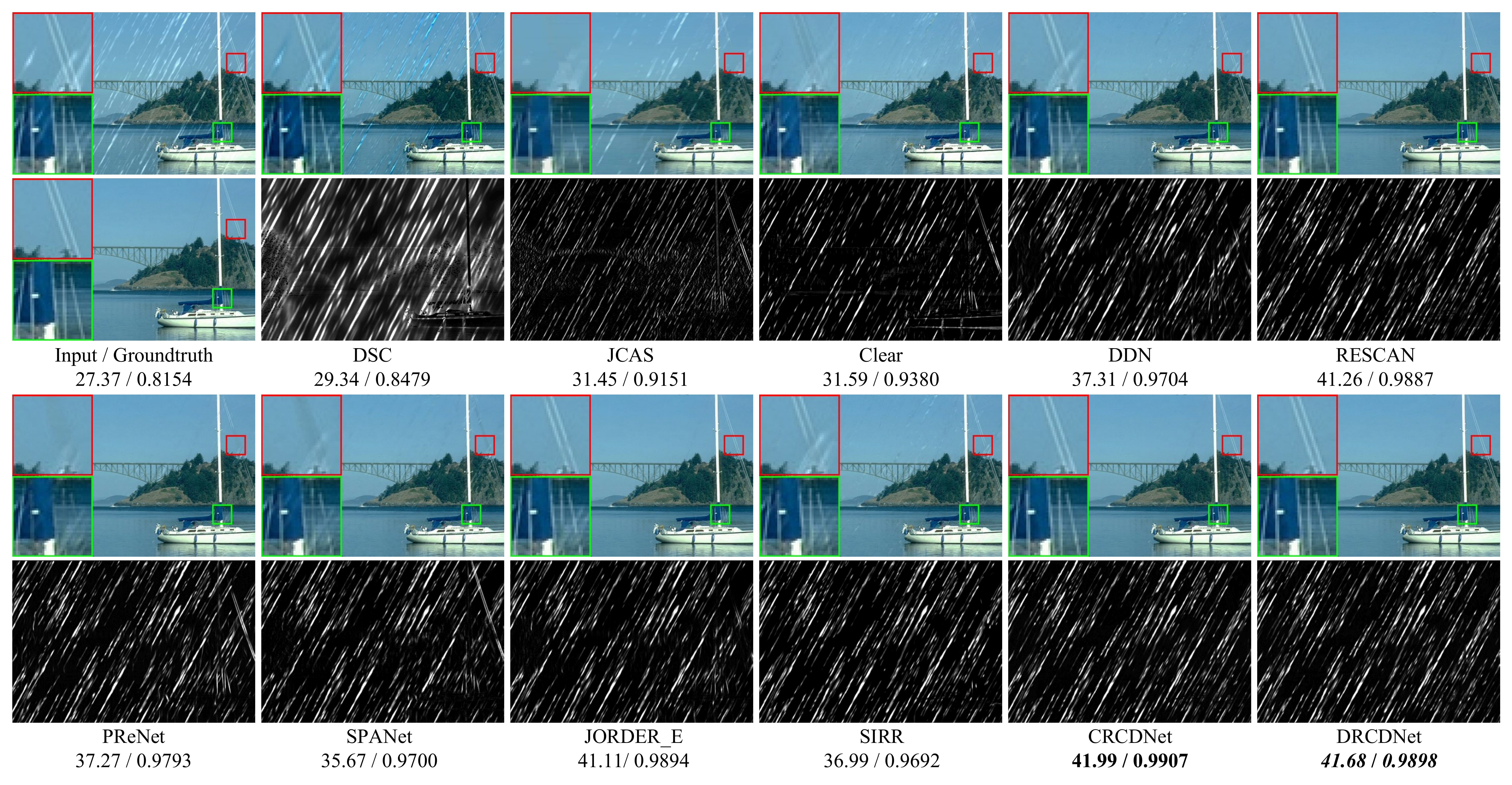}
  \end{center}
  \vspace{-7mm}
     \caption{(Training-test domain match) $1^{\text{st}}$ column: input rainy image from Rain100L (upper) and the  corresponding groundtruth (lower). $2^{\text{nd}}$-$12^{\text{th}}$ columns: derained results (upper) with two demarcated areas zoomed in 4 times for easy observation, and extracted rain layers (lower) by 11 competing methods. PSNR/SSIM listed below every derained result is for easy comparison. Bold and bold italic indicate top $1^{\text{st}}$ and $2^{\text{nd}}$ best results, respectively. }
  \label{figrainl}
   \vspace{-6mm}
\end{figure*}


\vspace{-2mm}\subsection{Training-Test Domain Match Experiments}
In this section, we evaluate the proposed CRCDNet and DRCDNet in the case that the rain types of testing data are consistent with that of training data, based on the benchmark datasets including  Rain100L, Rain100H, and Rain1400.


Fig.~\ref{figrainl} illustrates the deraining performance of all competing methods on a rainy image from Rain100L.
As shown, for background recovery, traditional model-based DSC and JCAS leave obvious rain streaks, and deep derainers lose certain useful image textures. However, the proposed CRCDNet and DRCDNet perform better in sufficiently removing the rain streaks and finely preserving image details. \emph{Moreover, we can easily observe that the rain layers extracted by CRCDNet and DRCDNet both contain fewer unexpected background details, which validates the reliability of embedding the RCD prior constraints into the network design.\footnote{{\textcolor{black}{More visual experimental results on Rain100H and Rain1400 are provided in the supplemental file.}}}}



\begin{table}[t]
\centering
\caption{Training-test domain match case: Average PSNR and SSIM comparisons on three datasets. Bold and bold italic indicate top $1^{\text{st}}$ and $2^{\text{nd}}$ best results, respectively.}\vspace{-1mm}
\setlength{\tabcolsep}{3.6pt}
\begin{tabular}{c|cc|cc|cc}
\Xhline{0.9pt}
  Datasets & \multicolumn{2}{@{}c@{}}{Rain100L}&\multicolumn{2}{|c@{}}{Rain100H}&\multicolumn{2}{|c@{}}{Rain1400} \\
\Xhline{0.9pt}
  Metrics & PSNR$\uparrow$ & SSIM$\uparrow$ & PSNR$\uparrow$ & SSIM$\uparrow$  & PSNR$\uparrow$ & SSIM$\uparrow$ \\
\hline
  Input & 26.90 & 0.8384 & 13.56 & 0.3709 & 25.24 & 0.8097 \\
\hline
  DSC\cite{Yu2015Removing} & 27.34 & 0.8494 & 13.77 & 0.3199  & 27.88 &0.8394 \\
\hline
  JCAS\cite{Gu2017Joint}  & 28.54 & 0.8524 & 14.62 & 0.4510 &26.20 & 0.8471 \\
\hline
  Clear\cite{Fu2017Clearing} &30.24 & 0.9344 & 15.33 & 0.7421 & 26.21& 0.8951 \\
\hline
  DDN\cite{Fu2017Removing}& 32.38 & 0.9258 & 22.85 & 0.7250 & 28.45 & 0.8888 \\
\hline
  RESCAN\cite{li2018recurrent}   & 38.52& 0.9812 &29.62 & 0.8720 &32.03& 0.9314\\
\hline
  PReNet\cite{ren2019progressive}& 37.54& 0.9795 &30.08& \textit{\textbf{0.9050}} & 32.09& 0.9418\\
\hline
  SPANet\cite{wang2019spatial} & 35.33 & 0.9694 &25.11 & 0.8332 & 29.85& 0.9148 \\
\hline
  JORDER\_E\cite{Yang2019Joint} &37.89 &0.9803& 30.21&0.8957 &32.00 & 0.9347 \\
\hline
  SIRR\cite{wei2019semi} & 32.37 & 0.9258 & 22.47 & 0.7164 & 28.44 & 0.8893\\
\hline
CRCDNet & \textbf{40.00} & \textbf{0.9860} & \textbf{31.28} & \textbf{0.9093} & \textbf{33.04} & \textbf{0.9472}\\
\hline
DRCDNet& \emph{\textbf{39.66}} &\textit{\textbf{0.9852}}&\textit{\textbf{30.50}}&0.8974 &\textit{\textbf{33.03}}&\textit{\textbf{0.9466}}\\
\Xhline{0.9pt}
\end{tabular}
\vspace{-6mm}
\label{tabsyn}
\end{table}


Table~\ref{tabsyn} reports the average PSNR and SSIM computed on the entire testing data of each synthesized dataset. It is seen that in this training-testing domain match case, our CRCDNet attains significant deraining performance on each evaluation dataset and DRCDNet performs comparable to CRCDNet.


\begin{figure*}[t]
  \begin{center}
 \vspace{-2mm}
     \includegraphics[width=0.94\linewidth]{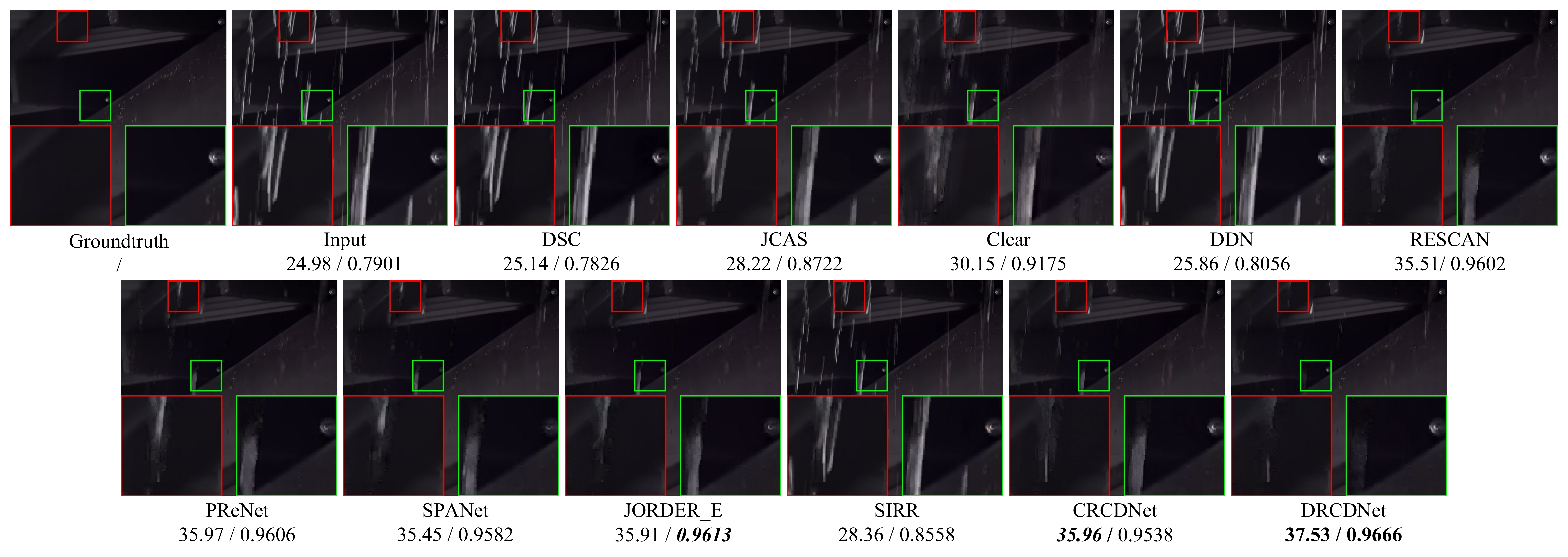}
  \end{center}
  \vspace{-7mm}
     \caption{(Training-test domain mismatch case) From left/upper to right/lower: groundtruth and input rainy image from the testing samples of SPA-Data, derained results of traditional DSC and JCAS, generalized results of all DL based competing methods trained on Rain100L.}
  \label{figlspa}
   \vspace{-3mm}
\end{figure*}
\begin{table*}[t]
\centering
\caption{Training-test domain mismatch case: average PSNR and SSIM comparisons on Dense10 and Sparse10. DL based methods are trained on Rain100H. Bold and bold italic indicate top $1^{\text{st}}$ and $2^{\text{nd}}$ best results, respectively.}
\vspace{-2.5mm}
\setlength{\tabcolsep}{4.45pt}
\begin{tabular}{c|c|ccccccccccccc}
\Xhline{0.9pt}
Datasets & Metrics& Input &DSC  &JCAS &Clear &DDN  & RESCAN &PReNet  &SPANet  &JORDER\_E  & SIRR &CRCDNet &DRCDNet\\
\Xhline{0.9pt}
\multirow{2}*{Dense10} & PSNR$\uparrow$

&19.17	&20.85	&19.93	&19.10	&21.38	&21.81	&21.91	&\emph{\textbf{22.11}}	&21.66	&21.23	&22.07	&\textbf{22.47}\\
&SSIM$\uparrow$ &0.8495	&0.8811	&0.8694	&0.8600	&0.8965	&0.9073	&0.9236	&0.9170	&0.9093	&0.8925	&\emph{\textbf{0.9241}}	&\textbf{0.9255}\\
\hline
\multirow{2}*{Sparse10} &PSNR$\uparrow$
&25.42	&26.37	&26.38	&24.45	&27.83	&28.73	&29.02	&27.55	&27.66	&27.48	&\emph{\textbf{29.07}}	&\textbf{29.28}\\
&SSIM$\uparrow$  &0.8956	&0.8989	&0.9043	&0.8785	&0.9249	&0.9337	&0.9412	&0.9301	&0.9257	&0.9181	&\emph{\textbf{0.9422}}	&\textbf{0.9431}\\
\Xhline{0.9pt}
\end{tabular}
\label{tabtodensesparse}
\vspace{-3mm}
\end{table*}
\begin{table*}
\vspace{-0mm}
\begin{center}
\caption{Training-test domain mismatch case: average PSNR and SSIM comparisons on the testing data of SPA-Data. }
\vspace{-2.5mm}
\setlength{\tabcolsep}{6.3pt}
\begin{tabular}{c|cccccccccccc}
\Xhline{0.9pt}
&   Input &DSC&JCAS&Clear&DDN & RESCAN & PReNet &SPANet & JORDER\_E & SIRR & CRCDNet &DRCDNet\\
  \Xhline{0.9pt}
    \multicolumn{13}{c}{\scriptsize{Data (training/testing):  Rain100L/SPA-Data; ~Difficulty: high.}}\\
    \hline
   PSNR$\uparrow$   &34.15   &34.83   &34.95   & 32.66   &34.66   &34.70  &34.91 &\textit{\textbf{35.13}} &35.04  &34.66 &34.88 &\textbf{35.23} \\
 SSIM$\uparrow$    &0.9269   &0.9410   &\textbf{0.9451}    & 0.9420   &0.9346  &0.9376 &0.9407 &\textit{\textbf{0.9443}} &0.9405 &0.9350 &0.9377 &0.9407\\

   \hline
     \multicolumn{13}{c}{\scriptsize{Data (training/testing):  Rain100L+Rain1400/SPA-Data; ~Difficulty: high.}}\\
    \hline
PSNR$\uparrow$   &34.15   &34.83   &34.95 &31.48  &34.67  &33.98  &34.65  &32.25    &33.98   &34.51  &\textit{\textbf{35.26}}  &\textbf{35.53}\\
SSIM$\uparrow$   &0.9269   &0.9410   &0.9451  &0.9357 &0.9410 &0.9432 &0.9411 &0.9393   &0.9413  &0.9336  &\textit{\textbf{0.9455}} &\textbf{0.9512}\\
\Xhline{0.9pt}
\end{tabular}
\label{tabtospa}
\normalsize
\end{center}
\vspace{-3mm}
\end{table*}

\begin{table*}
\vspace{-2mm}
\begin{center}
\caption{Training-test domain mismatch case: average BRISQUE and NIQE comparisons on Internet-Data. }\label{tabtowangreal} 
\vspace{-2.5mm}
\setlength{\tabcolsep}{5.5pt}
\begin{tabular}{c|cccccccccccc}
\Xhline{0.9pt}
&   Input &DSC&JCAS&Clear&DDN & RESCAN & PReNet &SPANet & JORDER\_E  & SIRR & CRCDNet &DRCDNet\\
 \Xhline{0.9pt}
    \multicolumn{13}{c}{\scriptsize{Data (training/testing):  Rain100H/Internet-Data; ~Difficulty: high.}}\\
    \hline
  BRISQUE$\downarrow$ & 28.52 &25.52 &38.03 &31.76 &27.23 & 28.30 &\textit{\textbf{ 26.81}} &27.42 &27.64 & 27.78 &28.32 &\textbf{26.05}\\
  NIQE$\downarrow$ & 5.1039 &5.0539  & 5.3634 &5.2781 &4.8501& \textit{\textbf{4.6730}} &4.6765 & 4.9052 &4.6869 & 4.8276 &4.7677 &\textbf{4.6606}\\
    \hline
     \multicolumn{13}{c}{\scriptsize{Data (training/testing):  Rain100L+Rain100H/Internet-Data; ~Difficulty: high.}}\\
    \hline
BRISQUE$\downarrow$ & 28.52 & 25.52 &38.03   &32.39 &27.05 & 25.80 & \textit{\textbf{25.01}} &26.74 &25.48 & 26.44 &26.00 & \textbf{23.35}\\
NIQE$\downarrow$ & 5.1039 &5.0539  &5.3634  &5.3539 &4.8806 & \textit{\textbf{4.4377}} &4.4630 & 4.8273 &4.5247 &4.8204 &4.6082 &\textbf{4.4207} \\
\Xhline{0.9pt}
\end{tabular}
\normalsize
\end{center}
\vspace{-6mm}
\end{table*}

\vspace{-3mm}\subsection{Training-Test Domain Mismatch Experiments}\label{mismatch}
Here we evaluate the CRCDNet and DRCDNet in the case that rain types are inconsistent between training and testing.

\vspace{0mm}
\noindent\textbf{Performance Comparison on Dense10 and Sparse10.} We first adopt Dense10 and Sparse10 to evaluate the generalization capability of all DL competing methods trained on Rain100H. From the quantitative results in Table~\ref{tabtodensesparse}, we can find that CRCDNet is still competing to the SOTA and DRCDNet even obtains higher PSNR and SSIM. \emph{This tells us that the proper embedding of prior constraints are helpful to alleviate the over-fitting issue and the dynamic inference model would make the space for rain layer estimation tighter and then help further \textcolor{black}{improve} the generalization performance.}\footnote{{\textcolor{black}{More visual experimental results are provided in the supplemental file.}}}





\vspace{0mm}
\noindent\textbf{Performance Comparison on SPA-Data.} This real dataset is composed of complicated rain patterns, diverse shooting scenes, and rich background details. All these factors bring great challenges to accurate rain layer extraction and the model generalization performance on the dataset. Fig.~\ref{figlspa} displays the reconstructed images where deep methods are trained on Rain100L. Clearly, the proposed DRCDNet performs better on both rain removal and detail preservation. 

Table~\ref{tabtospa} provides the quantitative comparisons under different testing scenarios. 
As for the generalization case from Rain100L to SPA-Data, although DRCDNet achieves the higher PSNR and SSIM than CRCDNet, due to the simplicity of rain types in Rain100L and the complexness of rainy samples in SPA-Data, the generalization performance of DRCDNet is not prominent. However, by utilizing Rain100L and Rain1400 with 14 rain types as training data, the generalization performance of DRCDNet is largely improved.



\noindent\textbf{Performance Comparison on Internet-Data.} 
Table~\ref{tabtowangreal} listed the quantitative comparisons on the real Internet-Data where all the DL based deraining models are trained on Rain100H and both Rain100L and Rain100H, respectively. As seen, our DRCDNet consistently achieves the lowest BRISQUE and NIQE, showing better generalization performance.\footnote{\textcolor{black}{More experiments are provided in the supplemental file.}}





\begin{figure*}[t]
  \begin{center}
     \includegraphics[width=0.95\linewidth]{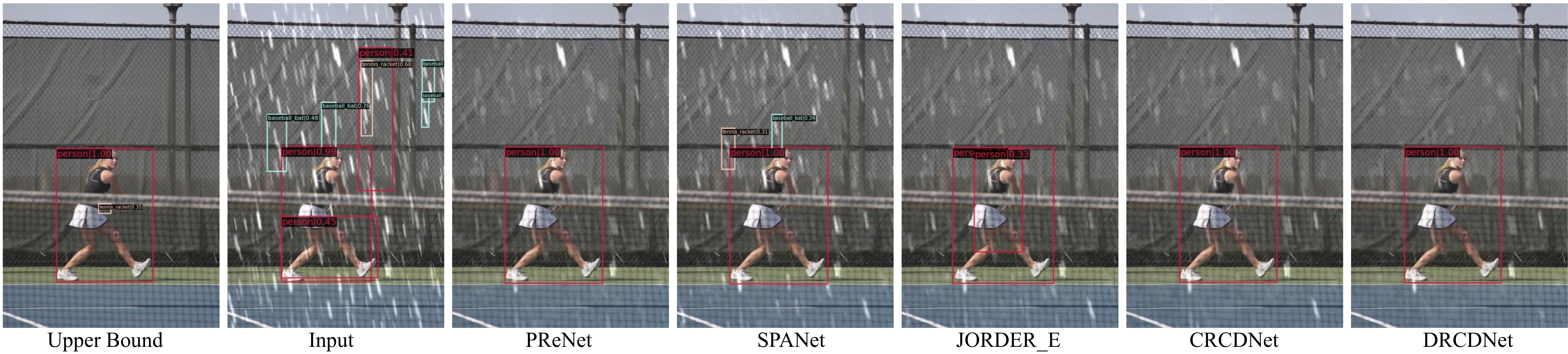}
  \end{center}
  \vspace{-6mm}
     \caption{\color{black}{{Object detection results of Faster RCNN on the synthesized COCO val2017 dataset. All the DL based deraining methods are trained on Rain100L.}}}
  \label{figdet}
    \vspace{-3mm}
\end{figure*}
\begin{figure*}[t]
  \begin{center}
     \includegraphics[width=0.95\linewidth]{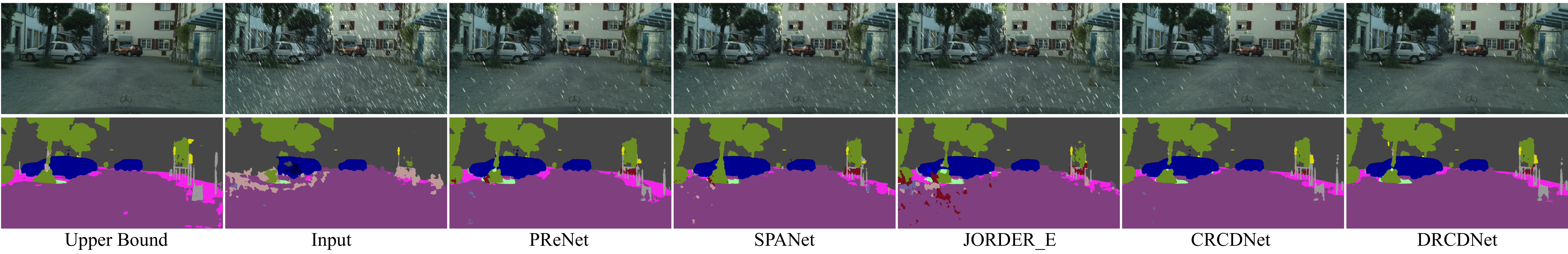}
  \end{center}
  \vspace{-6mm}
     \caption{\color{black}{{Visual results of joint deraining and semantic segmentation on the synthesized Cityscapes validation dataset. All the DL based deraining methods are trained on Rain100L. For each method, the first row is the generalized deraining result and the second row is the corresponding segmentation result on with the PSPNet segmentation network.}}}
  \label{figseg}
    \vspace{-5mm}
\end{figure*}
\begin{table*}[t]
\centering
\caption{\textcolor{black}{Object detection results (\%) on the synthesized COCO val2017~\cite{lin2014microsoft} with the metric as mAP@[.5, .95]. For DL based methods, the models are trained on Rain100L. Bold and bold italic indicate top $1^{\text{st}}$ and $2^{\text{nd}}$ best results, respectively.}}
\vspace{-2.5mm}
\setlength{\tabcolsep}{4.4pt}
\textcolor{black}{\begin{tabular}{c|cccccccccccc|c}
\Xhline{0.9pt}
Methods &  Input &DSC  &JCAS &Clear &DDN  & RESCAN &PReNet  &SPANet  &JORDER\_E  & SIRR &CRCDNet &DRCDNet & Upper Bound\\
\Xhline{0.9pt}
Faster RCNN &30.7&31.3&32.0&34.7 &35.1 &37.2 &37.6 &36.7 &37.6 &35.2 &\textit{\textbf{38.1}} &\textbf{38.3} &42.1\\
Mask RCNN &31.7 &32.5&33.4&36.1 &36.8 &38.8 &39.2 &36.9 &39.1 &36.9  &\textit{\textbf{39.7}} &\textbf{39.9} &44.5\\
YOLOv3 &14.4 &15.6&16.3 &18.0 &18.5 &19.8 &20.6 &19.9 &20.3 &18.4  &\textit{\textbf{20.8}} &\textbf{21.0} &22.2\\
\Xhline{0.9pt}
\end{tabular}}
\label{tabdet}
\vspace{-4mm}
\end{table*}
\begin{table*}[t]
\centering
\caption{\textcolor{black}{Semantic image segmentation results (\%) on the synthesized Cityscapes validation set~\cite{cordts2016cityscapes} with the metric as MIOU.}}
\vspace{-2.5mm}
\setlength{\tabcolsep}{4.4pt}
\textcolor{black}{\begin{tabular}{c|cccccccccccc|c}
\Xhline{0.9pt}
Methods &Input &DSC  &JCAS &Clear &DDN  & RESCAN &PReNet  &SPANet  &JORDER\_E  & SIRR  &CRCDNet &DRCDNet  &Upper Bound\\
\Xhline{0.9pt}
FCN &33.6 &34.8&37.1&46.6 &52.7 &66.0 &66.8 &66.0 &\textit{\textbf{67.0}} &51.0  &\textit{\textbf{67.0}} &\textbf{68.5} &73.4\\
PSPNet &33.6&36.2&38.6&48.1 &54.1 &69.8 &70.3 &68.1 &70.4 &52.2 &\textit{\textbf{70.9}} &\textbf{72.6} &78.5\\
DeepLabv3+&34.0&38.8&42.2&53.1 &58.1 &71.8 &72.1 &71.7 &72.3&55.9  &\textit{\textbf{72.6}} &\textbf{74.2} &79.8\\
\Xhline{0.9pt}
\end{tabular}}
\label{tabseg}
\vspace{-4mm}
\end{table*}

\vspace{-7.5mm}
\textcolor{black}{\subsection{Downstream Tasks}}
\textcolor{black}{In this subsection, to comprehensively substantiate the effectiveness of our proposed method in rain-removed image restoration, we conduct a series of experiments on the downstream tasks, including object detection and semantic segmentation to investigate the potential of different deraining methods in helping improve the high-level vision performance.}

\vspace{0mm}
\noindent\textcolor{black}{\textbf{Object Detection.}
For the objection detection task, we select the widely-adopted benchmark COCO val2017~\cite{lin2014microsoft}} \textcolor{black}{which} \textcolor{black}{consists of 5,000 images with bounding box annotations. Three popular detection algorithms are adopted, including Faster RCNN~\cite{ren2015faster}, Mask RCNN~\cite{he2017mask}, and YOLOv3~\cite{redmon2018YOLOv3}. Following~\cite{Fu2017Clearing,Yang2019Joint}, we synthesize the corresponding 5,000} \textcolor{black}{rainy images (called synthesized COCO val2017) by using Photoshop},\footnote{\url{https://www.photoshopessentials.com/photo-effects/rain/}}\textcolor{black}{which contains various rain types with different directions, scales, and magnitudes. To execute the downstream task, we firstly utilize all the comparing deraining methods to restore the rain-removed results of the synthesized COCO val2017. Then we adopt the publicly available pre-trained models~\cite{mmdet2020} of Faster RCNN, Mask RCNN, and YOLOv3 to perform the object detection task on these restored images. For quantitative comparison, we evaluate the mean Average Precision (mAP) averaged for IOU $\in$ [0.5 : 0.05 : 0.95] (COCO’s standard metric, simply denoted as mAP@[.5, .95]).}

\textcolor{black}{Fig.~\ref{figdet} presents the objection detection results of Faster RCNN on the rain-removed images obtained by different deraining methods. The ``Upper Bound'' represents the detection result of Faster RCNN on the corresponding clean rain-free image. All the DL based deraining methods are trained on Rain100L. It is obvious that rain streaks adversely degrade the detection accuracy and lead to the fake detection of the target. In addition, due to the corruption of rain streaks, all the methods cannot detect the tennis racket. However, compared with other deraining baselines, our proposed CRCDNet and DRCDNet achieve better visual effects in rain removal as well as detail preservation, which finely boosts the detection results.}

\textcolor{black}{Table~\ref{tabdet} lists the average quantitative results on the synthesized COCO val2017 of the three detection methods, \emph{i.e.}, Faster RCNN, Mask RCNN, and YOLOv3. As seen, the proposed CRCDNet and DRCDNet consistently help these three different detection algorithms obtain higher detection accuracy and achieve about 8\% improvement for mAP@[.5, .95] over the original input. This result substantiates that our proposed method indeed has the capability to accomplish the better restoration of rain-removed images and then help improve the performance of the downstream task in rainy weather conditions, which should be meaningful for practical applications. Besides, we can find that DRCDNet outperforms CRCDNet, showing the effectiveness of the proposed dynamic rain kernel inference mechanism.}

\vspace{0mm}
\noindent\textbf{\textcolor{black}{Semantic Segmentation.}}
\textcolor{black}{For the semantic segmentation task, we adopt the Cityscapes validation set~\cite{cordts2016cityscapes} as the benchmark, including 500 images with pixel-level annotations. Similar to the synthesized COCO val2017, we synthesize the rainy version of the Cityscapes validation set with Photoshop. FCN~\cite{long2015fully}, PSPNet~\cite{zhao2017pyramid}, and  DeepLabv3+~\cite{chen2018encoder} are utilized as the segmentation methods. The corresponding pre-trained models are available at~\cite{mmseg2020}. The commonly-used Mean Intersection over Union (MIOU) is taken as the performance metric for quantitative evaluation.}

\textcolor{black}{Based on an image selected from the synthesized Cityscapes validation set, Fig.~\ref{figseg} shows the visual segmentation results of PSPNet on the rain-removed images obtained by different rain-removal methods.  It is clearly observed that the existence of rain streaks severely corrupts the image details, leading to the bad segmentation result. Attributed to the stronger deraining capability and the better generalization potential, our proposed CRCDNet and DRCDNet 
restore more credible contents with more details, which effectively promote the semantic segmentation with higher accuracy, approaching the ``Upper Bound'' corresponding to the rain-free scenario.}

\textcolor{black}{Table~\ref{tabseg} reports the MIOU of different segmentation methods on the synthesized Cityscapes validation set. 
We can easily observe that under these three segmentation algorithms, 
the segmentation accuracy of the rain-removed images restored by the proposed CRCDNet and DRCDNet consistently shows a significant improvement over that of the original rainy images input by about 40\%, and DRCDNet always obtains the highest scores. In addition, DRCDNet has about 2\% relative improvement over CRCDNet, showing the role of the proposed dynamic rain kernel mechanism in helping improving the generalization performance.
}
\begin{table}[t]
\vspace{-0mm}
\begin{center}
\caption{Comparison of network parameters and test running time for the input image with size 515 $\times$ 512 on GPU.}
\vspace{-2.5mm}
\setlength{\tabcolsep}{3.0pt}
\begin{tabular}{c|c|c|c|c|c}
\Xhline{0.9pt}
Methods & Clear&DDN & RESCAN & PReNet &SPANet \\
  \Xhline{0.9pt}
 Parameter \# &754,691   &57,369   &149,823    &168,963 &283,716\\
    \hline
Time (Seconds) &0.50 &0.61 &0.61 &0.21 &0.42 \\
\hline
\hline
Methods & JORDER\_E & SIRR & CRCDNet &DRCDNet & /\\
  \Xhline{0.9pt}
 Parameter \# &4,169,024   &58,578   &2,858,546 &2,251,406 & /\\
\hline
Time (Seconds) &3.12  &3.65 &0.84 &0.71 & /\\

\Xhline{0.9pt}
\end{tabular}
\label{tabpara}
\normalsize
\end{center}
\vspace{-8mm}
\end{table}

\vspace{1mm}
\subsection{Network Parameters and Inference Time}
Table~\ref{tabpara} presents the comparisons including network parameters and average inference time on an  NVIDIA  GeForce  GTX1080Ti  GPU. This shows that the proposed CRCDNet and DRCDNet are comparable to other competing methods.

\vspace{-3mm}
\section{Conclusion and Future Work}\label{sec:conclusion}
In this paper, we have proposed a novel interpretable network architecture, called RCDNet, specifically for the single image rain removal task. As compared with most of current deep derainers, the peculiarity is that we explicitly embed the intrinsic rain convolutional dictionary (RCD) prior model of rain streaks into deep networks. Besides, 
each module in RCDNet has its own specific physical meanings, and is correspondent to the implementation operators of the algorithm designed for solving the RCD model. This makes the network have easily visualized interpretation for all its module elements and thus facilitates its easy analysis for what happens in the network. Furthermore, considering that the rain patterns of training data are inconsistent with testing data in most real scenarios, we have carefully designed a dynamic rain kernel inference mechanism and correspondingly built an interpretable DRCDNet, which can dynamically infer the corresponding rain kernels complying with diverse rain types of testing rainy images. This helps shrink the space for estimating rain layer and makes the network capable of being finely generalized to testing data even with rain patterns different from training data. All these superiorities have been comprehensively substantiated by a series of experiments, including model verification, network visualization, rain kernel visualization, training/test domain match/mismatch evaluations. Besides, the extracted elements through the end-to-end learning by the network, like the diverse rain kernels, are potentially useful for other related tasks on rainy images.

\textcolor{black}{If the degraded images captured in rainy weather also contain non-streaking rain types, such as mist which is often caused by heavy rains, our proposed RCD prior may not be able to finely represent this degradation form. How to finely execute the joint removal of rain streak, rain mist, and raindrops is a more challenging-yet-meaningful practical problem and deserves further exploration in the future.}

\ifCLASSOPTIONcaptionsoff
  \newpage
\fi
\vspace{-4mm}
\bibliography{refs}

\clearpage

\setcounter{table}{0} 
\setcounter{figure}{0}
\setcounter{equation}{0}
\setcounter{section}{0}

\section*{Supplementary Material}

In this supplementary material, we provide more details on the optimization algorithm, network design, and parameter settings in our experiments. Besides, we show more analysis on model capability and provide more ablation studies. Furthermore, we utilize more representative rainy images with various rain patterns to demonstrate more experimental results for performance comparisons and model verification.

\section{More details of optimization algorithm}
In this section, we provide a detailed derivation for the optimization algorithm given in Section 4.2 of the main text.\footnote{In Section 6 of the main text, we have adopted the similar optimization algorithm for dynamic RCD model.} The expression of the original optimization problem is:
\begin{equation}\label{suppo2}
\min_{{\mM,\mB}}\left\|\mO-\mB-\mK \otimes \mM\right\|_{F}^{2}+\lambda_{1} p_{1}(\mM)+\lambda_{2} p_{2}(\mB),
\end{equation}
where $\mK\in\mathbb{R}^{k\times k\times 3\times N}$ and $\mM\in\mathbb{R}^{H\times W\times N}$. $k \times k$ is the size of rain kernels representing the repetitive local patterns of rain streaks. $\lambda_{1}$ and $\lambda_{2}$ are trade-off parameters. $p_{1}(\cdot)$ and $p_{2}(\cdot)$ mean the penalty functions (i.e., regularizers) to deliver the prior structures of $\mM$ and $\mB$, respectively.

As explained in the main text, we prefer to build a new algorithm for solving the problem (\ref{suppo2}) through alternately updating $\mM$ and $\mB$ by the proximal gradient technique~\cite{beck2009fast}. The details are as follows:

\textbf{Updating $\mM$}: The rain maps $\mM$ can be updated by solving:
\begin{equation}\label{suppargqm}
\mM^{(s)} = \arg\min_{\mM}Q_{1}\left(\mM,\mM^{(s-1)}\right),
\end{equation}
where $\mM^{(s-1)}$ is the updated result obtained in the last iteration, and $Q_{1}(\mM,\mM^{(s-1)})$ is a quadratic approximation of the objective function  (\ref{suppo2}) with respect to $\mM$ \cite{beck2009fast}, expressed as:
\small
\begin{equation}\label{suppqm}
\begin{split}
Q_{1}\!\left(\!\mM\! ,\mM^{(s-1)}\!\right) \!\!&=\! f\!\!\left(\!\mM^{(s-1)}\!\right) +\frac{1}{2\eta_{1}}\left\|\mM\!-\!\mM^{(s-1)} \right\|_F^2  \!\!\\+&\left\langle \! \mM\!-\!\mM^{(s-1)}, \nabla\! f\left(\!\mM^{(s-1)}\!\right) \!\right\rangle
\!+\!\lambda_{1} p_{1}\left( \mM \right)\!,
\end{split}
\end{equation}
\normalsize
where $f\left(\mM^{(s-1)}\right)\!=\! \left\|\mO\!-\!\mB^{(s-1)}\!- \!\mK\otimes\mM^{(s-1)}\right\|_{F}^{2}$ and $\eta_{1}$ is the stepsize parameter.  It is easy to prove that the  problem (\ref{suppargqm}) is equivalent to:
\begin{equation}\label{suppminq}
  \!\min_{\mM} \frac{1}{2} \left\| \mM \!-\! \left(\! \mM^{(s-1)}\!\!-\! \eta_{1}\nabla f\left(\mM^{(s-1)}\right)  \!\right) \right\|_F^2 + \lambda_{1}\eta_{1} p_{1}\left( \mM \right).
\end{equation}
Corresponding to general regularization terms \cite{donoho1995noising}, the solution of  Eq. (\ref{suppminq}) is:
\begin{equation}\label{suppsolm}
 \mM^{(s)} = \mbox{prox}_{\lambda_{1}\eta_{1}}\left(\! \mM^{(s-1)} \!-\! \eta_{1}\nabla f\left(\mM^{(s-1)}\right)  \!\right).
\end{equation}
Moreover, by  substituting
\small
\begin{equation}\label{suppgradf}
  \nabla f\left(\mM^{(s-1)}\right)=\mK\otimes^{T}\left(\mathcal{K} \otimes \mathcal{M}^{(s-1)} +\mB^{(s-1)}-\mO\right),
\end{equation}
\normalsize
where $\otimes^T$ denotes the transposed convolution,\footnote{The operation $\otimes^T$ can be directly executed by the function as ``torch.nn.ConvTransposed2d" in PyTorch~\cite{paszke2017automatic}.} we can obtain the updating formula for $\mM$ as:\footnote{It can be proved that, with  small enough $\eta_1$ and $\eta_2$, Eq. (\ref{suppupdatem}) and Eq. (\ref{suppupdateb}) can both lead to the reduction of objective function (\ref{suppo2})~\cite{beck2009fast}. \label{suppfn:repeat}}
\small
\begin{equation}\label{suppupdatem}
  \mM^{(s)}=
  \mbox{prox}_{\lambda_{1}\eta_{1}}\left(\mM^{(s-1)}
  - \eta_{1}\mK\!\otimes^{T}\!\left(\mathcal{K}\! \otimes \!\mathcal{M}^{(s-1)}\!+\!\mB^{(s-1)}\!-\!\mO\!\right) \!\right)\!,
\end{equation}
\normalsize
where $\mbox{prox}_{\lambda_{1}\eta_1}(\cdot)$ is the proximal operator dependent on the regularization term $p_{1}(\cdot)$ with respect to $\mM$. Instead of being derived from manually-designed regularizer as in traditional prior-based methods, the form of the implicit proximal operator $\mbox{prox}_{\lambda_{1}\eta_1}(\cdot)$ can be expressed through a convolutional network module and automatically adapted from training data in an end-to-end manner.

\vspace{2mm}
\textbf{Updating $\mB$}:
Similarly, the quadratic approximation of the objective function (\ref{suppo2}) with respect to $\mB$  is:
\small
\begin{equation}\label{suppQproblemb}
\begin{split}
Q_{2}\!\left(\!\mB,\mB^{(s-1)}\!\right)\!&=\! g\left(\!\mB^{(s-1)}\!\right) \!+\!\!\frac{1}{2\eta_{2}}\left\|\mB-\mB^{(s-1)} \right\|_F^2\\
+&\left\langle \mB-\mB^{(s-1)}, \nabla g\left(\!\mB^{(s-1)}\!\right) \right\rangle\!+ \lambda_{2} p_{2}\left( \mB \right),
\end{split}
\end{equation}
\normalsize
where $g\left(\mB^{(s-1)}\right)\!=\! \left\|\mO\!-\!\mB^{(s-1)}- \mK\otimes\mM^{(s)}\right\|_{F}^{2}$ and $\eta_{2}$ is the stepsize parameter. Then the equivalent optimization problem is:
\small
\begin{equation}\label{suppsubProblem2b}
  \min_{\mB}\frac{1}{2}\left\| \mB - \left( \mB^{(s-1)} \!-\! \eta_{2}\nabla g\left(\mB^{(s-1)}\right)  \!\right) \right\|_F^2 + \lambda_{2}\eta_{2} p_{2} \left( \mB \right).
\end{equation}\normalsize
With $\nabla g\left(\!\mB^{(s-1)}\!\right) = \mathcal{K} \otimes \mathcal{M}^{(s)}+\mB^{(s-1)}-\mO$, it is easy to deduce that  the final updating rule for $\mB$ is\footref{fn:repeat}:
\small
\begin{equation}\label{suppupdateb}
\begin{split}
\mB^{(s)} \!=\!\mbox{prox}_{\lambda_{2}\eta_{2}}\left(\left(1-\eta_2\right) \mB^{(s-1)}
  \!+\!\eta_{2}\left(\!\mO\! -\mathcal{K}\otimes \mathcal{M}^{(s)}\right)\right),
\end{split}
\end{equation}
\normalsize
where $\mbox{prox}_{\lambda_{2}\eta_2}(\cdot)$ is the  proximal operator correlated to the regularization term $p_{2}(\cdot)$ with respect to $\mB$.

Here, Eq.~(\ref{suppminq}) and Eq.~(\ref{suppsubProblem2b}) correspond to Eq.~(9) and Eq.~(13) in the main text, respectively.
%
\section{More Analysis about Proximal Operators}
In the main text, when building the CRCDNet (see Section 5) and DRCDNet (see Section 6), we
adopt the deep residual network (ResNet) \cite{he2016deep} to represent the proximal operators $\mbox{prox}_{\lambda_{1}\eta_1}(\cdot)$ for updating $\mM$, $\mbox{prox}_{\lambda_{2}\eta_2}(\cdot)$ for updating $\mB$, and $\mbox{prox}_{\lambda_{3}\eta_3}(\cdot)$ for updating $\alpha$. This rationality can be verified from the following aspects:

 \begin{figure*}[t]
  \begin{center}
     \includegraphics[width=1\linewidth]{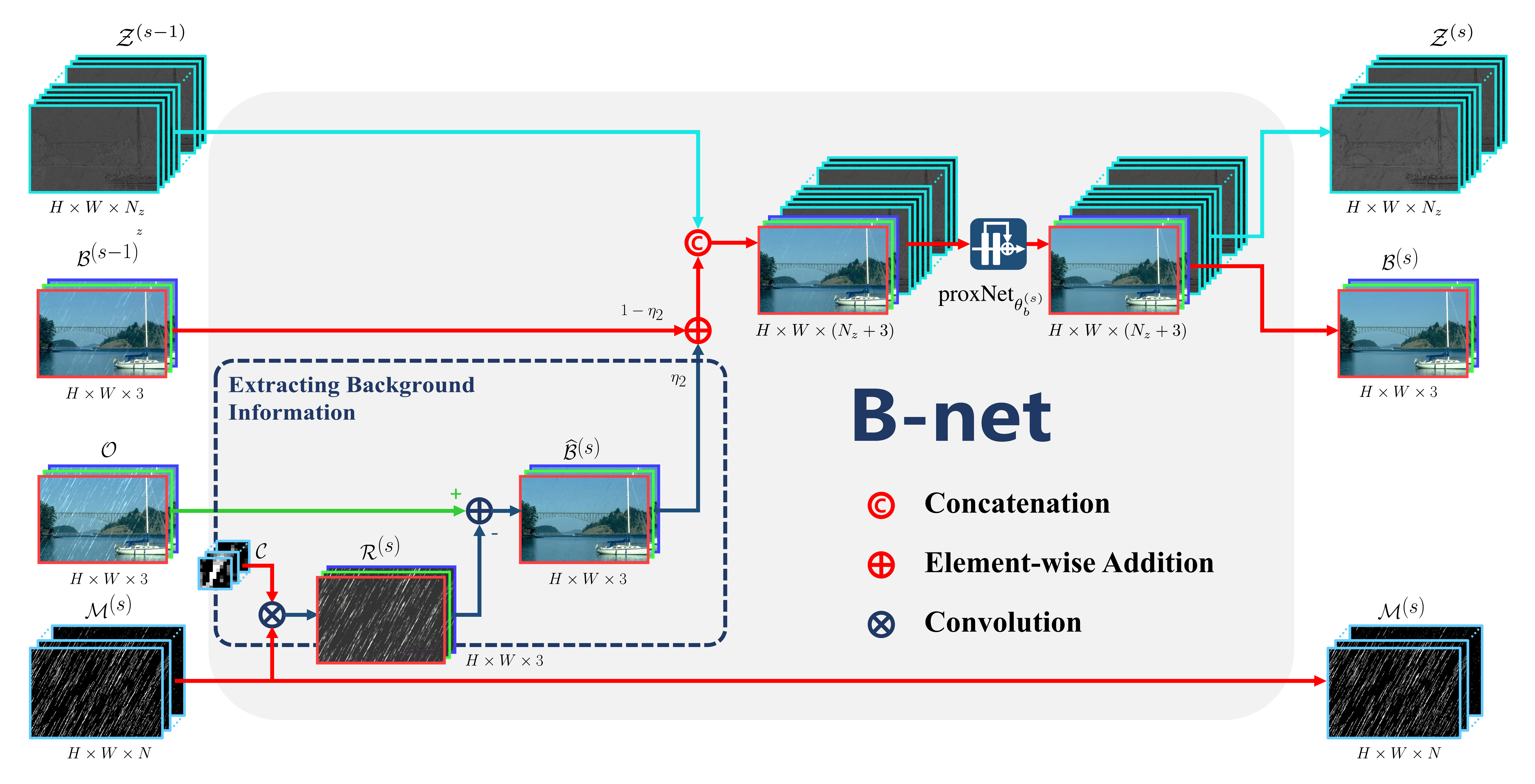}
  \end{center}
  \vspace{-1mm}
     \caption{Illustration of the B-net with channel concatenation operator at the $s^{\text{th}}$ stage. The images are better observed by zooming in on screen.}
  \label{suppfigflowrevise}
  \end{figure*}
First, the choice of ResNet has been proved to be effective for proximal operator in other applications such as dehazing and spectral fusion (Ref. [77] and [78] in paper); Second, the ResNet can achieve well fine-tuning effects along the unfolding network, which finely simulates the gradual amelioration idea of the iterative process, as shown in Figure 7 of the main paper; Third, the structure of proximal operator is similar to ResNet, and thus easy to design and optimize.

Specifically, for easy understanding, we take the updating module of rain map $\mM$ in CRCDNet as an example. As derived, the updating formula of rain map $\mM$ is as folllows (Ref. Eq.~(10) in paper):
\begin{equation}\label{suppsolm2}
\begin{split}
 \mM^{(s)} &= \mbox{prox}_{\lambda_{1}\eta_{1}p_{1}}\left(\! \mM^{(s-1)} \!-\! \eta_{1}\nabla f\left(\mM^{(s-1)}\right)  \!\right)\\
 &\doteq\text{argmin}_{\mM} \frac{1}{2} \!\left\| \mM \!-\!\! \widetilde{\mM} \right\|_F^2 \!+ \lambda_{1}\eta_{1} p_{1}\left( \mM \right),\\
\end{split}
\end{equation}
where $\widetilde{\mM}= \mM^{(s-1)}-\eta_{1}\nabla f\left(\mM^{(s-1)}\right) $. $p_{1}\left( \mM \right)$ is a regularizer to deliver the prior structure of $\mM$. When $p_{1}\left( \mM \right)=\parallel\mM\parallel_{1}$, representing the sparsity prior of $\mM$,
\begin{equation}\label{suppunfoldm}
\begin{split}
 \mM^{(s)}
 & = \text{argmin}_{\mM} \frac{1}{2} \!\left\| \mM \!-\!\! \widetilde{\mM} \right\|_F^2 \!+ \lambda_{1}\eta_{1} \parallel\mM\parallel_{1}\\
 &=\text{soft}\left(\widetilde{\mM},\eta_{1}\right)= \left(\mid\widetilde{\mM}\mid-\eta_{1}\right)_{+}\text{sign}(\widetilde{\mM})\\
 &=\widetilde{\mM}- \text{sign}(\widetilde{\mM}) \left(\text{min}\{\eta_{1},\mid\widetilde{\mM}\mid\}\right)\\
&\approx \text{ResNet}\left(\widetilde{\mM}\right)\\
&= \text{ResNet}\left(\! \mM^{(s-1)}\!\!-\! \eta_{1}\nabla f\left(\mM^{(s-1)}\right)  \!\right)\\
&= \text{proxNet}_{\theta_m^{(s)}}\left({\mathcal{M}}^{(s-1)} - \mathcal{G}^{(s)}\right),
\end{split}
\end{equation}
where $\text{soft}(u,a)\doteq \left\{\begin{matrix}
\text{sign}\left ( u \right )\left ( \left | u \right |-a \right ) \quad\text{if} \left | u \right |> a\\
\quad \quad \quad 0 \quad \quad \quad \quad \text{otherwise}
\end{matrix}\right.$ is the soft-thresholding function. The last equation in Eq.~(\ref{suppunfoldm}) is exactly the unfolding expression for updating $\mM$ presented as Eq.~(15) in the main text.
 
\section{More details of network design}
In this section, we present more details of the network design, including the channel concatenation operator, initialization at $s=0$, proximal networks, i.e., $\text{proxNet}_{\theta_m^{(s)}}(\cdot)$, $\text{proxNet}_{\theta_b^{(s)}}(\cdot)$, and $\text{proxNet}_{\theta_a^{(s)}}(\cdot)$.

\textbf{Channel concatenation operator.}
From Fig. 4 in the main text, the input tensor of $\text{proxNet}_{\theta_b^{(s)}}(\cdot)$ is $\left((1-\eta_2)\mathcal{B}^{(s-1)}+\eta_2\widehat{\mathcal{B}}^{(s)}\right)$ which has the same size of ${H\times W\times 3}$ as the to-be-estimated $\mathcal{B}$. Evidently, this is not beneficial for learning $\mathcal{B}$ since most of the previous updating information would be compressed due to not sufficiently specified number of channels. To better keep and deliver image features, we simply expand the input tensor at the $3^{\text{rd}}$ mode for more channels in experiments.

Specifically, we introduce an auxiliary variable $\mathcal{Z}$ with the size of ${H \times W \times N_{z}}$, put it behind the tensor $\left((1-\eta_2)\mathcal{B}^{(s-1)}+\eta_2\widehat{\mathcal{B}}^{(s)}\right)$ at the channel mode, and then obtain a new input tensor with the size of ${H \times W \times (N_{z}+3)}$ as the input of $\text{proxNet}_{\theta_b^{(s)}}(\cdot)$. Thus the number of input channels of $\text{proxNet}_{\theta_b^{(s)}}(\cdot)$ varies from 3 to $(N_{z}+3)$. For the output with the size ${H \times W \times (N_{z}+3)}$ of $\text{proxNet}_{\theta_b^{(s)}}(\cdot)$, we decompose it into two sub-tensors based on the channel mode: one sub-tensor corresponding to the first 3 channels is taken as the updated background $\mathcal{B}^{(s)}$ and the other one corresponding to remaining channels is regraded as $\mathcal{Z}^{(s)}$ with the size ${H \times W \times N_{z}}$. Please refer to Fig.~\ref{suppfigflowrevise} for better understanding.

\textbf{Initialize $\mathcal{B}^{(0)}$ and $\mathcal{Z}^{(0)}$}. In our network implementation, we set $\mathcal{M}^{(0)}=0$, and initialize $\mathcal{B}^{(0)}$ and $\mathcal{Z}^{(0)}$ by:
\begin{equation}\label{suppinibz}
\{\mathcal{B}^{(0)} ~| ~\mathcal{Z}^{(0)}\} = \mbox{proxNet}_{\theta_{b}^{(0)}}\left(\text{concat}\left(\mathcal{O},\mathcal{C}_{z}\otimes\mathcal{O} \right)\right),
\end{equation}
where $\mathcal{C}_{z}$ is the learnable convolutional filters with the size of ${k_{z} \times k_{z}\times 3 \times N_{z} }$. Refer to Eq.~(5) in the main text for better understanding of the convolutional operator. Actually, this has been ready-made in current popular deep learning (DL) framework such as Tensorflow\footnote{\url{https://tensorflow.google.cn/}} and PyTorch.\footnote{\url{https://pytorch.org/docs/stable/index.html}} The function $\text{concat}(\cdot)$ means the channel concatenation operator as illustrated before. Here the operator is executed between $\mathcal{O}$ and $\mathcal{C}_{z}\otimes\mathcal{O}$.  $\mbox{proxNet}_{\theta_{b}^{(0)}}(\cdot)$ is a deep residual network (ResNet)~\cite{he2016deep} with the same structure as $\text{proxNet}_{\theta_{b}^{(s)}}(\cdot)$ $(s=1,\cdots,S)$, but with different parameter as ${\theta_{b}^{(0)}}$. Clearly, the output tensor of $\mbox{proxNet}_{\theta_{b}^{(0)}}(\cdot)$ is also with the size of ${H \times W \times (N_{z}+3)}$, and can be decomposed into $\mathcal{B}^{(0)}$ with the size of ${H \times W \times 3}$  and $\mathcal{Z}^{(0)}$ with the size of $H \times W \times {N_{z} }$  based on the channel mode, as depicted above. In this manner, we keep the numbers of input and output channels through $\mbox{proxNet}_{\theta_{b}^{(0)}}(\cdot)$ consistent, and take the corresponding fine-tuned results of $\mathcal{O}$ and $\mathcal{C}_{z}\otimes\mathcal{O}$ with the ResNet as $\mathcal{B}^{(0)}$ and $\mathcal{Z}^{(0)}$, respectively. Intuitively, this is simple and reasonable.
  \begin{figure}[t]
  \begin{center}
     \includegraphics[width=1\linewidth]{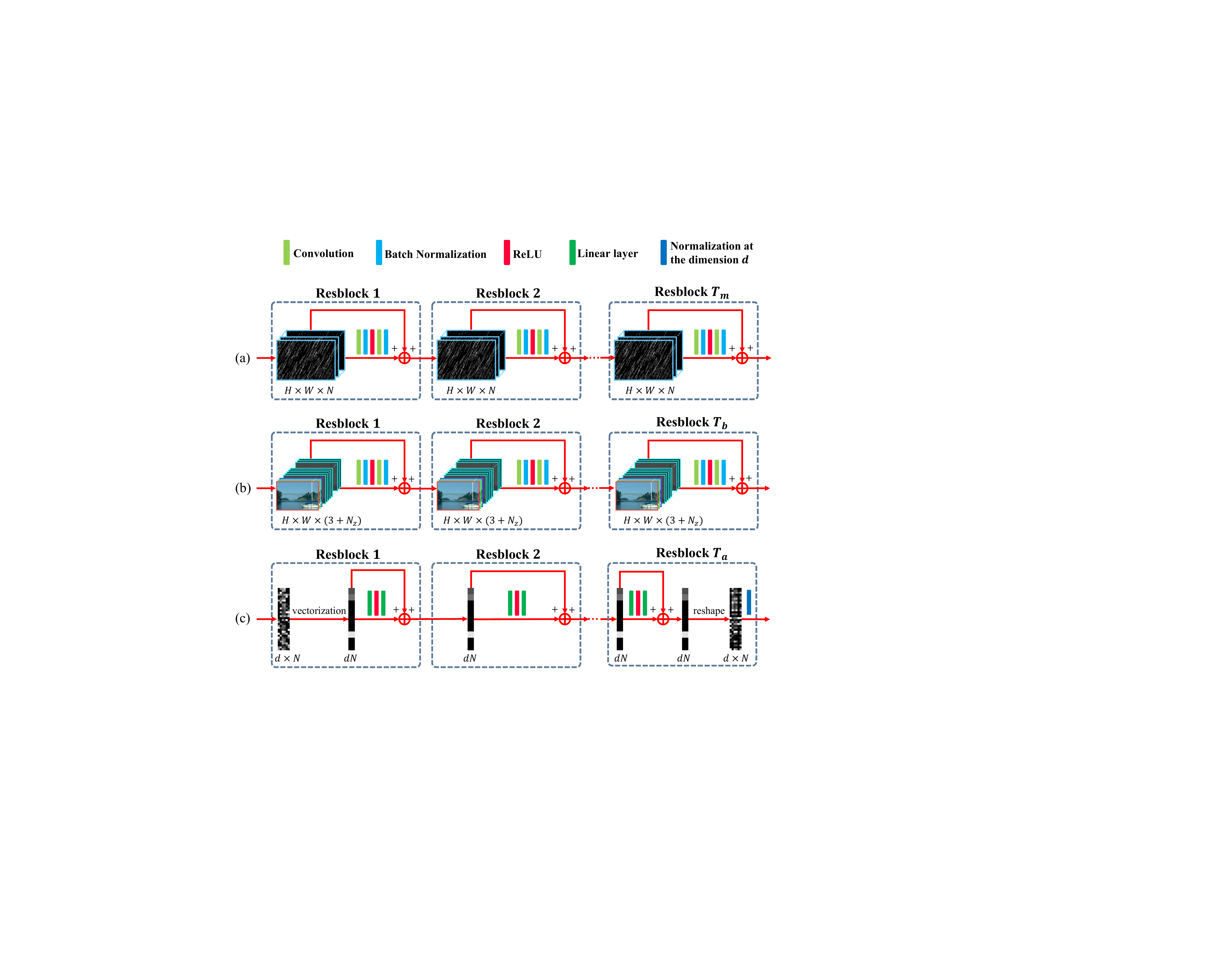}
  \end{center}
  \vspace{-2mm}
     \caption{(a) The exploited ResNet for the proximal network $\text{proxNet}_{\theta_m^{(s)}}(\cdot)$. (b) The exploited ResNet for the proximal network $\text{proxNet}_{\theta_b^{(s)}}(\cdot)$. (c) The exploited ResNet for the proximal network $\text{proxNet}_{\theta_a^{(s)}}(\cdot)$.}
  \label{suppmbres}
  \end{figure}

\textbf{Proximal Network and ResNet.} As stated before, we adopt the ResNet to build the proximal network $\text{proxNet}_{\theta_{m}^{(s)}}(\cdot)$, $\text{proxNet}_{\theta_{b}^{(s)}}(\cdot)$, and $\text{proxNet}_{\theta_a^{(s)}}(\cdot)$ $(s=1,\cdots,S)$. Please refer to Fig.~\ref{suppmbres} for detailed illustration of the used ResNet for $\text{proxNet}_{\theta_m^{(s)}}(\cdot)$, $\text{proxNet}_{\theta_b^{(s)}}(\cdot)$, and $\text{proxNet}_{\theta_a^{(s)}}(\cdot)$. For simplicity, in all our experiments, throughout all stages, the number $T_{m}$ of Resblocks in $\text{proxNet}_{\theta_m^{(s)}}(\cdot)$ and $T_{b}$ in  $\text{proxNet}_{\theta_b^{(s)}}(\cdot)$ is fixed with the same setting as $T$. However, due to the involvement of linear layer in $\text{proxNet}_{\theta_a^{(s)}}(\cdot)$, we only set $T_a=1$ in DRCDNet for fewer network parameters.


In CRCDNet, all the parameters involved can be automatically fit from training data in an end-to-end manner, including $\{\theta_m^{(s)},\theta_b^{(s)}\}_{s=1}^{S}$, rain kernels $\mK$, $\eta_1$, and $\eta_2$.

In DRCDNet, all the involved parameters, including $\{\theta_m^{(s)},\theta_b^{(s)},\theta_a^{(s)}\}_{s=1}^{S}$, rain kernel dictionary $\mD$, $\eta_1$, $\eta_2$, and $\eta_3$, can be automatically learned from the training data pairs in an end-to-end manner.

\section{More implementation details}
We use PyTorch~\cite{paszke2017automatic} to implement our networks including CRCDNet and DRCDNet, on a PC equipped with an Intel (R) Core(TM) i7-8700K at 3.70GHZ and a NVIDIA GeForce GTX 1080Ti GPU. We adopt the Adam optimizer~\cite{Kingma2014Adam} with the batch size of 10 and the patch size of 64$\times$64. The initial learning rate is $1\times$$10^{-3}$ and divided by 5 every 25 epochs. The total epoch is 100. It is worth mentioning  that for all datasets in the main text, these parameter settings are the same. This guarantees an easy reproduction for our method and a rational verification for its generality.
%
\section{More ablation studies}
In this section, we utilize CRCDNet to provide more ablation studies based on Rain100L including 200 rainy/clean image pairs for training and 100 pairs for testing~\cite{Yang2019Joint}. Two performance metrics are employed, including peak-signal-to-noise ratio (PSNR)~\cite{Huynh2008Scope} and structure similarity (SSIM)~\cite{wang2004image}. As the human visual system is sensitive to the Y channel of a color image in YCbCr space, we compute PSNR and SSIM based on this luminance channel.
\subsection{Loss function}
As stated in the main text, we adopt the mean square error (MSE)~\cite{Jing2012Removing} for the learned background and the rain layer at every stage as the training objective function:
\begin{equation}\label{suppLoss}
  L = \sum_{s=0}^{S}\rho_{s}\left\|\mathcal{B}^{(s)}\!-\!\mathcal{B} \right\|_F^2\!+\!\sum_{s=1}^{S}\gamma_{s}\left\| {\mathcal{O}\!-\!\mathcal{B}\!-\!\mathcal{R}}^{(s)} \right\|_F^2,
\end{equation}
where $\mathcal{B}^{(s)}$ and $\mathcal{R}^{(s)}$ denote the derained result and extracted rain layer at the $s^{\text{th}}$ stage, respectively. $\rho_{s}$ and $\gamma_{s}$ are tradeoff parameters.

Here, we choose the channel concatenation number $N_{z}$ as 32. The size $k$ of rain kernels is 9, and $k_{z}=3$. Setting the number $T$ of Resblocks in each ResNet at every stage as 4 and the stage number $S$ as 17, we study the effect of loss function with different parameter settings of $\rho_{s}$ and $\gamma_{s}$ on rain removal performance of CRCDNet, as depicted in Table~\ref{supptabLoss}. From Version 1, we can find that even when we only adopt a single loss on the final derained result $\widehat{\mathcal{B}}$, our method can still significantly outperform other comparison methods on Rain100L, as reported in Table~\ref{supptabsyn}. By comparing Versions 1 and 3, we can see that the performance can be further improved by imposing supervision loss on the final extracted rain layer $\mathcal{R}^{(S)}$. Besides, intra-stage supervision for the results $\mathcal{B}^{(s)}$ and $\mathcal{R}^{(s)}$ is helpful for enabling the network to be evolved to a better direction. This can be easily concluded by comparing Versions 1 and 2 (Versions 3 and 4). Furthermore, taking Version 1 as a benchmark, from Versions 2, 3, and 4, we observe that intra-stage supervision loss for rain layer plays more important role in deraining performance than that for background layer. This also reflects the effectiveness of M-net. Based on the analysis aforementioned, we thus present a Version 5 and take it as our final loss function in all subsequent experiments where we set $\rho_{S}=\gamma_{S}=1$ to make the outputs at the final stage play a dominant role, and other parameters as 0.1 to help find the correct parameter at each stage.
\begin{table*}[htp]
\centering
\caption{Rain removal performance on Rain100L of the proposed CRCDNet with different loss functions. Note that the tradeoff parameter not mentioned in each version is 0 by default. For example, for Version 1, $\rho_{s}=0$ $(s=0,\cdots,S-1)$ and $\gamma_{s}=0$ $(s=1,\cdots,S)$.}
\footnotesize
\setlength{\tabcolsep}{1.6pt}
\begin{tabular}{c|c|c|cc}
\hline
Version & Parameter setting & Loss function & PSNR & SSIM \\

\hline
1 & $\rho_S=1$ & $L= \left\|\mathcal{B}^{(S)}\!-\!\mathcal{B} \right\|_F^2$ & 39.90 & 0.9855\\
\hline
2 & \tabincell{c}{$\rho_S=1$\\$\rho_s=0.1(s\neq S)$} & $L= \left\|\mathcal{B}^{(S)}\!-\!\mathcal{B} \right\|_F^2+ \sum_{s=0}^{S-1}0.1\left\|\mathcal{B}^{(s)}\!-\!\mathcal{B} \right\|_F^2$ & 39.93 & 0.9857\\
\hline
3 & $\rho_S=\gamma_S=1$ & $L= \left\|\mathcal{B}^{(S)}\!-\!\mathcal{B} \right\|_F^2 + \left\| {\mathcal{O}\!-\!\mathcal{B}\!-\!\mathcal{R}}^{(S)} \right\|_F^2$ & 39.94 & 0.9860\\
\hline
4 & \tabincell{c}{$\rho_S=\gamma_S=1$\\$\gamma_s=0.1(s\neq S)$} & $L= \left\|\mathcal{B}^{(S)}\!-\!\mathcal{B} \right\|_F^2 + \left\| {\mathcal{O}\!-\!\mathcal{B}\!-\!\mathcal{R}}^{(S)} \right\|_F^2+\sum_{s=1}^{S-1}0.1\left\| {\mathcal{O}\!-\!\mathcal{B}\!-\!\mathcal{R}}^{(s)} \right\|_F^2$& 39.98& 0.9860\\
\hline
5 & \tabincell{c}{$\rho_S=\gamma_S=1$\\$\rho_s=\gamma_s=0.1(s\neq S)$} & $L= \left\|\mathcal{B}^{(S)}\!-\!\mathcal{B} \right\|_F^2 + \left\| {\mathcal{O}\!-\!\mathcal{B}\!-\!\mathcal{R}}^{(S)} \right\|_F^2+ \sum_{s=0}^{S-1}0.1\left\|\mathcal{B}^{(s)}\!-\!\mathcal{B} \right\|_F^2+\sum_{s=1}^{S-1}0.1\left\| {\mathcal{O}\!-\!\mathcal{B}\!-\!\mathcal{R}}^{(s)} \right\|_F^2$& \textbf{40.00} & \textbf{0.9860}\\
\hline
\end{tabular}
\label{supptabLoss}
\end{table*}
\subsection{Network architecture}
From Fig.~\ref{suppfigflowrevise}, the key factors affecting our network architecture include: Resblocks number $T$ involved in each ResNet, stage number $S$, and channel concatenation number $N_{z}$. As aforementioned, $\text{proxNet}_{\theta_m^{(s)}}(\cdot)$ and $\text{proxNet}_{\theta_b^{(s)}}(\cdot)$ have the same Resblocks number $T$ and it keeps the same among all stages. In the following, we evaluate the effect of these factors on the CRCDNet.
\begin{table}[t]
\centering
\caption{Effect of Resblocks number $T$, involved in the ResNet $\text{proxNet}_{\theta_m^{(s)}}(\cdot)$ and $\text{proxNet}_{\theta_b^{(s)}}(\cdot)$ of Fig.~\ref{suppmbres}, on the performance of CRCDNet.}\vspace{-3mm}
\footnotesize
\setlength{\tabcolsep}{8.8pt}
\begin{tabular}{c|c|c|c|c|c}
\hline
$T$ & $T$=1& $T$=2 & $T$=3 & $T$=4 & $T$=5\\
\hline
PSNR & 39.04 &39.52 &39.80 &40.00 &39.98\\
\hline
SSIM  &0.9833 &0.9848 &0.9856 &0.9860 &0.9859\\
\hline
\end{tabular}
\label{supptabT}
\end{table}
\begin{table}[t]
\centering
\caption{Effect of stage number $S$ on the performance of CRCDNet.}\vspace{-3mm}
\footnotesize
\setlength{\tabcolsep}{1.6pt}
\begin{tabular}{c|c|c|c|c|c|c|c|c}
\hline
Stage No. & $S$=0 & $S$=2 & $S$=5 & $S$=8 & $S$=11 & $S$=14& $S$=17 & $S$=20\\
\hline
PSNR & 35.93 &38.46 &39.35  &39.60 &39.81 &39.90 &40.00 &39.91\\
\hline
SSIM  &0.9689 &0.9813 &0.9842 &0.9850 &0.9855 &0.9858 &0.9860 & 0.9858\\
\hline
\end{tabular}
\label{supptabS}
\end{table}
\begin{figure}[t]
  \begin{center}
     \includegraphics[width=0.9\linewidth]{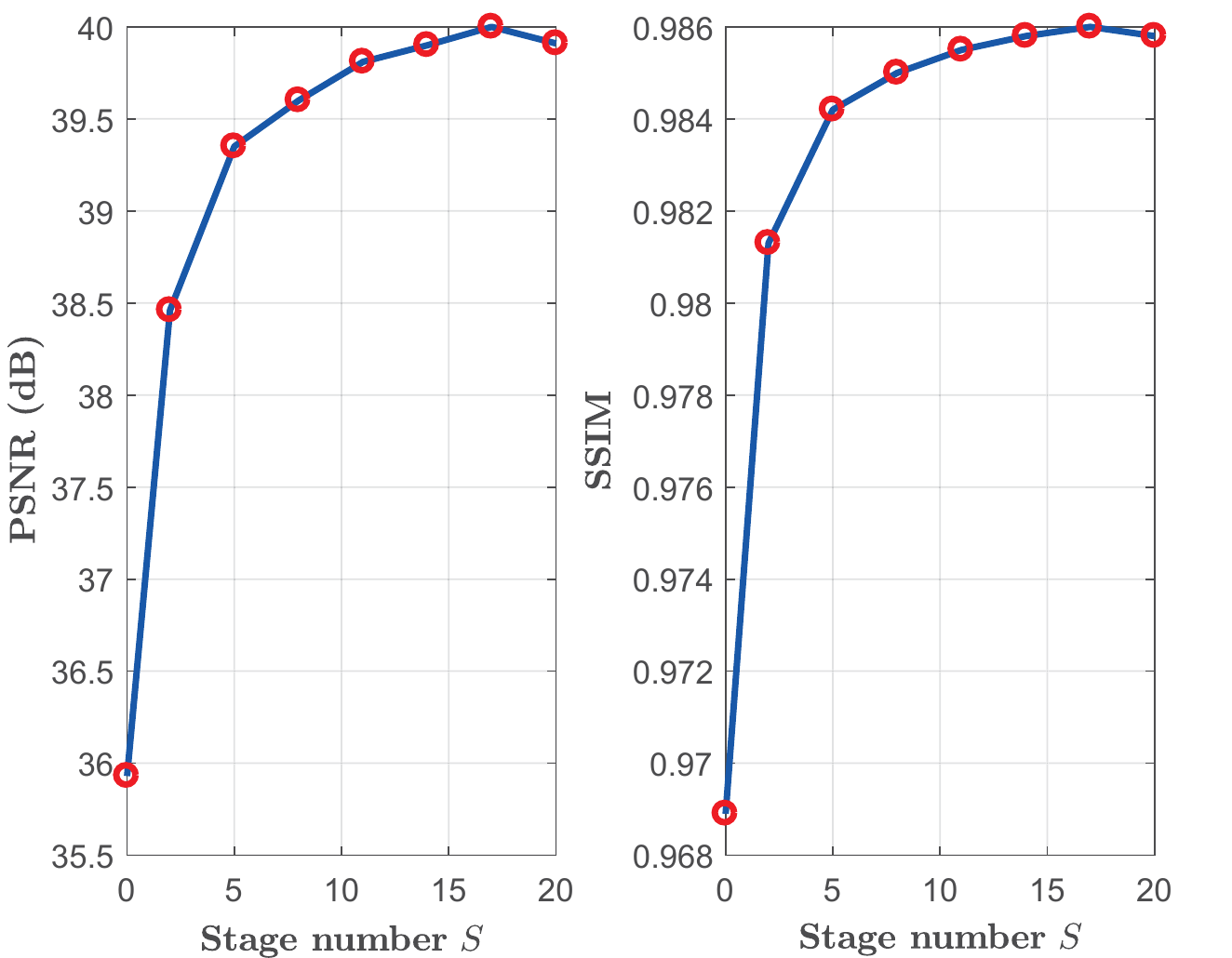}
  \end{center}
  \vspace{-3mm}
     \caption{Average PSNR and SSIM with different stage number $S$.}
  \label{suppfigstage}
\end{figure}
\begin{table}[t]
\centering
\caption{Effect of channel concatenation number $N_{z}$ on the performance of CRCDNet.}\vspace{-3mm}
\footnotesize
\setlength{\tabcolsep}{1.7pt}
\begin{tabular}{c|c|c|c|c|c|c|c|c}
\hline
 $N_{z}$ &  $N_{z}$=0 &  $N_{z}$=2 &  $N_{z}$=8 &  $N_{z}$=14 &  $N_{z}$=20 &  $N_{z}$=26 &  $N_{z}$=32 &$N_{z}$=38\\
\hline
PSNR & 38.99 &39.13 &39.36  &39.69 &39.72 &39.82 &40.00 &39.98\\
\hline
SSIM  &0.9830 &0.9836 &0.9843 &0.9851 &0.9853 &0.9856 &0.9860 &0.9859\\
\hline
\end{tabular}
\label{supptabNz}
\vspace{-5mm}
\end{table}
\begin{table}[t]
\centering
\caption{Hyper-parameter settings for CRCDNet and DRCDNet. ``/'' means that the parameter is not involved in CRCDNet.}\vspace{-3mm}
\scriptsize
\setlength{\tabcolsep}{0.5pt}
\begin{tabular}{c|c|c|c}
\hline
\multirow{2}{*}{Parameters} & \multirow{2}{*}{Notations} &  \multicolumn{2}{c}{Settings}\\
\cline{3-4}
                                               &   &CRCDNet & DRCDNet \\

\hline
$k \times k $ & Size of rain kernel &$9\times9$ &$9\times9$\\
\hline
$N$  & Rain kernels No. &32 &6\\
\hline
$d$ & Rain kernels No. in dictionary $D$ & / & 32\\
\hline
$T$ & Resblocks No. in $\text{proxNet}_{\theta_m^{(s)}}(\cdot)$ and $\text{proxNet}_{\theta_b^{(s)}}(\cdot)$ &4&4\\
\hline
$T_a$ & Resblocks No. in  $\text{proxNet}_{\theta_a^{(s)}}(\cdot)$ & / &1\\
\hline
$S$ & Stages No. & 17 & 11\\
\hline
$N_z$ & Channel concatenation No. & 32 & 32\\
\hline
\end{tabular}
\label{supptabpara}
\end{table}
  \begin{figure*}[htp]
  \begin{center}
     \includegraphics[width=1\linewidth]{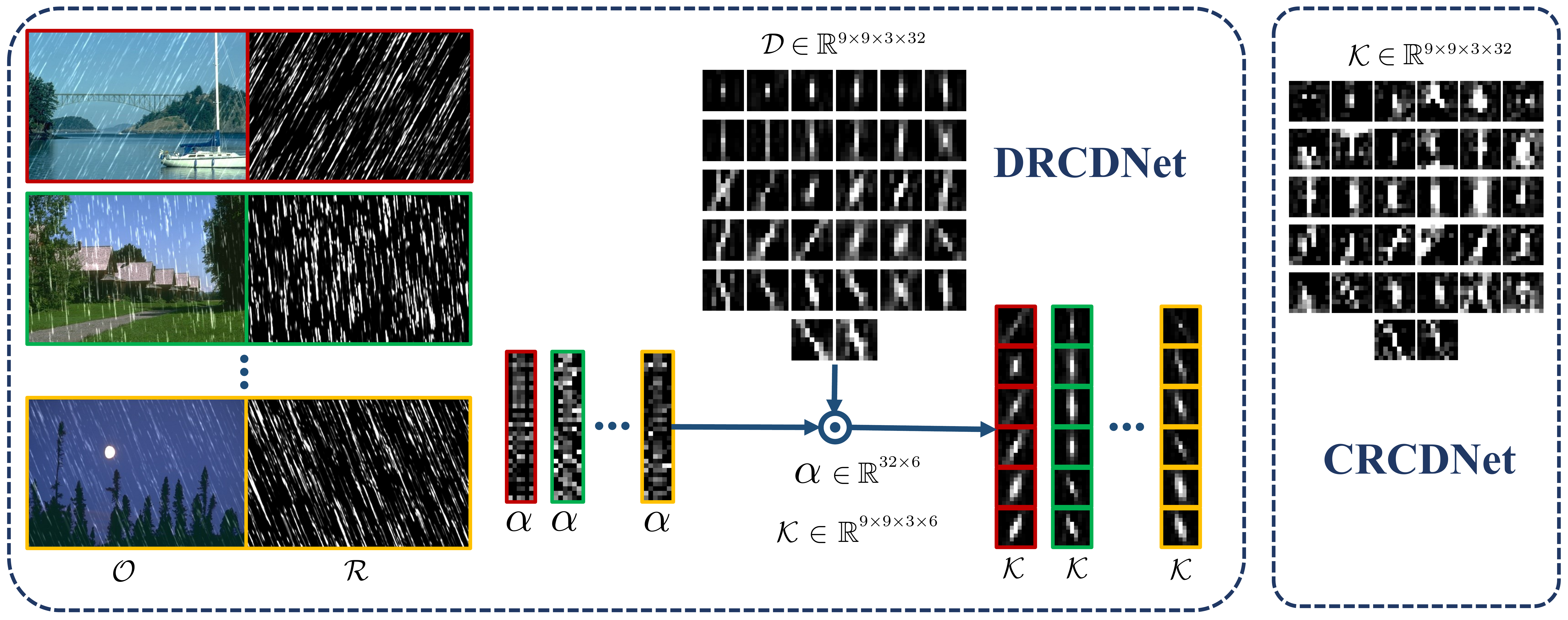}
  \end{center}
  \vspace{-2mm}
     \caption{Rain kernel modelling mechanism comparisons of CRCDNet and DRCDNet. Specifically, the rain kernel $\mK$ for CRCDNet and the dictionary $\mD$ for DRCDNet are both learned on the entire Rain100L training set. In CRCDNet, different rainy images always share the same rain kernel $\mK$. However, for DRCDNet, the adaptive learning of sample-variant $\alpha$ achieves the dynamic prediction of rain kernel $\mK=\mD\alpha$ which is correspondent to the rain types of every testing rainy image. For example, for the rainy image $\mO$ with rain streaks in the direction of left (marked as red box), the inferred rain kernels $\mK$ correspondingly present local patterns to the left (marked as red box).}
  \label{suppfigkernelcompare}
  \end{figure*}

\textbf{Resblocks number $T$ at each stage.}
Setting $S$ as 17 and $N_{z}$ as 32, we separately select Resblocks number $T$ as 1, 2, 3, 4, and 5. The quantitative comparison is presented in Table~\ref{supptabT}. It can be easily observed that more Resblocks usually bring higher average PSNR and SSIM. However, larger $T$ would make gradient propagation more difficult, explaining the fact that the performance with $T=5$ is a little inferior to the case of $T=4$. Hence, we select $T$ as 4 below.

\textbf{Stage number $S$.}
Table~\ref{supptabS} and Fig.~\ref{suppfigstage} clearly present the effect of stage number $S$ on the rain removal performance of CRCDNet. Here, $S=0$ means that the initialization $\mathcal{B}^{(0)}$ is directly regraded as the recovery result. Taking $S=0$ as a baseline, it is seen that only with 2 stages, our method achieves significant rain removal performance, which validates the essential role of the proposed M-net and B-net. We also observe that when $S=20$, its performance is slightly lower than that of $S=17$ since larger $S$ would make gradient propagation more difficult as explained before. Based on such observation, we set $S$ as 17 throughout all our experiments.

\textbf{Channel concatenation number $N_{z}$.} By setting $T$ as 4 and $S$ as 17, Table~\ref{supptabNz} lists the PSNR and SSIM obtained by our proposed algorithm with different channel concatenation number $N_{z}$. It is easy to see that larger channel concatenation number $N_{z}$ is indeed helpful for passing more effective image features and thus brings higher PSNR and SSIM. The fact confirms the rationality of the proposed concatenation operator in the network design. Besides, the case $N_{z}=38$ has a litter lower performance than the case $N_{z}=32$ since larger $N_{z}$ with more parameters makes the network training difficult. We thus choose $N_{z}$ as 32.

\section{More Explanations about The Proposed CRCDNet and DRCDNet}
\textbf{Parameter settings.} We list the key parameter settings for CRCDNet and DRCDNet as shown in Table~\ref{supptabpara}. Apart from these hyper-parameters, other settings for DRCDNet are kept the same as those of CRCDNet, for example, the loss function as described in Section 5. It should be noted that to keep the almost same network parameters with CRCDNet, the total stage number $S$ in DRCDNet is set as 11.

\noindent\textbf{Rain kernel modelling mechanism.} Fig.~\ref{suppfigkernelcompare} visually illustrates the rain kernel modelling mechanism for CRCDNet and DRCDNet. Specifically, for CRCDNet, all the input rainy samples always share the same rain kernel $\mK$ with size $9\times9\times3\times32$, which has no sample-variant characteristic at the testing phase. However, for DRCDNet, every input rainy image has its own specific rain kernel $\mK$ with size $9\times9\times3\times6$, which is achieved by the adaptive learning of sample-variant $\alpha$. Besides, as compared with the rain kernels $\mK$ in CRCDNet, we can easily find that the rain kernels in the dictionary $\mathcal{D}$ are fairly diverse and have more sharper shapes. Such adaptive rain kernel prediction mechanism makes DRCDNet have potential to obtain better generalization capability.
\section{More Experimental Results}
In this section, based on the datasets in the main text, we demonstrate more experimental results including training-test domain match experiments and training-test domain mismatch experiments for comprehensively validating the effectiveness of CRCDNet and DRCDNet, respectively.

\textbf{Benchmark Datasets.} Eight datasets are adopted, including 5 synthesized ones and 3 real ones. The synthetic ones are achieved by manually adding rain streaks to clear background images. SPA-Data is semi-automatically generated where the background images are estimated based on multiple concessive frames of rainy videos captured in real scenes. Internet-Data and MPID\_Rain+Mist(R) are both collected from web and they consist  of unlabeled samples with complicated rain types. 


\textbf{Comparison methods.} The comparison methods include: model-based DSC~\cite{Yu2015Removing},\footnote{\url{https://sites.google.com/view/taixiangjiang/\%E9\%A6\%96\%E9\%A1\%B5/state-of-the-art-methods}}
and JCAS~\cite{Gu2017Joint};\footnote{\url{https://sites.google.com/site/shuhanggu/home}} deep learning (DL)-based Clear~\cite{Fu2017Clearing},\footnote{\url{https://xueyangfu.github.io/projects/tip2017.html}}
DDN~\cite{Fu2017Removing},\footnote{\url{https://xueyangfu.github.io/projects/cvpr2017.html}}
RESCAN~\cite{li2018recurrent},\footnote{\url{https://github.com/XiaLiPKU/RESCAN}}
PReNet~\cite{ren2019progressive},\footnote{\url{https://github.com/csdwren/PReNet}}
SPANet~\cite{wang2019spatial},\footnote{\url{https://stevewongv.github.io/derain-project.html}} JORDER\_E~\cite{Yang2019Joint},\footnote{\url{https://github.com/flyywh/}} and semi-supervised-based SIRR~\cite{wei2019semi}.\footnote{\url{https://github.com/wwzjer/Semi-supervised-IRR}}
\subsection{Training-test Domain Match Experiments}
\begin{figure*}[t]
  \begin{center}
     \includegraphics[width=1\linewidth]{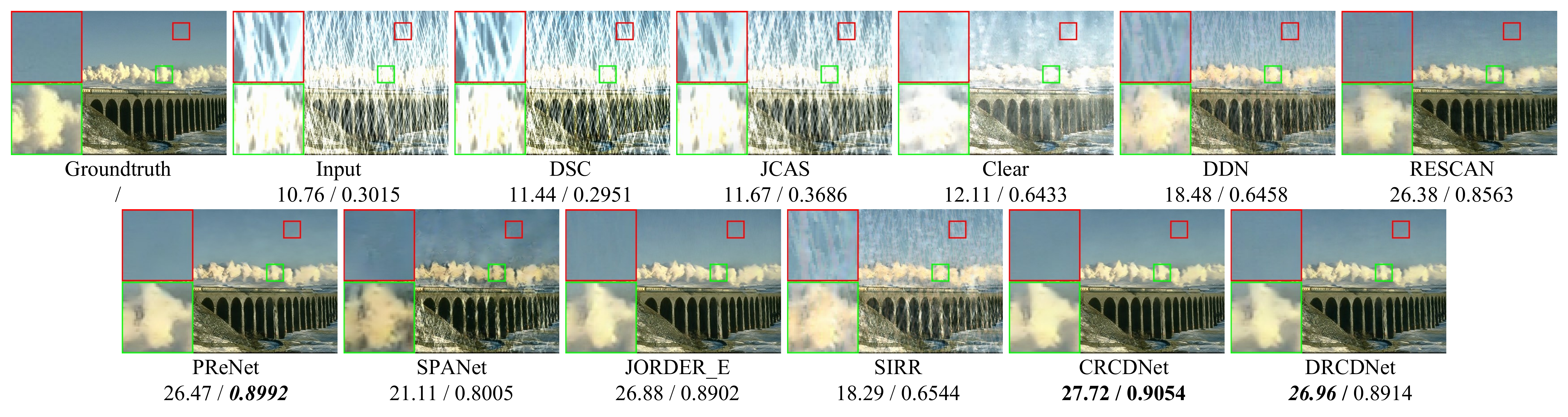}
  \end{center}
  \vspace{-5mm}
     \caption{\textcolor{black}{(Training-test domain match) From left/upper to right/lower: groundtruth and rainy image from Rain100H, derained results of all competing methods.}}
  \label{suppfig11rainh}
  \vspace{-2mm}
\end{figure*}

\begin{figure*}[htp]
  \begin{center}
  \vspace{-1mm}
     \includegraphics[width=1.0\linewidth]{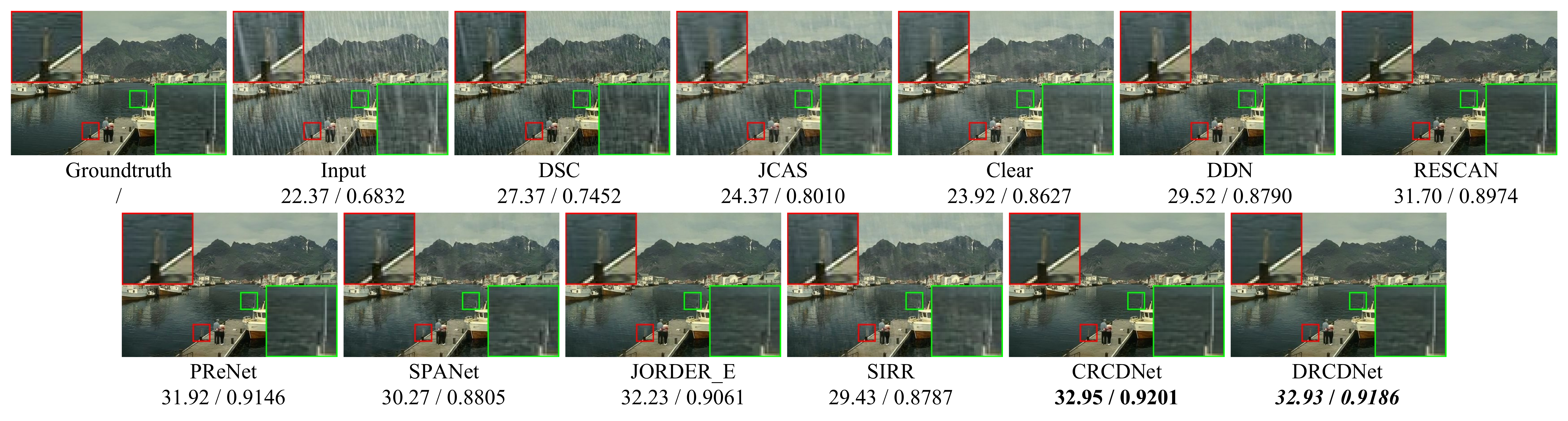}
  \end{center}
  \vspace{-5mm}
     \caption{\textcolor{black}{(Training-test domain match case) Rain removal performance comparisons on a rainy image from Rain1400. The best performed results are highlighted in bold. The images are better observed by zooming in on screen.}}
  \label{supprain1400}
    \vspace{-2mm}
\end{figure*}

\textcolor{black}{\textbf{Performance Comparison on Rain100H.}
Fig.~\ref{suppfig11rainh} presents the derained results of all competing methods on one typical rainy image from Rain100H. It is easy to see that the rain removal performance of most comparison methods is adversely degraded by \emph{heavy rains}. Comparatively, the proposed CRCDNet, as well as DRCDNet, still achieve better visual quality and quantitative measures.}

\textbf{Performance comparisons on Rain1400.} Fig.~\ref{supprain1400} depicts the rain removal comparison results on the rainy image selected from Rain1400. From it, we can easily conclude that as compared with other competing methods, our proposed CRCDNet not only removes more rain streaks but also preserves background details better.


Table~\ref{supptabsyn} reports the quantitative results of all competing methods. It is seen that our CRCDNet attains the best deraining performance among all competing methods on each dataset. Moreover, even we only adopt the single loss as the objective function, that is, Version 1 in Table~\ref{supptabLoss}, our network RCDNet$_{1}$ still has a dominant deraining performance.

\textbf{Performance comparisons on SPA-Data.} We then analyze the performance of all competing methods on the real SPA-Data from~\cite{wang2019spatial} contains 638492 rainy/clean image pairs for training and 1000 testing ones.

\begin{figure*}[htp]
  \begin{center}
     \includegraphics[width=1.005\linewidth]{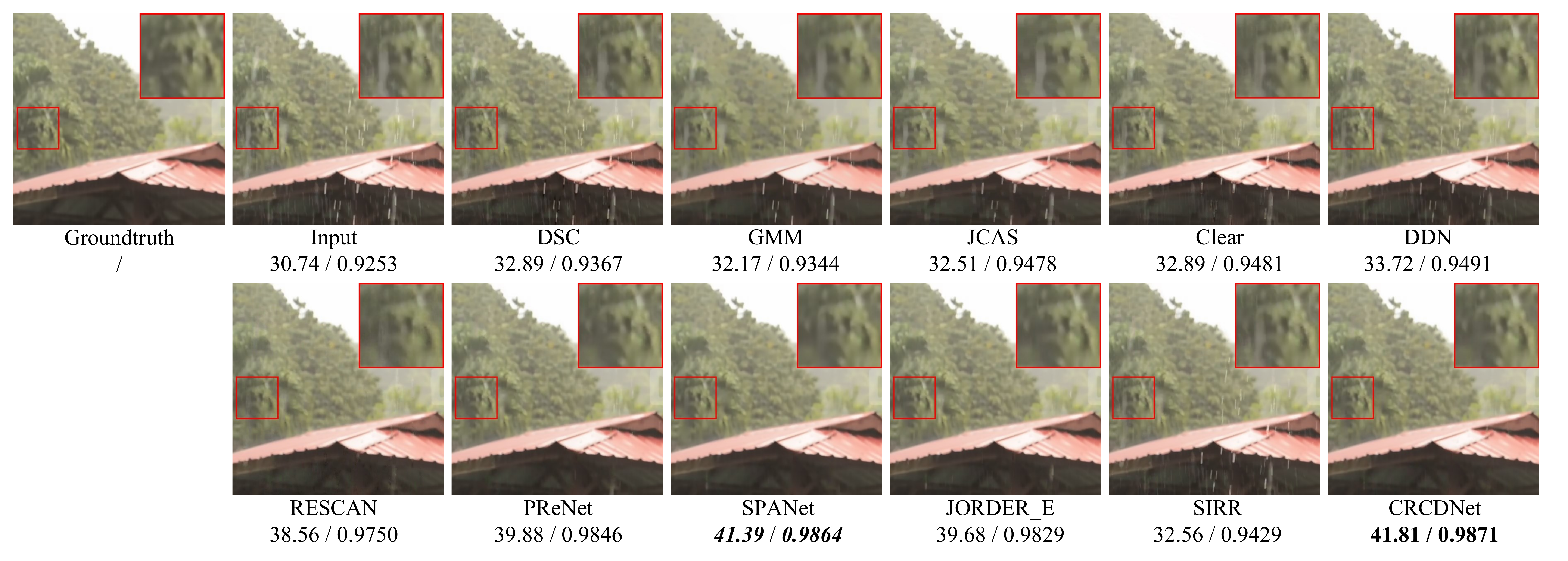}
  \end{center}
  \vspace{-5mm}
     \caption{(Training-test domain match case) Rain removal performance comparisons on a real rainy image with various (long/thin/heavy) rain streaks from SPA-Data. The images are better observed by zooming in on screen.}
  \label{suppwang}
\end{figure*}

\begin{figure*}[!htp]
  \begin{center}
  \vspace{-0mm}
     \includegraphics[width=1.017\linewidth]{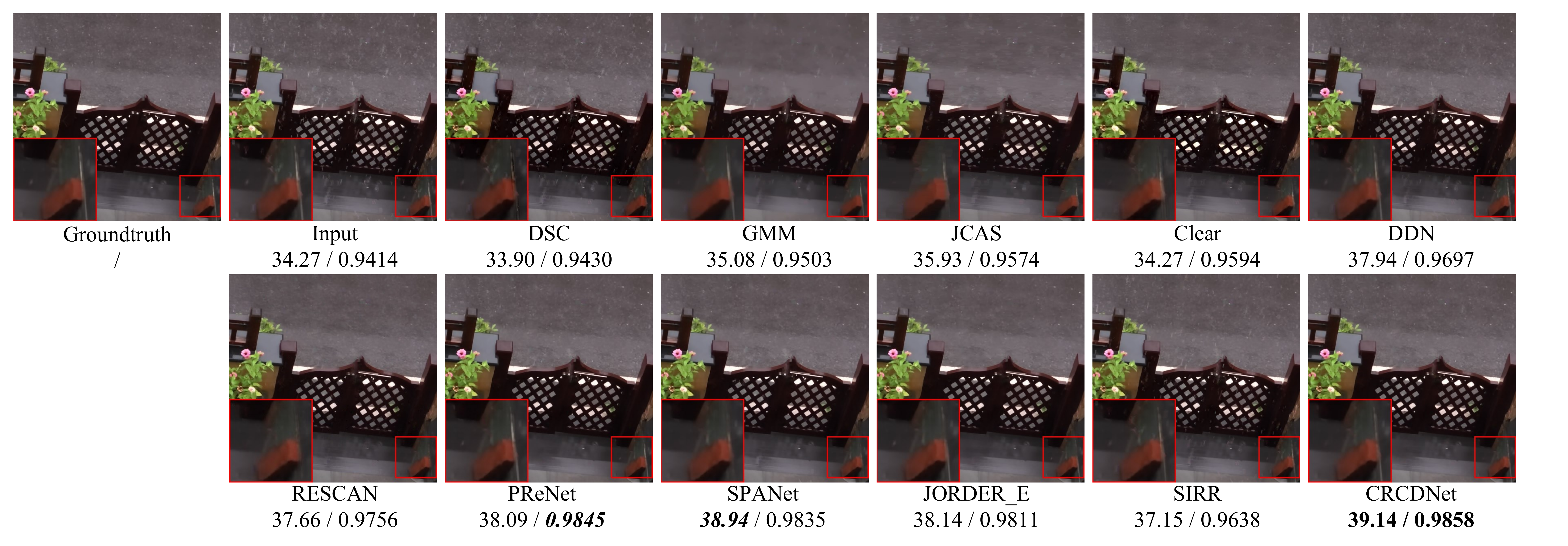}
  \end{center}
  \vspace{-5mm}
     \caption{(Training-test domain match case) Rain removal performance comparisons on a real rainy image with light rain streaks from SPA-Data. The images are better observed by zooming in on screen.}
  \label{suppwang2}
\end{figure*}

Fig.~\ref{suppwang} and Fig.~\ref{suppwang2} show two real rainy images with very different rain patterns from SPA-Data to comprehensively evaluate the rain removal performance of all competing methods, both visually and quantitatively. As observed, traditional model-based DSC and JCAS leave obvious rain streaks in the derained results. Among DL-based methods, some cannot perfectly remove all rain streaks such as DDN, PReNet, JORDER\_E, and SIRR, while some blur image details, including Clear and SPANet. For RESCAN, it leaves some rain streaks as shown in Fig.~\ref{suppwang} while adversely loses background textures as presented in Fig.~\ref{suppwang2}. Our CRCDNet, however, can always achieve an evident superior performance than other methods under different rain types.

Table~\ref{supptabwang} compares the derained results on SPA-Data of all competing methods quantitatively. It is easy to see that even for such complex and diverse rain patterns, the proposed RCDNet$_{1}$ with the simplest loss function still significantly outperforms other comparison methods.
\begin{table}[t]
\centering
\caption{Training-test domain match case: Average PSNR and SSIM comparisons on three datasets. Bold and bold italic indicate top $1^{\text{st}}$ and $2^{\text{nd}}$ best results, respectively.}\vspace{-3mm}
\setlength{\tabcolsep}{3.6pt}
\begin{tabular}{c|cc|cc|cc}
\hline
  Datasets & \multicolumn{2}{@{}c@{}}{Rain100L}&\multicolumn{2}{|c@{}}{Rain100H}&\multicolumn{2}{|c@{}}{Rain1400} \\
\hline
  Metrics & PSNR$\uparrow$ & SSIM$\uparrow$ & PSNR$\uparrow$ & SSIM$\uparrow$  & PSNR$\uparrow$ & SSIM$\uparrow$ \\
\hline
  Input & 26.90 & 0.8384 & 13.56 & 0.3709 & 25.24 & 0.8097 \\
\hline
  DSC\cite{Yu2015Removing} & 27.34 & 0.8494 & 13.77 & 0.3199  & 27.88 &0.8394 \\
\hline
  JCAS\cite{Gu2017Joint}  & 28.54 & 0.8524 & 14.62 & 0.4510 &26.20 & 0.8471 \\
\hline
  Clear\cite{Fu2017Clearing} &30.24 & 0.9344 & 15.33 & 0.7421 & 26.21& 0.8951 \\
\hline
  DDN\cite{Fu2017Removing}& 32.38 & 0.9258 & 22.85 & 0.7250 & 28.45 & 0.8888 \\
\hline
  RESCAN\cite{li2018recurrent}   & 38.52& 0.9812 &29.62 & 0.8720 &32.03& 0.9314\\
\hline
  PReNet\cite{ren2019progressive}& 37.54& 0.9795 &30.08& \textit{\textbf{0.9050}} & 32.09& 0.9418\\
\hline
  SPANet\cite{wang2019spatial} & 35.33 & 0.9694 &25.11 & 0.8332 & 29.85& 0.9148 \\
\hline
  JORDER\_E\cite{Yang2019Joint} &37.89 &0.9803& 30.21&0.8957 &32.00 & 0.9347 \\
\hline
  SIRR\cite{wei2019semi} & 32.37 & 0.9258 & 22.47 & 0.7164 & 28.44 & 0.8893\\
  \hline
RCDNet$_{1}$ & \emph{\textbf{39.90}} & \textit{\textbf{0.9855}} & \textit{\textbf{30.91}} & 0.9037 &  \textit{\textbf{32.78}} & \textit{\textbf{0.9446}}\\
\hline
CRCDNet & \textbf{40.00} & \textbf{0.9860} & \textbf{31.28} & \textbf{0.9093} & \textbf{33.04} & \textbf{0.9472}\\
\hline
\end{tabular}
\label{supptabsyn}
\end{table}
\begin{table}[t]
\centering
\caption{Training-test domain match case: Average PSNR and SSIM comparisons on SPA-Data~\cite{wang2019spatial}.}\vspace{-2mm}
\footnotesize
\setlength{\tabcolsep}{2.7pt}
\begin{tabular}{c|cccccc}
\hline
 Methods &Input  &DSC  &JCAS &Clear &DDN  & RESCAN\\
\hline
 PSNR$\uparrow$  &34.15     &34.95    &34.95  &34.39   &36.16  &38.11\\
\hline
 SSIM$\uparrow$  &0.9269    &0.9416   &0.9453   &0.9509   &0.9463 &0.9707  \\
\hline
 Methods  &PReNet  &SPANet  &JORDER\_E  &SIRR  &RCDNet$_{1}$ &RCDNet\\
\hline
 PSNR$\uparrow$    &40.16     &40.24    &40.78& 35.31 &\textit{\textbf{40.99}}    &{\textbf{41.47}}\\
\hline
 SSIM$\uparrow$    &\emph{\textbf{0.9816}}  &0.9811  &0.9811   &0.9411  & \textit{\textbf{0.9816}}   &\textbf{{0.9834}}\\
\hline
\end{tabular}
\label{supptabwang}
\end{table}
\begin{figure*}[t]
  \begin{center}
     \includegraphics[width=1\linewidth]{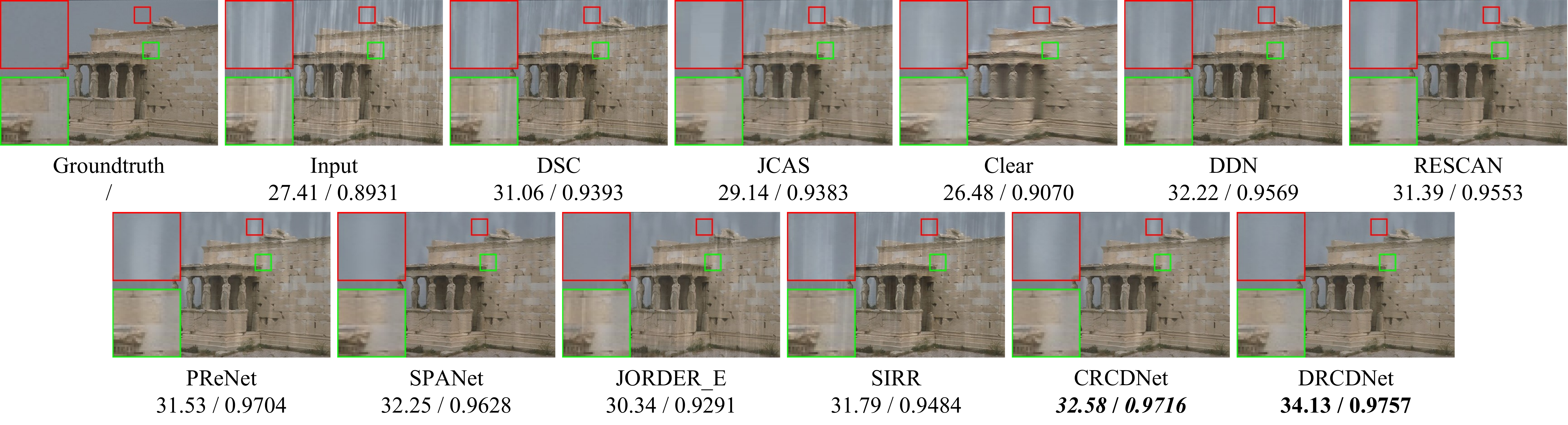}
  \end{center}
  \vspace{-5mm}
     \caption{\textcolor{black}{(Training-test domain mismatch case) From left/upper to right/lower: groundtruth and input rainy image from Dense10, derained results of traditional DSC and JCAS, generalized results of all deep competing methods trained on Rain100H.}}
  \label{suppfig13hdense10}
    \vspace{-3mm}
\end{figure*}
 \begin{figure*}[t]
  \begin{center}
     \includegraphics[width=1\linewidth]{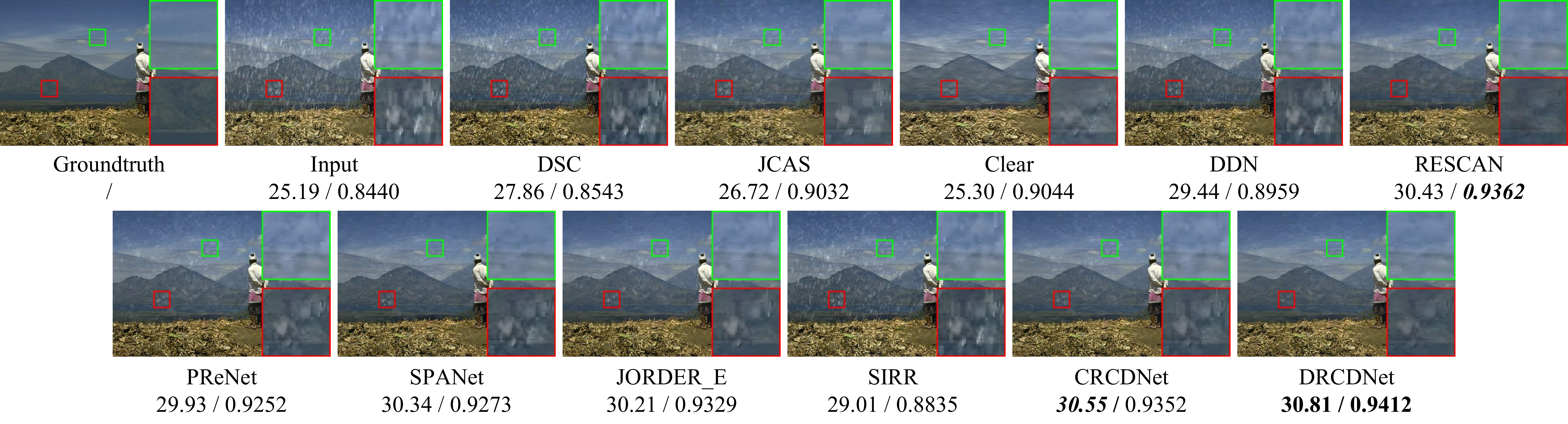}
  \end{center}
  \vspace{-4mm}
     \caption{(Training-test domain mismatch case) From left/upper to right/lower: groundtruth and input rainy image from Sparse10, derained results of traditional DSC and JCAS, generalized results of all deep competing methods trained on Rain100H, each with two demarcated areas zoomed in 4 times for easy observation. PSNR/SSIM results are included for easy reference.}
  \label{suppfightosparse10}
\end{figure*}
\begin{figure*}[t]
  \begin{center}
     \includegraphics[width=1\linewidth]{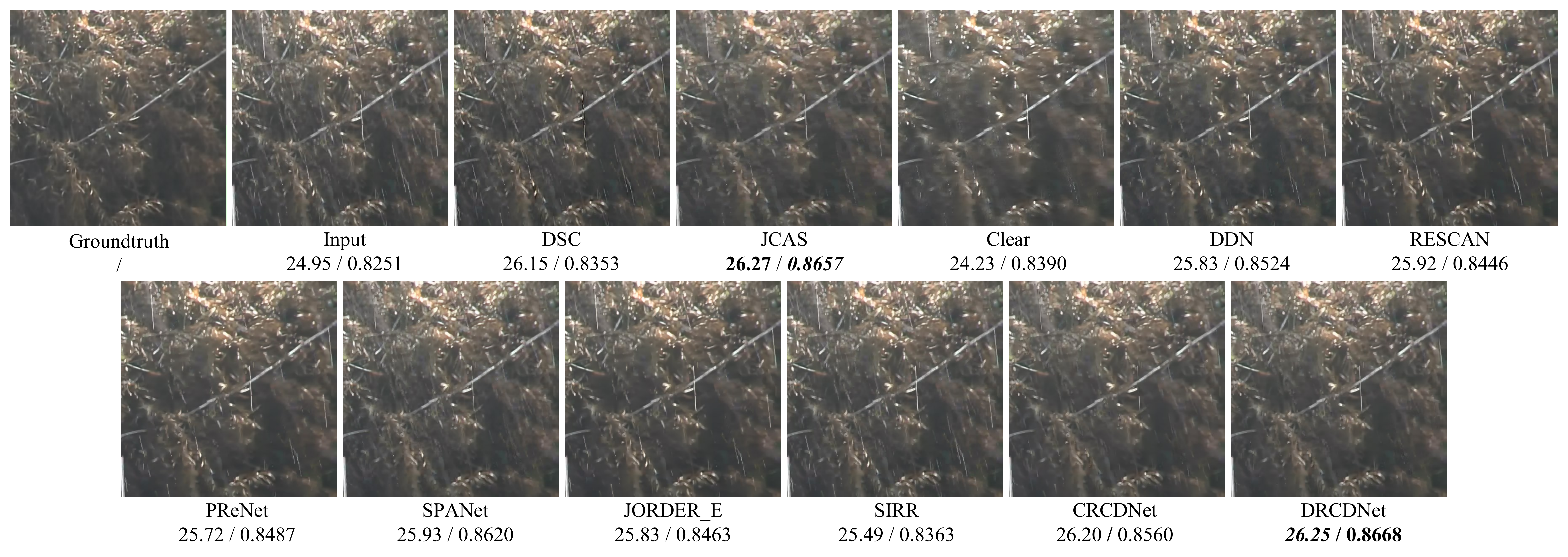}
  \end{center}
  \vspace{-4mm}
     \caption{(Training-test domain mismatch case) From left/upper to right/lower: groundtruth and input rainy image from SPA-Data, derained results of traditional DSC and JCAS, generalized results of all DL based competing methods trained on the training data of both Rain100L and Rain1400. PSNR/SSIM results are included for easy reference.}
  \label{suppfigl1400tospa}
\end{figure*}
\subsection{Training-test Domain Mismatch Experiments}

{\textcolor{black}{\textbf{Performance Comparison on Dense10.} Fig.~\ref{suppfig13hdense10} shows the derained results on the input rainy image from Dense10 which has quite different rain types from Rain100H, where all the DL competing methods are trained on Rain100H. In such an obvious domain mismatch testing case, almost all the approaches fail to remove mass rain streaks. However, with the flexible rain kernel prediction mechanism, our DRCDNet gets better visual effect than CRCDNet.}}

\textbf{Performance Comparison on Sparse10.} Fig.~\ref{suppfightosparse10} shows the derained results of all comparison derainers where the input rainy image is selected from Sparse10 and all DL competing methods are trained on Rain100H. Under the circumstance of such an obvious domain mismatch testing, almost all the comparison approaches fail to remove mass rain streaks. However, with the flexible rain kernel prediction mechanism, our DRCDNet gets better visual effect, which even evidently outperforms the traditional optimization model based SOTA methods, i.e., DSC and JCAS. It is worthy to be indicated that, at the testing phase only forward computations are implemented in DRCDNet. This can thus be seen as an unsupervised learning/inference manner, but without complicated iterative optimizations as in DSC and JCAS.

\textbf{Performance Comparison on SPA-Data.} Fig.~\ref{suppfigl1400tospa} displays the reconstructed images where every deep derainer is trained on the training data of both Rain100L and Rain1400. Clearly, as compared with other DL based baselines, the proposed DRCDNet performs better on both rain removal and detail preservation, which substantiates the favorable effectiveness of the proposed dynamic RCD model for rain streaks in this cross domain scenario.

\begin{figure*}[t]
  \begin{center}
     \includegraphics[width=1\linewidth]{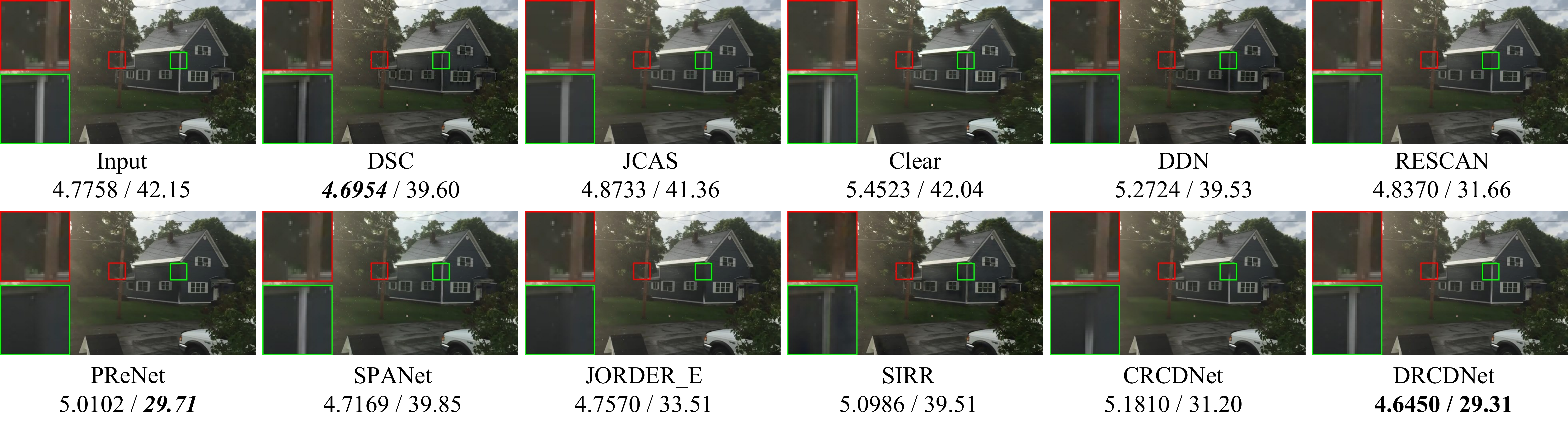}
  \end{center}
  \vspace{-4mm}
     \caption{\textcolor{black}{Training-test domain mismatch case: From left/upper to right/lower: input rainy image from Internet-Data, derained results of traditional DSC and JCAS, generalized results of all DL based competing methods trained on the training data of both Rain100L and Rain100H, each with two demarcated areas zoomed in 4 times for easy observation. NIQE/BRISQUE results are included for easy reference.}}
  \label{suppfig15lhwang}
    \vspace{-3mm}
\end{figure*}

\textcolor{black}{\textbf{Performance Comparison on Internet-Data.} Fig.~\ref{suppfig15lhwang} shows the derained results on a typical rainy sample from the real Internet-Data where deep derainers are trained on both Rain100L and Rain100H. As seen,  traditional DSC and JCAS, and some DL ones leave obvious rains in the restored backgrounds, and most of algorithms seriously blur the image details. Comparatively, our DRCDNet achieves the lowest BRISQUE and NIQE.}

\begin{figure*}[t]
  \begin{center}
     \includegraphics[width=1\linewidth]{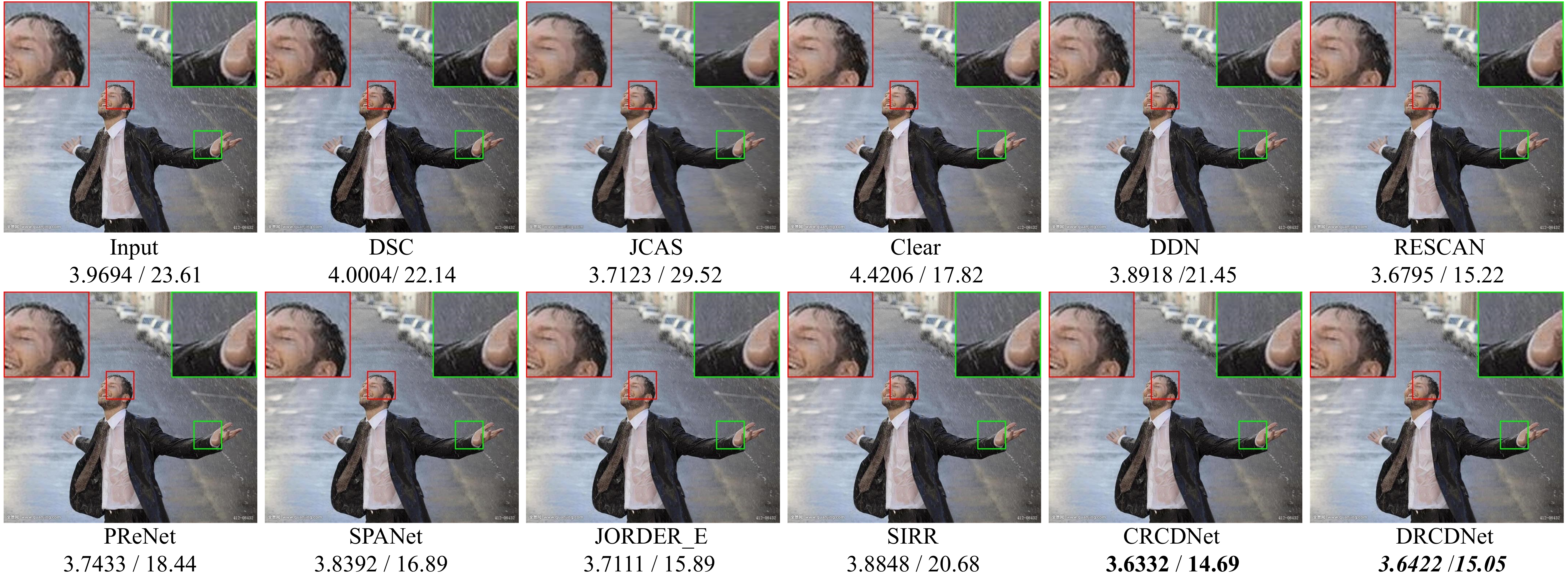}
  \end{center}
  \vspace{-4mm}
     \caption{(Training-test domain mismatch case): From left/upper to right/lower: input rainy image from MPID\_Rain+Mist(R), derained results of traditional DSC and
JCAS, generalized results of all DL based competing methods trained on  Rain100L. NIQE/BRISQUE results are included for easy reference.}
  \label{suppfigltompidmist}
\end{figure*}

\textbf{Performance Comparison on MPID\_Rain+Mist(R).} Fig.~\ref{suppfigltompidmist} shows the derained results on a typical rainy sample from the real MPID\_Rain+Mist(R) where deep derainers are trained on Rain100L. As seen, our proposed CRCDNet and DRCDNet achieve better visual quality. Table~\ref{supptabmpid} provides the quantitative comparison. Due to the large domain gap between Rain100L and MPID\_Rain+Mist(R), almost all the competing methods achieve relatively higher BRISQUE and NIQE. This is still room for further improving the generalization performance on real rain mist scenarios.

\section{More Model Verification}
\subsection{Model Verification for CRCDNet}
In this section, we provide more deraining results of CRCDNet based on more samples with diverse rain patterns and visualize the learned rain kernels, so as to fully verify the underlying mechanism of the proposed CRCDNet.

\begin{figure*}[t]
  \begin{center}
  \vspace{-2mm}
     \includegraphics[width=1.0\linewidth]{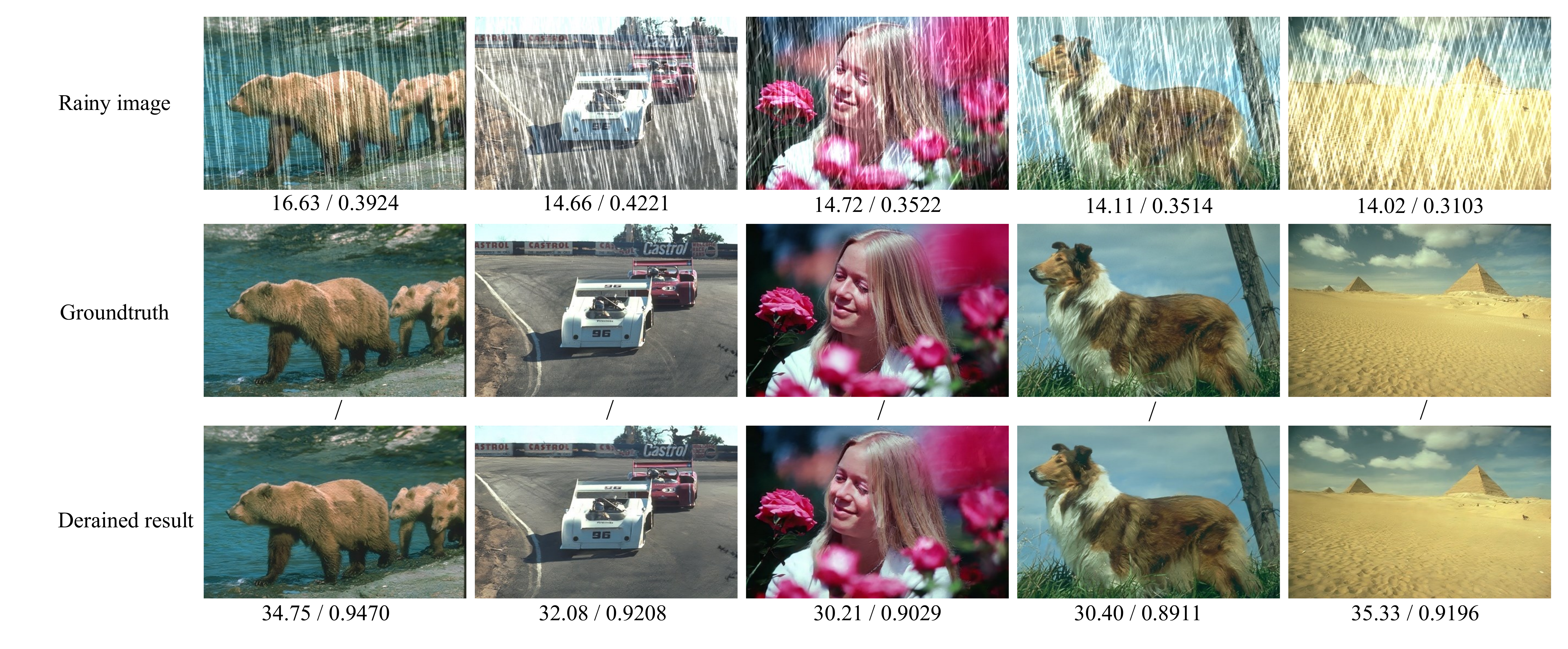}
  \end{center}
  \vspace{-3mm}
     \caption{(Training-test domain match case) (a) The first row: 5 representative rainy images with different rain densities and directions involved in Rain100H. (b) The second row: the corresponding groundtruth. (c) The third row: The derained results of our CRCDNet. PSNR/SSIM are listed below the corresponding images for easy reference.}
  \label{suppverifyh}
  \end{figure*}

\begin{figure*}[t]
  \begin{center}
     \includegraphics[width=1\linewidth]{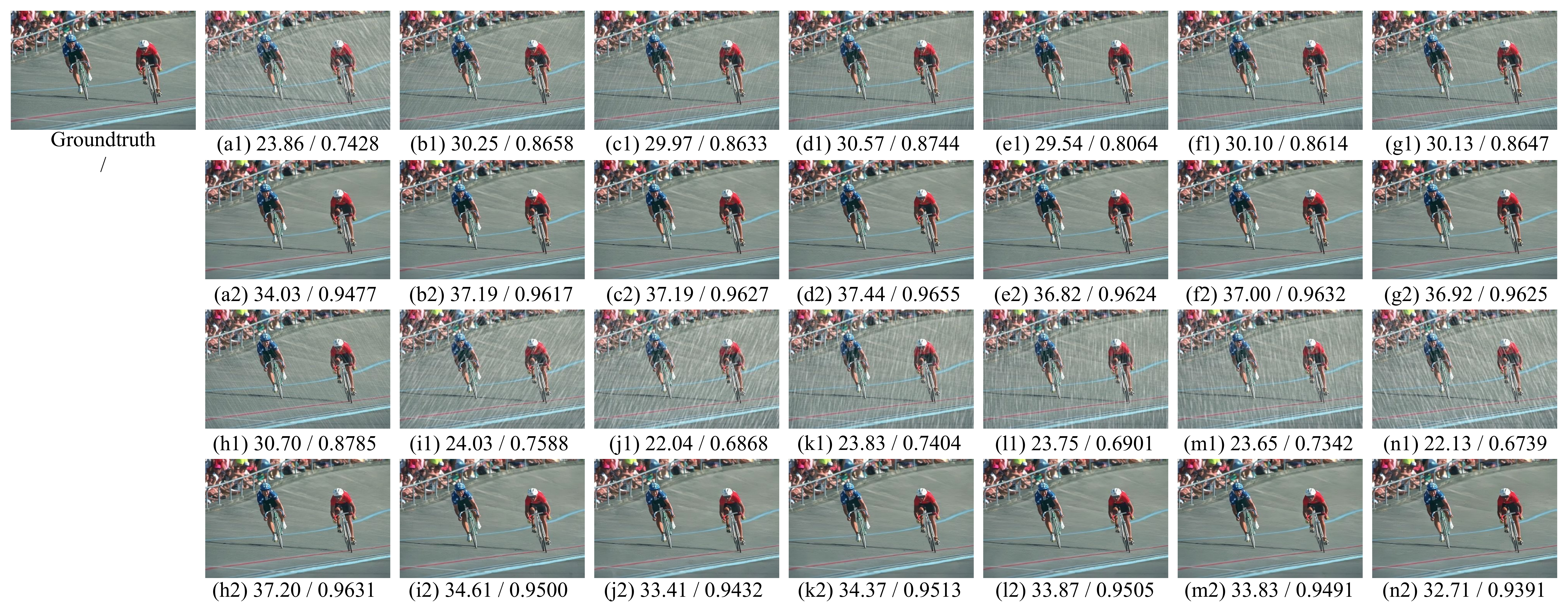}
  \end{center}
  \vspace{-3mm}
     \caption{(Training-test domain match case) (a1)-(n1) The input rainy images with 14 kinds of different rain streak orientations and magnitudes from Rain1400. (a2)-(n2) The corresponding derained results of the CRCDNet. The 14 inputs share the same groundtruth as displayed in the first column. PSNR/SSIM are listed below the corresponding images for easy reference. The images are better observed by zooming in on screen.}
  \label{suppverify1400}
  \end{figure*}

\begin{figure*}[t]
\begin{center}
  \vspace{-2mm}
     \includegraphics[width=1.0\linewidth]{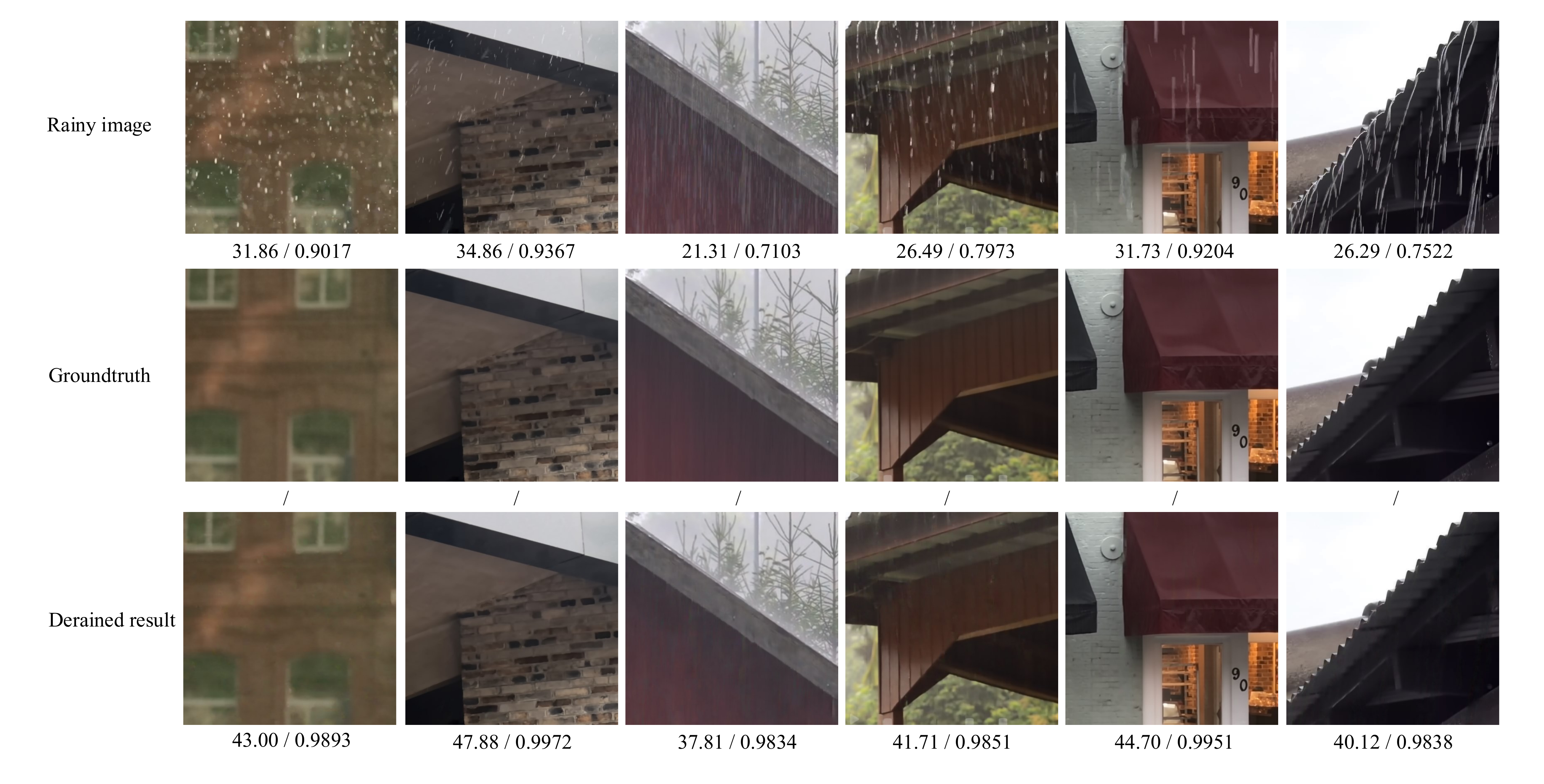}
  \end{center}
  \vspace{-3mm}
     \caption{(Training-test domain match case) (a) The first row: 6 rainy images representing different rain patterns from small raindrops to long/wide/thick strips involved in SPA-Data. (b) The second row: the corresponding groundtruth. (c) The third row: The derained results of our CRCDNet. PSNR/SSIM are listed below the corresponding images for easy reference. The images are better observed by zooming in on screen.}
  \label{suppverifywang}
  \end{figure*}

\begin{figure*}[t]
\vspace{-3mm}
\centering
\subfigure[Rain100H]{
\begin{minipage}[t]{0.15\linewidth}
\centering
\includegraphics[width=0.8in]{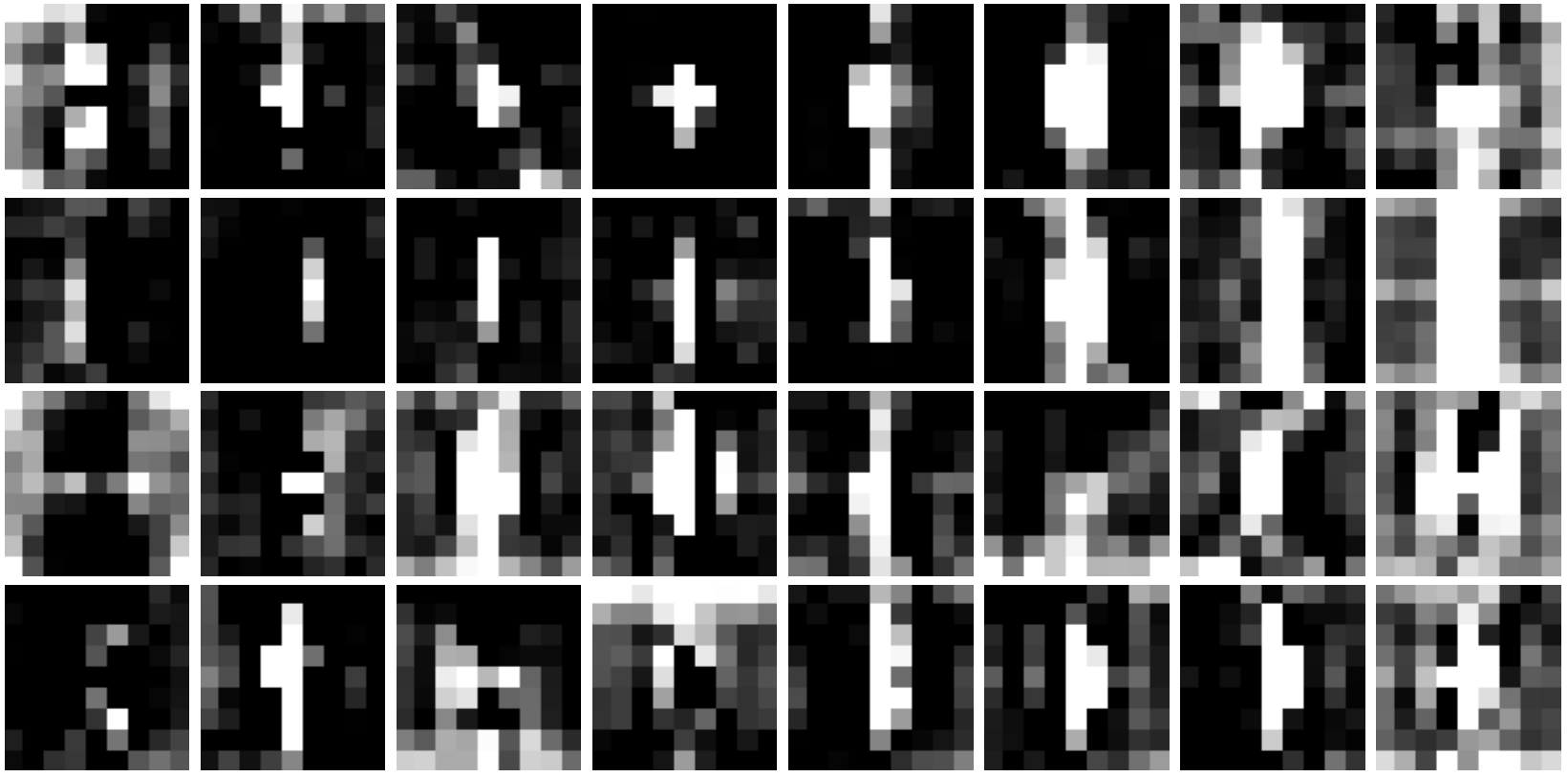}
\end{minipage}%
}%
\subfigure[Rain1400]{
\begin{minipage}[t]{0.15\linewidth}
\centering
\includegraphics[width=0.8in]{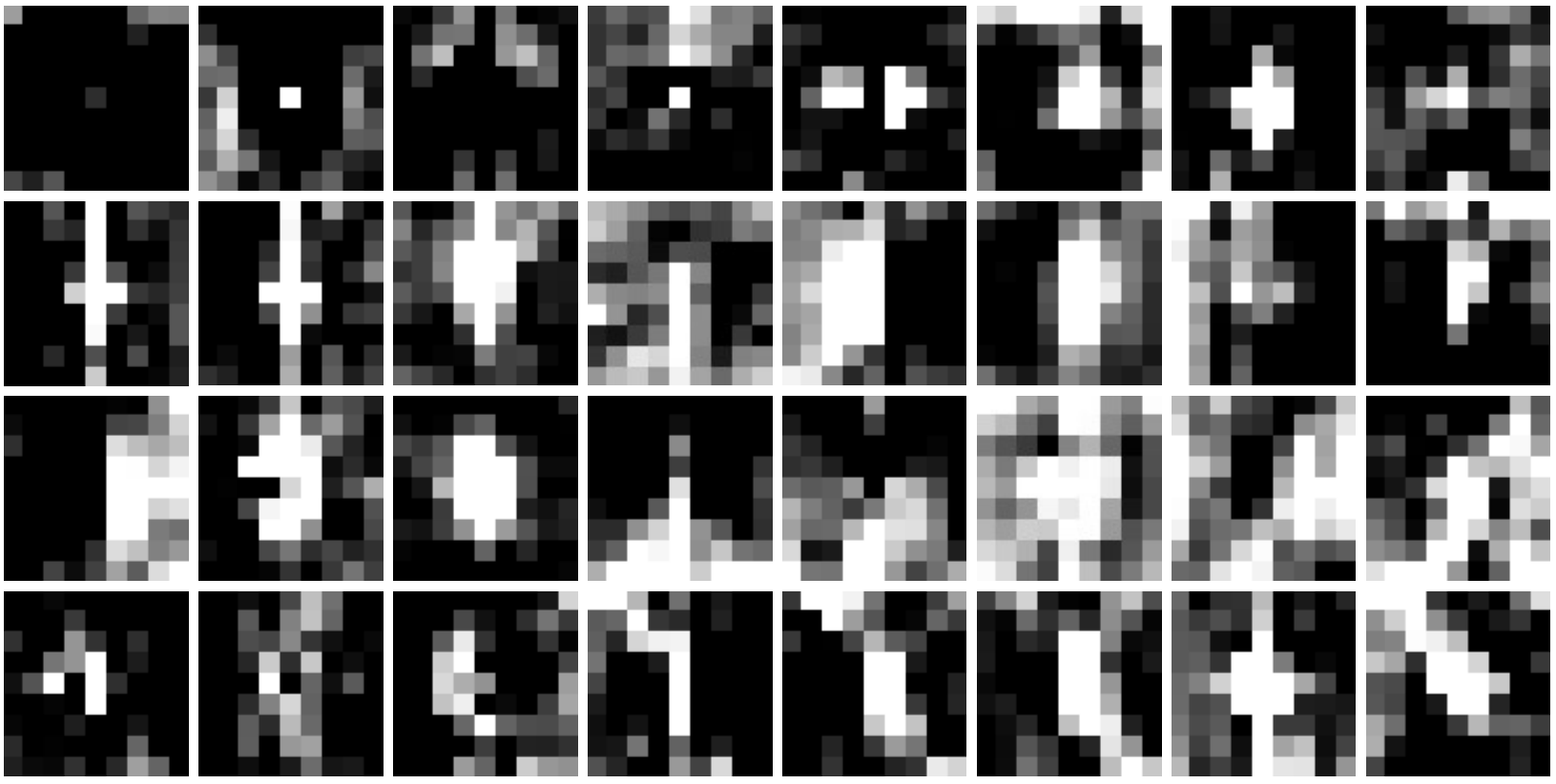}
\end{minipage}%
}%
\subfigure[SPA-Data]{
\begin{minipage}[t]{0.15\linewidth}
\centering
\includegraphics[width=0.8in]{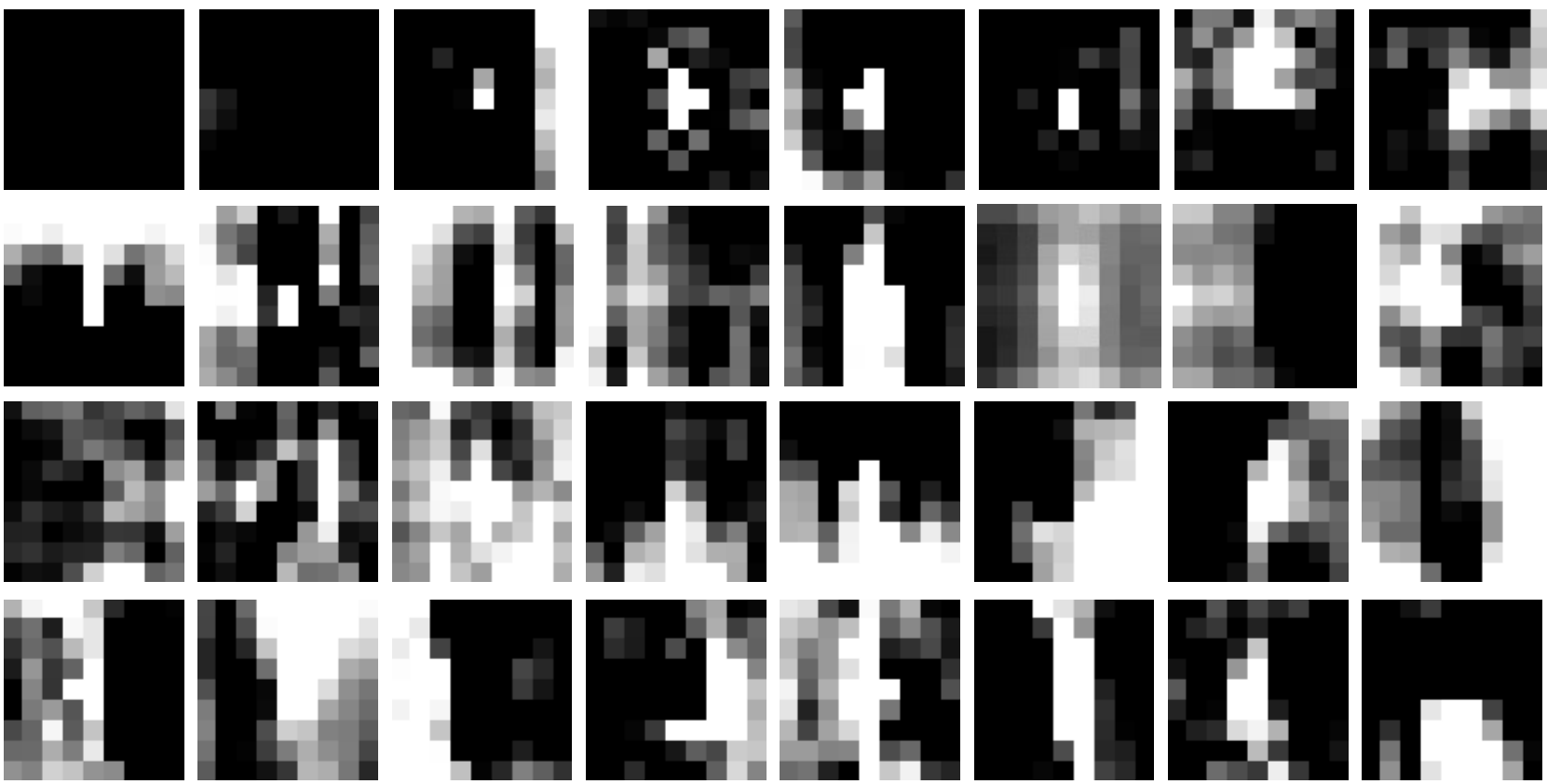}
\end{minipage}
}%
\centering
\vspace{-3mm}
\caption{Rain kernels ${\mathcal{K}}=\{\mathcal{K}_{n}\}_{n=1}^{N=32}$ learned from Rain100H, Rain1400, and SPA-Data, respectively, by the proposed CRCDNet.}
\label{suppverifyc}
\end{figure*}

\textbf{Training-test domain match case.} For the three datasets, including Rain100H with 5 types of rain streaks, Rain1400 with 14 kinds of ones, and SPA-Data, we correspondingly select several rainy images with representative rain patterns as shown in Fig.~\ref{suppverifyh}, Fig.~\ref{suppverify1400}, and Fig.~\ref{suppverifywang}, respectively. From the visual and quantitative demonstration, we can easily observe that even for such complex rain patterns, CRCDNet always attains excellent deraining performance. This substantiates the good flexibility and generality of our method in the  training-test domain match case. Besides, the corresponding rain kernels for these datasets extracted by our methods are shown in Fig.~\ref{suppverifyc}. Clearly, it is observed that the shapes of rain kernels are different among the three datasets, and the rain kernels learned from Rain1400 and SPA-Data are more diverse than that learned from Rain100H. This fits faithfully with the diversity of these datasets. It confirms that our method is indeed capable of automatically learning diverse rain kernels that are potentially useful for the related tasks on rainy images.

\begin{table*}
\vspace{-0mm}
\begin{center}
\caption{Training-test domain mismatch case: average BRISQUE and NIQE comparisons on MPID\_rain+mist. }\label{supptabtowangreal} 
\vspace{-1.5mm}
\setlength{\tabcolsep}{5.5pt}
\begin{tabular}{c|cccccccccccc}
\Xhline{0.9pt}
&   Input &DSC&JCAS&Clear&DDN & RESCAN & PReNet &SPANet & JORDER\_E  & SIRR & CRCDNet &DRCDNet\\
 \Xhline{0.9pt}
    \multicolumn{13}{c}{\scriptsize{Data (training/testing):  Rain100L/MPID\_Rain+Mist(R); ~Difficulty: high.}}\\
    \hline
  BRISQUE$\downarrow$ & 23.05 &22.54 &30.94 &24.89 &23.09 & 23.02 &23.01 &22.73&\textbf{\textit{22.46}} & 23.23 &\textbf{22.11} &22.61\\
  NIQE$\downarrow$ & 3.4086 &3.4537  & 3.8667 &3.6528 &3.3935& 3.3479 &3.3323 &3.3339 &3.3314 &3.3985 &\textit{\textbf{3.3065}} &\textbf{3.3021}\\
\Xhline{0.9pt}
\end{tabular}
\label{supptabmpid}
\normalsize
\end{center}
\vspace{-0mm}
\end{table*}
\subsection{Model Verification for DRCDNet}
\begin{figure*}[t]
  \begin{center}
     \includegraphics[width=1.0\linewidth]{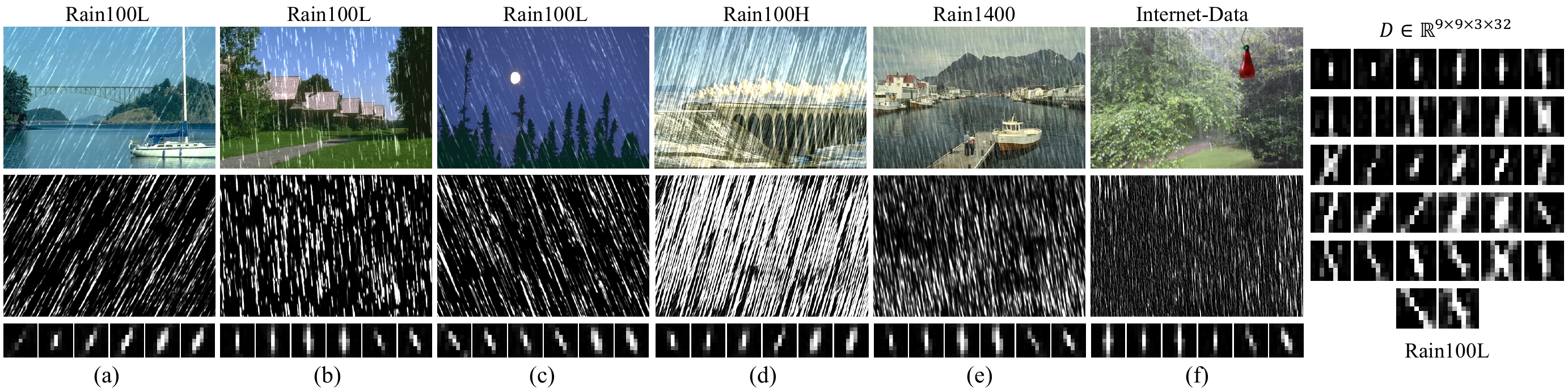}
  \end{center}
  \vspace{-5mm}
     \caption{The left: for each column in (a)-(f), ($1^{\text{st}}$ row) rainy images, ($2^{\text{nd}}$ row)  the extracted rain layers,  and ($3^{\text{rd}}$ row)  the corresponding rain kernels dynamically predicted by DRCDNet. The right:  the rain kernel dictionary ${\mD}$ learned by DRCDNet based on Rain100L training set. Especially, the rainy images in (a)-(c) are from Rain100L testing set and the ones in (d)-(f) are from other testing set.}
  \label{suppfigverdrcd}
\end{figure*}

\textbf{Generalize to synthesized data and Internet-Data.}
By training DRCDNet on Rain100L, the learned rain kernel dictionary $\mathcal{D}$ is shown at the right of Fig.~\ref{suppfigverdrcd}. With the trained model on Rain100L, we test typical rainy samples from different sources, including training/test domain match cases (a)-(c) and mismatch cases (d)-(f). As shown in each column of (a)-(f), the extracted rain layers ($2^{\text{nd}}$ row) contain less background details, and the inferred rain kernels ($3^{\text{rd}}$ row) are finely in accordant with the rain patterns (e.g., directions, scales, thickness) in input rainy images ($1^{\text{st}}$ row). Besides, we can also observe that the rain kernels ($3^{\text{rd}}$ row) for every testing sample are not simply selected from $\mathcal{D}$, and they are adaptively inferred by DRCDNet, even with new rain patterns not in $\mathcal{D}$. This not only validates the effectiveness of the dynamic RCD modelling manner for rain layer, but also reflects the advantages of the DRCDNet over adaptive inference. Compared with CRCDNet where rain kernels are always fixed as the dictionary $\mK$ for different testing rainy images, such adaptive prediction mechanism makes DRCDNet have potential to obtain better generalization capability.
  \begin{figure*}[!htp]
  \begin{center}
     \includegraphics[width=0.7\linewidth]{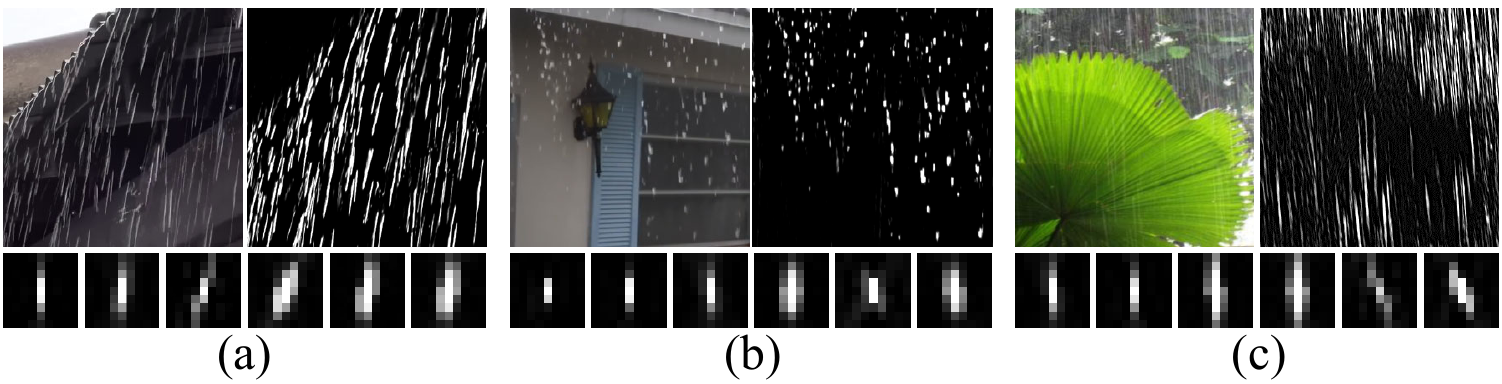}
  \end{center}
  \vspace{-6mm}
     \caption{ (Training-test domain mismatch case) $1^{\text{st}}$ row: real rainy images from SPA-Data and extracted rain layers. $2^{\text{nd}}$ row: rain kernels dynamically predicted  by the proposed DRCDNet trained on synthetic Rain100L.}
  \label{suppdynamickernel}
\end{figure*}

\textbf{Generalize to SPA-Data.} Fig.~\ref{suppdynamickernel} presents the generalization results from Rain100L to SPA-Data for DRCDNet. As seen, rain kernels are adaptively inferred according to the rain types of input rainy images. Besides, since the intrinsic RCD prior model of rain streaks is explicitly encoded into each stage of DRCDNet, the extracted rain layer at any stage of DRCDNet always complies with the physical constraints. Then the learned rain layer inclines to have better reliability and accuracy.

\end{document}